\documentclass[12pt]{article}
\usepackage{latexsym}
\usepackage{amsmath,amsfonts}
\usepackage{times}
\allowdisplaybreaks[4]

\catcode`\@=11

\newcount\hour
\newcount\minute
\newtoks\amorpm \hour=\time\divide\hour by 60\minute
=\time{\multiply\hour by 60 \global\advance\minute by-\hour}
\edef\standardtime{{\ifnum\hour<12 \global\amorpm={am}%
        \else\global\amorpm={pm}\advance\hour by-12 \fi
        \ifnum\hour=0 \hour=12 \fi
        \number\hour:\ifnum\minute<10
        0\fi\number\minute\the\amorpm}}
\edef\militarytime{\number\hour:\ifnum\minute<10
0\fi\number\minute}

\def\draftlabel#1{{\@bsphack\if@filesw {\let\thepage\relax
   \xdef\@gtempa{\write\@auxout{\string
      \newlabel{#1}{{\@currentlabel}{\thepage}}}}}\@gtempa
   \if@nobreak \ifvmode\nobreak\fi\fi\fi\@esphack}
        \gdef\@eqnlabel{#1}}
\def\@eqnlabel{}
\def\@vacuum{}
\def\marginnote#1{}
\def\draftmarginnote#1{\marginpar{\raggedright\scriptsize\tt#1}}
\def\draft{
        \pagestyle{plain}
        \overfullrule=2pt
        \oddsidemargin -.5truein
        \def\@oddhead{\sl \phantom{\today\quad\militarytime} \hfil
        \smash{\Large\sl DRAFT} \hfil \today\quad\militarytime}
        \let\@evenhead\@oddhead
        \let\label=\draftlabel
        \let\marginnote=\draftmarginnote
        \def\ps@empty{\let\@mkboth\@gobbletwo
        \def\@oddfoot{\hfil \smash{\Large\sl DRAFT} \hfil}
        \let\@evenfoot\@oddhead}
        \def\@eqnnum{(\theequation)\rlap{\kern\marginparsep\tt\@eqnlabel}%
        \global\let\@eqnlabel\@vacuum}  }

\newcommand{\rf}[1]{(\ref{#1})}
\renewcommand{\theequation}{\thesection.\arabic{equation}}
\renewcommand{\thefootnote}{\fnsymbol{footnote}}
\newcommand{\newsection}{    
\setcounter{equation}{0}\section}

\def\appendix#1{\addtocounter{section}{1}\setcounter{equation}{0}
\renewcommand{\thesection}{\Alph{section}}
\section*{Appendix \thesection\protect\indent \parbox[t]{11.15cm}{#1}}
\addcontentsline{toc}{section}{Appendix \thesection\ \ \ #1}}

\def\Asf{{\sf A}}
\def\Bsf{{\sf B}}

\def\asf{{\sf a}}
\def\bsf{{\sf b}}
\def\csf{{\sf c}}
\def\esf{{\sf e}}
\def\Nsf{{\sf N}}

\def\nline{\,\nabla\kern -0.7em\raise0.2ex\hbox{/}\,\,}
\def\yline{\,y\kern -0.47em /}
\def\aline{\,a\kern -0.49em /}
\def\parline{\,\partial\kern -0.55em /\,\,}

\newcommand{\Mo}{\mathbb{M}}
\newcommand{\No}{\mathbb{N}}
\newcommand{\Po}{\mathbb{P}}

\newcommand{\Zo}{\mathbb{Z}}

\def\be{\begin{equation}}
\def\ee{\end{equation}}
\def\beq{\begin{eqnarray}}
\def\eeq{\end{eqnarray}}

\def\Esm{{\scriptscriptstyle E}}
\def\Rsm{{\scriptscriptstyle R}}
\def\Lsm{{\scriptscriptstyle L}}

\def\smpt{{\scriptscriptstyle [2]}}
\def\smp3{{\scriptscriptstyle [3]}}

\def\smpn{{\scriptscriptstyle [n]}}
\def\Thsm{{\scriptscriptstyle {\rm Th}}}

\def\nbf{{\bf n}}
\def\pbf{{\bf p}}

\def\Jbf{{\bf J}}

\def\Lbf{{\bf L}}
\def\Mbf{{\bf M}}
\def\Nbf{{\bf N}}
\def\Pbf{{\bf P}}

\def\Xbf{{\bf X}}
\def\Vbf{{\bf V}}

\def\ibf{{\bf i}}
\def\iibf{{\bf ii}}
\def\iiibf{{\bf iii}}
\def\ivbf{{\bf iv}}

\def\AA{{\cal A}}

\def\II{{\cal I}}

\def\XX{{\cal X}}

\def\Nsf{{\sf N}}

\def\half{\frac{1}{2}}

\def\alphab{\bar{\alpha}}

\def\Nb{{\bar{N}}}

\def\Vb{{\bar{V}}}
\def\VVb{{\bar{\cal V}}}
\def\vb{{\bar{v}}}

\def\ab{{\bar{a}}}

\def\bas{{\rm bas}}
\def\irm{{\rm i}}

\def\dyn{{\rm dyn}}
\def\minrm{{\rm min}}

\def\diff{{\rm diff}}

\def\for{{\rm for}}
\def\fix{{\rm fix}}

\def\free{{\rm free}}
\def\sign{{\rm sign}}
\def\sol{{\rm sol}}
\def\tr{{\rm tr}}
\def\prj{{\rm prj}}
\def\on-shell{{\rm on-shell}}
\def\on-sh{{\rm on-sh}}

\def\ub{\bar{u}}
\def\vb{\bar{v}}

\def\betach{\check{\beta}}

\def\noinbf#1{\noindent {\bf #1}}

\begin{document}


\begin{flushright}
FIAN-TD-2022-02  \ \ \ \ \ \ \ \\
arXiv: 2206.13268 \ V2
\end{flushright}

\vspace{1cm}

\begin{center}

{\Large \bf Interacting massive and massless arbitrary spin

\medskip
fields in 4d flat space}

\vspace{2.5cm}

R.R. Metsaev\footnote{ E-mail: metsaev@lpi.ru }

\vspace{1cm}

{\it Department of Theoretical Physics, P.N. Lebedev Physical
Institute, \\ Leninsky prospect 53,  Moscow 119991, Russia }

\vspace{3cm}

{\bf Abstract}

\end{center}

Massive and massless arbitrary integer spin fields propagating in four-dimensional flat space are studied. The massive and massless fields are treated by using a light-cone gauge helicity basis formalism.
Cubic cross-interactions between massive and massless fields and cubic interactions between massive fields are investigated. We introduce a classification of such cubic interactions  and using our classification we build all cubic interaction vertices. Realization of generators of the Poincar\'e algebra on space of interacting fields is  found. As a by-product, some illustrative examples of light-cone form for 3-point invariant amplitudes of massive and massless fields are also discussed.

\vspace{3cm}


\newpage
\renewcommand{\thefootnote}{\arabic{footnote}}
\setcounter{footnote}{0}

\section{ \large Introduction}

The light-cone gauge formulation of relativistic dynamics originally proposed in Ref.\cite{Dirac:1949cp} turned out to useful for a study of many aspects of quantum field theory and string theory. For example, we mention the application of this approach for the investigation of ultraviolet finiteness of $N=4$ supersymmetric Yang-Mills theory in Refs.\cite{Brink:1982wv,Mandelstam:1982cb}.
The light-cone gauge approach offers also certain simplifications of approaches  to building string field theory \cite{Kaku:1974zz,Green:1983hw}. Use of the light-cone approach for the study of superfield formulation of IIB 10d supergravity and 11d supergravity may be found in Refs.\cite{Green:1982tk,Metsaev:2004wv}.

The extensive use of the light-cone gauge approach for the study of higher-spin massless fields begun in Ref.\cite{Bengtsson:1983pd}, where the cubic vertices for the Yang-Mills-like interactions of higher-spin massless fields in $R^{3,1}$ space were built. The full list of cubic vertices for arbitrary integer spin massless fields in $R^{3,1}$ space was obtained later on in Ref.\cite{Bengtsson:1986kh}. Our aim in this paper is to extend the results in Ref.\cite{Bengtsson:1986kh} to the case of  arbitrary integer spin massive and massless fields propagating in $R^{3,1}$ space. Namely, in this paper, we are interested in cubic vertices for the cross-interactions between massive and massless fields and cubic vertices for the interactions between massive fields. We provide a classification for such cubic interactions and, using our classification, we build all cubic vertices for the massive and massless fields.%
\footnote{ Light-cone gauge parity-even cubic vertices for arbitrary spin massive and massless fields in $R^{d-1,1}$, $d\geq  4$, were studied in Ref.\cite{Metsaev:2005ar,Metsaev:2007rn}. Lorentz covariant parity-even cubic vertices for arbitrary spin massive and massless on-shell TT fields in $AdS_d$, $d\geq 4$, were considered in Ref.\cite{Joung:2012rv}. BRST-BV parity-even cubic vertices for arbitrary spin massive and massless fields in $R^{d-1,1}$, $d\geq 4$, were obtained in Ref.\cite{Metsaev:2012uy}.}
To our knowledge, our classification for the cubic vertices of massive fields is novel and has not previously been reported in the literature.
We note that, in Ref.\cite{Bengtsson:1986kh}, to build cubic vertices, the higher-spin massless fields were considered by using the light-cone gauge helicity basis formalism. As it is well known, the light-cone gauge helicity basis formalism for massless fields admits the straightforward generalization to the case of massive fields. We use such generalized light-cone gauge helicity basis formalism for building cubic vertices that involve massive fields.

This paper is organized as follows.

In Sec.\ref{sec-02}, we review a light-cone gauge helicity basis formulism for massive and massless fields propagating in $R^{3,1}$.
Also we review a field-theoretical realization of the Poincar\'e algebra on space of the massive and massless fields.

In Sec.\ref{sec-03}, we discuss $n$-point interaction vertices. We present restrictions on $n$-point interaction vertices which are obtained by using kinematical symmetries of the Poincar\'e algebra.

In Sec.\ref{sec-04}, we investigate cubic vertices. We start with the presentation of restrictions on the cubic vertices which are obtained by considering kinematical and dynamical symmetries of the Poincar\'e algebra. After that, we introduce a light-cone gauge dynamical principle and formulate our complete system of equations which allows us to fix all solutions for cubic vertices uniquely.

In Sec.\ref{sec-05}, we introduce harmonic representatives of the cubic vertices. We reformulate our complete system of equations for the cubic vertices in terms of the harmonic vertices. After that we introduce meromorphic vertices which are in one-to-one correspondence with the harmonic vertices. We reformulate all our equations for the harmonic vertices in terms of the meromorphic vertices.

In Sec.\ref{sec-06}, we discuss our classification for the cubic interactions.

In Sec.\ref{sec-07}, we discuss the interactions between two massless fields and one massive field, while, in
Secs.\ref{sec-08},\ref{sec-09}, we discuss the interactions between one massless field and two massive fields. Secs.\ref{sec-10}, \ref{sec-11} are devoted to the interactions between three massive fields. In Secs.\ref{sec-07}-\ref{sec-11}, we present the explicit expressions for the meromorphic vertices and discuss some illustrative examples of 3-point invariant amplitudes in the light-cone frame.

Our conclusions are summarized in Sec.\ref{concl}.

In Appendix A, we describe our notation and various identities we use in our study  of cubic vertices. In Appendix B, we discuss various helpful realizations for the spin operators of massive fields. In Appendix C, we discuss the incorporation of internal symmetries in our approach. In Appendix D, we describe the hermitian conjugation rules for our vertices.  Appendix E is devoted to technical details of the derivation of equations for the harmonic vertices and the meromorphic vertices. Appendix F is devoted to review of 3-point invariant amplitudes in the light-cone frame. In Appendix G, we make comments on our classification of cubic vertices.  Appendix H is devoted to the derivation of the meromorphic vertices.

\newsection{ \large Light-cone gauge helicity basis formalism for free massive and massless fields }\label{sec-02}

\noindent {\bf Light-cone frame of Poincar\'e algebra $iso(3,1)$}. We follow a method proposed in Ref.\cite{Dirac:1949cp}. According to this method, the problem of building a new light-cone gauge formulation of a dynamical system amounts to the problem of building the respective light-cone gauge solution for commutators of an underlying symmetry algebra.
For the massive and massless fields propagating in $R^{3,1}$ space, the underlying    symmetries are associated with the Poincar\'e algebra $iso(3,1)$. Therefore, let us  review the well known light-cone gauge realization of the Poincar\'e algebra on a space of the massive and massless fields in $R^{3,1}$ space.

In $R^{3,1}$ space, the Poincar\'e algebra $iso(3,1)$ is spanned by the four translation generators denoted as $P^\mu$ and the six generators of the $so(3,1)$ Lorentz algebra denoted as $J^{\mu\nu}$. We use the following non-trivial commutators of the Poincar\'e algebra:
\be \label{18042022-man-01}
[P^\mu,\,J^{\nu\rho}]=\eta^{\mu\nu} P^\rho - \eta^{\mu\rho} P^\nu\,,
\qquad {} [J^{\mu\nu},\,J^{\rho\sigma}] = \eta^{\nu\rho} J^{\mu\sigma} + 3\hbox{ terms}\,,
\ee
where $\eta^{\mu\nu}$ stands for the mostly positive Minkowski metric.
Let $x^\mu$, $\mu=0,1,2,3$, be the Lorentz basis coordinates. The light-cone frame  coordinates $x^\pm$, $x^\Rsm$, $x^\Lsm$ are defined then as
\be \label{18042022-man-02}
x^\pm \equiv \frac{1}{\sqrt{2}}(x^3  \pm x^0)\,,\qquad
\qquad x^\Rsm \equiv \frac{1}{\sqrt{2}}(x^1 + \irm x^2)\,,\qquad x^\Lsm \equiv \frac{1}{\sqrt{2}}(x^1 - \irm x^2)\,.
\ee
The $x^+$ is considered as a light-cone evolution parameter. In the light-cone frame, the $so(3,1)$ Lorentz algebra vector $X^\mu$ is represented as $X^+,X^-,X^\Rsm$, $X^\Lsm$, while a scalar product of two vectors $X^\mu$ and $Y^\mu$ is represented as
\be  \label{18042022-man-03}
\eta_{\mu\nu}X^\mu Y^\nu = X^+Y^- + X^-Y^+ + X^\Rsm Y^\Lsm + X^\Lsm Y^\Rsm\,.
\ee
Relation \rf{18042022-man-03} implies that the non-vanishing elements of $\eta_{\mu\nu}$ are given by $\eta_{+-} = \eta_{-+}=1$, $\eta_{\Rsm\Lsm} = \eta_{\Lsm\Rsm} = 1$. This implies the following interrelations for covariant and contravariant components of the vector $X^\mu$: $X^+=X_-$, $X^-=X_+$, $X^\Rsm=X_\Lsm$, $X^\Lsm=X_\Rsm$.

In the light-cone frame, the generators of the Poincar\'e algebra can be split into two groups:
\beq
\label{18042022-man-04} && \hspace{-1.6cm}
P^+,\quad
P^\Rsm,\hspace{0.6cm}
P^\Lsm,\hspace{0.5cm}
J^{+\Rsm},\quad
J^{+\Lsm},\quad
J^{+-},\quad
J^{\Rsm\Lsm}, \quad
\hbox{ kinematical generators};
\\
\label{18042022-man-05}  && \hspace{-1.6cm}
P^-, \quad
J^{-\Rsm}, \quad
J^{-\Lsm}, \quad
\hspace{5.2cm} \hbox{ dynamical generators}.
\eeq
Needless to say that, in the light-cone frame, commutation relations of the Poincar\'e algebra are obtained from the ones in \rf{18042022-man-01} by using  $\eta^{\mu\nu}$ which has non-vanishing elements $\eta^{+-}=\eta^{-+}=1$, $\eta^{\Rsm\Lsm}=\eta^{\Lsm\Rsm}=1$. The Poincar\'e algebra generators are assumed to satisfy the following hermitian conjugation rules:
\be \label{18042022-man-06}
P^{\pm \dagger} = P^\pm, \qquad \ \
P^{\Rsm\dagger} = P^\Lsm, \qquad
J^{\Rsm\Lsm\dagger} =  J^{\Rsm\Lsm}\,,\quad
J^{+-\dagger} = - J^{+-}, \quad
J^{\pm \Rsm\dagger} = -J^{\pm \Lsm}\,.
\ee
We now discuss the light-cone gauge helicity basis formalism for arbitrary spin massive and massless fields.

\noindent {\bf Massless fields}. In the light-cone frame, a spin-$s$ massless field is described by complex-valued helicity basis fields $\phi_\lambda(x)$, $\lambda=\pm s$, where the label $\lambda$ stands for a helicity, while the argument $x$ stands for the space-time coordinates $x^\pm$, $x^{\Rsm,\Lsm}$ \rf{18042022-man-02}. The hermitian conjugation rule is given by
\be \label{18042022-man-07}
(\phi_\lambda(x))^\dagger = \phi_{-\lambda}(x)\,.
\ee

\noindent {\bf Massive spin-$s$ field}. In the light-cone frame, a mass-$m$ and spin-$s$  field can be described by the following set of complex-valued helicity basis fields:
\beq
\label{18042022-man-08} && \phi_{m,s;n}(x)\,, \hspace{1cm} n= 0,\pm 1,\ldots, \pm s\,,
\\
\label{18042022-man-09} && (\phi_{m,s;n}(x))^\dagger = \phi_{m,s;-n}(x)\,.
\eeq
To obtain the description of the massive field in an easy--to--use form,  we introduce the bosonic spinor-like creation operators $u$, $v$  and the respective annihilation operators $\ub$, $\vb$ defined by the relations
\beq
\label{18042022-man-10} && [\ub,u]=1\,,\qquad [\vb,v] = 1\,, \qquad v^\dagger = \vb\, \qquad u^\dagger = \ub\,,\qquad\ub|0\rangle=0\,,\qquad \vb|0\rangle=0\,.\qquad\qquad
\eeq
Often, we refer to $u$ and $v$ as oscillators. Using the oscillators, we collect fields \rf{18042022-man-08} into a ket-vector,
\be \label{18042022-man-12}
|\phi_{m,s}(x)\rangle = \sum_{n=-s}^s \frac{ u^{s+n} v^{s-n} }{ \sqrt{(s+n)!(s-n)!} } \phi_{m,s;n}(x)|0\rangle\,.
\ee

Below, we prefer to use momentum space fields which are obtainable from the ones above discussed by making the Fourier transform with respect to the spatial coordinates $x^-$, $x^\Rsm$, and $x^\Lsm$,
\beq
\label{18042022-man-13} && \phi_\lambda(x) = \int \frac{ d^3p }{ (2\pi)^{3/2} } e^{\irm(\beta x^- + p^\Rsm x^\Lsm  + p^\Lsm x^\Rsm)} \phi_\lambda(x^+,p)\,,
\\
\label{18042022-man-14} && |\phi_{m,s}(x)\rangle = \int \frac{ d^3p }{ (2\pi)^{3/2} } e^{\irm (\beta x^- + p^\Rsm x^\Lsm  + p^\Lsm x^\Rsm)} |\phi_{m,s}(x^+,p)\rangle\,, \qquad d^3p \equiv d\beta dp^\Rsm dp^\Lsm\,,\qquad
\eeq
where the argument $p$ stands for the momenta $\beta$, $p^\Rsm$, $p^\Lsm$. In terms of the fields $\phi_\lambda(x^+,p)$, $\phi_{m,s;n}(x^+,p)$, the hermicity conditions \rf{18042022-man-07},\rf{18042022-man-09} are represented as
\be \label{18042022-man-15}
(\phi_\lambda(x^+,p))^\dagger = \phi_{-\lambda}(x^+,-p)\,,\qquad (\phi_{m,s;n}(x^+,p))^\dagger = \phi_{m,s;-n}(x^+,-p)\,.
\ee

\noindent {\bf Realization of the Poincar\'e algebra on massless and massive fields}. The realization of the Poincar\'e algebra \rf{18042022-man-01} in terms of differential operators acting on the massless field $\phi_\lambda(p)$ and the massive field $|\phi_{m,s}(x^+,p)\rangle$ takes the following well known form:
\beq
&& \hbox{ \it Realizations on space of $\phi_\lambda(x^+,p)$ and $|\phi_{m,s}(x^+,p)\rangle$}:
\nonumber\\[-7pt]
\label{18042022-man-16}  && P^\Rsm = p^\Rsm\,,  \quad P^\Lsm = p^\Lsm\,,   \quad   \hspace{0.8cm} P^+=\beta\,,\quad
P^- = p^-\,, \quad p^- \equiv - \frac{2p^\Rsm p^\Lsm+m^2}{2\beta}\,,\qquad
\\
\label{18042022-man-17}  && J^{+\Rsm}= \irm x^+ P^\Rsm + \partial_{p^\Lsm}\beta\,, \hspace{1.6cm} J^{+\Lsm}= \irm x^+ P^\Lsm + \partial_{p^\Rsm}\beta\,, \
\\
\label{18042022-man-18}  && J^{+-} = \irm x^+P^- + \partial_\beta \beta\,,  \hspace{1.6cm} J^{\Rsm\Lsm} =  p^\Rsm\partial_{p^\Rsm} - p^\Lsm\partial_{p^\Lsm} + M^{\Rsm\Lsm}\,,
\\
\label{18042022-man-19}  && J^{-\Rsm} = -\partial_\beta p^\Rsm + \partial_{p^\Lsm} p^- + M^{\Rsm\Lsm} \frac{p^\Rsm}{\beta} + \frac{1}{\beta} M^\Rsm\,,
\\
\label{18042022-man-20}  && J^{-\Lsm} = -\partial_\beta p^\Lsm + \partial_{p^\Rsm} p^- - M^{\Rsm\Lsm}\frac{p^\Lsm}{\beta} +  \frac{1}{\beta} M^\Lsm\,,
\eeq
where, for partial derivatives, we use the notation
\be \label{18042022-man-21}
\partial_\beta\equiv \partial/\partial \beta\,, \quad \partial_{p^\Rsm}\equiv \partial/\partial p^\Rsm\,, \qquad \partial_{p^\Lsm}\equiv \partial/\partial p^\Lsm\,,
\ee
while quantities $M^{\Rsm\Lsm}$, $M^\Rsm$, $M^\Lsm$ appearing in \rf{18042022-man-18}-\rf{18042022-man-20} are defined as
\beq
&& \hspace{-4cm}  \label{18042022-man-22}  \hbox{\it for massless field:} \hspace{1cm}
 M^{\Rsm\Lsm}=\lambda\,, \qquad M^\Rsm = 0\,, \qquad M^\Lsm = 0\,;
\\
\label{18042022-man-25} && \hspace{-4cm} \hbox{\it for massive field:} \qquad M^{\Rsm\Lsm} = \half (N_u - N_v\big) \,,\hspace{1cm} N_u \equiv u \ub\,, \qquad  N_v \equiv v \vb\,,
\nonumber\\
&& M^\Rsm = \frac{m}{\sqrt{2}} u \vb \,, \qquad  M^\Lsm = - \frac{m}{\sqrt{2}} v \ub\,.
\eeq
We note the following commutators and hermitian conjugation rules for operators  \rf{18042022-man-25}:
\beq
\label{18042022-man-26} && [M^\Rsm,M^\Lsm] = - m^2 M^{\Rsm\Lsm}\,,\qquad [M^{\Rsm\Lsm},M^\Rsm ] = M^\Rsm\,, \qquad [M^{\Rsm\Lsm},M^\Lsm ] = - M^\Lsm\,,\qquad %
\\
\label{18042022-man-28} && M^{\Rsm\Lsm\dagger} = M^{\Rsm\Lsm}\,, \hspace{2.4cm} M^{\Rsm\dagger} = - M^\Lsm\,.
\eeq
In Appendix B, we discuss alternative helpful realizations for the operators $M^\Rsm$, $M^\Lsm$, $M^{\Rsm\Lsm}$.

\noinbf{Field-theoretical form of Poincar\'e algebra}.
To quadratic order in the light-cone gauge fields, a field-theoretical representation of the Poincar\'e algebra generators for the spin-$s$, $s=|\lambda|$,  massless field and the spin-$s$ massive field takes the following respective forms:
\beq
\label{18042022-man-29} G_\smpt & = & 2 \int d^3p\,\, \beta \phi_s^\dagger G_\diff \phi_s\,, \hspace{2.1cm} \hbox{ for massless field}\,;
\\
\label{18042022-man-30} G_\smpt & = & \int d^3p\,\,  \beta \langle \phi_{m,s}| G_\diff |\phi_{m,s}\rangle\,,\hspace{1.2cm} \hbox{ for massive field},
\eeq
where a notation $G_\smpt$ is used for the field-theoretical representation of the generators \rf{18042022-man-04},\rf{18042022-man-05}, while
$G_\diff$ stands for the differential operators described  in
\rf{18042022-man-16}-\rf{18042022-man-28}.

The massless field $\phi_\lambda$ and the massive fields $\phi_{m,s;n}$ satisfy the following respective Poisson-Dirac equal-time commutators:
\beq
\label{18042022-man-31} && \hspace{-2cm} [\phi_\lambda(x^+,p),\phi_{\lambda'}(x^+,p')] = \frac{1}{2\beta}\delta_{\lambda+\lambda',0}\delta^3(p+p') \,,
\\
\label{18042022-man-32} &&\hspace{-2cm} [\phi_{m,s;n}(x^+,p),\phi_{m,s';n'}(x^+,p')] = \frac{1}{2\beta} \delta_{s,s'}\delta_{n+n',0} \delta^3(p+p')\,.
\eeq
By using formulas in \rf{18042022-man-29}-\rf{18042022-man-32}, we can easily check the standard equal-time commutators between the fields and the  generators of the Poncar\'e algebra,
\be \label{18042022-man-33}
[ \phi_\lambda,G_\smpt\,] =  G_\diff \phi_\lambda \,, \qquad [|\phi_{m,s}\rangle,G_\smpt\,] =  G_\diff |\phi_{m,s}\rangle \,.
\ee

To conclude this section we note that the light-cone gauge action takes the form
\be \label{18042022-man-34}
S  = S_\free  + \int dx^+ P_{\rm int}^-\,,
\ee
where $S_\free$ is an action for free fields, while $P_{\rm int}^-$ is a light-cone gauge Hamiltonian describing interactions. The actions for the free spin-$s$, $s=|\lambda|$, massless field and the free spin-$s$ massive field take the following respective forms:
\beq
\label{18042022-man-35} && \hspace{-1.4cm} S_\free^{\rm massless} = \int dx^+ d^3p \,  \phi_s^\dagger \Box \phi_s\,, \hspace{1cm} S_\free^{\rm massive} =  \half \int dx^+ d^3p \,  \langle \phi_{m,s}|  \big( \Box - m^2) |\phi_{m,s}\rangle\,,
\\
\label{18042022-man-36} && \hspace{0.1cm} \Box \equiv   2\irm \beta \frac{\partial}{\partial x^+}  - 2p^\Rsm p^\Lsm\,.
\eeq
As in string theory, the incorporation of internal symmetries can be done by using the Chan--Paton method (see Appendix C).

\newsection{ \large Poincar\'e algebra kinematical restrictions on $n$-point dynamical generators} \label{sec-03}

In general, for theories of interacting massive and massless fields, the dynamical generators of the Poincar\'e algebra \rf{18042022-man-05} denoted as $G^\dyn$ take the following expansion:
\be \label{19042022-man-01}
G^\dyn = \sum_{n=2}^\infty G_\smpn^\dyn\,,
\ee
where $G_\smpn^\dyn$ \rf{19042022-man-01} is a functional
that has $n$ powers of massive and massless fields. For $n=2$, the dynamical and kinematical generators are given in \rf{18042022-man-29}, \rf{18042022-man-30}.%
\footnote{ We recall that, with the exception of $J^{+-}$, all kinematical generators \rf{18042022-man-04} are  quadratic in the fields. The kinematical generator $J^{+-}$ takes the following form $J^{+-} = J_0 + \irm x^+ P^-$, where  $J_0$ is also quadratic in the fields, while the $P^-$ is the dynamical generator appearing in \rf{18042022-man-05}.}
Restrictions on $G_\smpn^\dyn$, $n\geq 3$, imposed by the Poincar\'e algebra commutators between kinematical generators \rf{18042022-man-04} and dynamical generators \rf{18042022-man-05} we refer to as kinematical restrictions.
Our aim in this Section is to review the kinematical restrictions.
We start with kinematical restrictions which are obtained by using $P^\Rsm$, $P^\Lsm$, $P^+$ symmetries in \rf{18042022-man-04}. Namely, by using commutators between the kinematical generators $P^\Rsm$, $P^\Lsm$, $P^+$ and the dynamical generators, we get the following expressions for $G_\smpn^\dyn$, $n\geq 3$,
\beq
\label{19042022-man-02} && P_\smpn^- = \int\!\! d\Gamma_\smpn\,\,  \langle \Phi_\smpn  |\!\cdot\!| p_\smpn^-\rangle\,,
\\
\label{19042022-man-03} && J_\smpn^{-\Rsm} = \int\!\! d\Gamma_\smpn\,\,  \langle \Phi_\smpn |\!\cdot\!| j_\smpn^{-\Rsm}\rangle  +   \langle \Xbf_\smpn^\Rsm \Phi_\smpn|\!\cdot\! | p_\smpn^-\rangle \,,
\\
\label{19042022-man-04} && J_\smpn^{-\Lsm} = \int\!\! d\Gamma_\smpn\,\,  \langle \Phi_\smpn |\!\cdot\!| j_\smpn^{-\Lsm}\rangle +   \langle \Xbf_\smpn^\Lsm \Phi_\smpn |\!\cdot\! | p_\smpn^- \rangle \,,
\eeq
where we use the notation
\beq
\label{19042022-man-05} && d\Gamma_\smpn =  (2\pi)^3\delta^{(3)}(\sum_{a=1}^n p_a)\prod_{a=1}^n \frac{d^3p_a}{(2\pi)^{3/2} }\,, \qquad d^3 p_a = dp_a^\Rsm dp_a^\Lsm d\beta_a\,,
\\
\label{19042022-man-06} && \Xbf_\smpn^\Rsm =  - \frac{1}{n}\sum_{a=1}^n \partial_{p_a^\Lsm}\,, \hspace{1cm} \Xbf_\smpn^\Lsm = - \frac{1}{n}\sum_{a=1}^n\partial_{p_a^\Rsm}\,,
\\
\label{19042022-man-07} && \langle\Phi_\smpn|  \equiv \prod_b \phi_{\lambda_b}^\dagger(x^+,p_b)\prod_c \langle \phi_{m_c,s_c}(x^+, p_c )| \,,\qquad b \cup c = 1,\ldots,n\,, \qquad b \cap c = \emptyset\,, \qquad
\\
\label{19042022-man-08}  &&| p_\smpn^-\rangle = p_\smpn^- \prod_c |0\rangle_c \,, \qquad | j_\smpn^{-\Rsm}\rangle = j_\smpn^{-\Rsm} \prod_c |0\rangle_c \,,
\qquad | j_\smpn^{-\Lsm}\rangle = j_\smpn^{-\Lsm} \prod_c |0\rangle_c \,,
\eeq
and the indices $a,b,c=1,\ldots,n$ label massive or massless fields entering $n$-point dynamical generators. We refer to $p_\smpn^-$ as $n$-point interaction vertex, while $p_\smp3^-$ is referred to as cubic vertex.
Sometimes, $p_\smpn^-$ and $j_\smpn^{-\Rsm,\Lsm}$ \rf{19042022-man-08} will be denoted as $g_\smpn$ and referred to as densities.
The densities $g_\smpn$ depend on the momenta $p_a^\Rsm$, $p_a^\Lsm$, $\beta_a$, and oscillators $u_a$, $v_a$
$a=1,2,\ldots,n$,
\be \label{19042022-man-09}
g_\smpn = g_\smpn(p_a^\Rsm,p_a^\Lsm,\beta_a,u_a,v_a)\,, \qquad g_\smpn = p_\smpn^-,\quad j_\smpn^{-\Rsm},\quad j_\smpn^{-\Lsm}\,.
\ee
In \rf{19042022-man-02}-\rf{19042022-man-04}, the product $\langle \Phi_\smpn|\!\cdot\! | g_\smpn\rangle$ stands for the shortcut defined  as follows
\beq
&& \hspace{-1.5cm}\hbox{ \it for vertices involving only massive fields:}
\nonumber\\
\label{19042022-man-09-a1} && \hspace{-1cm} \langle \Phi_\smpn  |\!\cdot\!| p_\smpn^-\rangle \equiv \langle \Phi_\smpn | p_\smpn^-\rangle \,, \hspace{0.5cm} \langle \Phi_\smpn  |\!\cdot\!| j_\smpn^{-\Rsm}\rangle \equiv \langle \Phi_\smpn | j_\smpn^{-\Rsm}\rangle \,,\hspace{0.5cm} \langle \Phi_\smpn  |\!\cdot\!| j_\smpn^{-\Lsm}\rangle \equiv \langle \Phi_\smpn | j_\smpn^{-\Lsm}\rangle \,;
\\
&&\hspace{-1.5cm} \hbox{\it for vertices involving massless and massive fields:}
\nonumber\\
\label{19042022-man-09-a2} && \hspace{-1cm} \langle \Phi_\smpn  |\!\cdot\!| p_\smpn^-\rangle \equiv \langle \Phi_\smpn | p_\smpn^-\rangle \, + \langle \Phi_{\smpn,\II}| \II\, p_\smpn^-\rangle \,,
\\
\label{19042022-man-09-a3} && \hspace{-1cm}\langle \Phi_\smpn  |\!\cdot\!| j_\smpn^{-\Rsm}\rangle \equiv \langle \Phi_\smpn | j_\smpn^{-\Rsm}\rangle \, - \langle \Phi_{\smpn,\II} |\II\, j_\smpn^{-\Lsm}\rangle \,,
\\
\label{19042022-man-09-a4} && \hspace{-1cm}\langle \Phi_\smpn  |\!\cdot\!| j_\smpn^{-\Lsm}\rangle \equiv \langle \Phi_\smpn | j_\smpn^{-\Lsm}\rangle \, - \langle \Phi_{\smpn,\II}| \II\, j_\smpn^{-\Rsm}\rangle \,,
\eeq
where, the bra-vector $\langle \Phi_{\smpn,\II}|$ and the action of the operator $\II$ on the densities $g_\smpn$ are defined as
\beq
\label{19042022-man-09-a5} && \hspace{-0.7cm}\langle \Phi_{\smpn,\II}|\, \equiv \prod_b \phi_{-\lambda_b}^\dagger(x^+,p_b)\prod_c \langle \phi_{m_c,s_c}(x^+, p_c )| \,,\qquad b \cup c = 1,\ldots,n\,, \qquad b \cap c = \emptyset\,, \qquad\qquad
\\
\label{19042022-man-09-a6} && \hspace{-0.7cm} \II\, g_\smpn(p_a^\Rsm,p_a^\Lsm,\beta_a,u_a,v_a) = g_\smpn^* (\II\,p_a^\Rsm,\II\,p_a^\Lsm,\II\,\beta_a,\II\,u_a,\II\,v_a)\,,\hspace{0.3cm}
\\
&& \hspace{-0.7cm} \II\,p_a^\Rsm = - p_a^\Lsm\,,\hspace{0.3cm}  \II\,p_a^\Lsm = - p_a^\Rsm\,, \hspace{0.3cm}  \II\,\beta_a = -\beta_a\,, \hspace{0.3cm} \II\,u_a = v_a\,,\hspace{0.3cm} \II\,v_a = u_a\,.\qquad
\eeq
In \rf{19042022-man-09-a6} and throughout this paper, the asterisk $*$ stands for a complex conjugation.
Shortly speaking, $\langle \Phi_{\smpn,\II}|$ \rf{19042022-man-09-a5} is obtained by making the replacement $\lambda_b\rightarrow - \lambda_b$ in the expression  for $ \langle \Phi_\smpn|$ \rf{19042022-man-07}, while the action of the operator $\II$ on the densities $g_\smpn$ is given by
\be \label{19042022-man-09-a7}
\II\, g_\smpn(p_a^\Rsm,p_a^\Lsm,\beta_a,u_a,v_a) = g_\smpn^*(-p_a^\Lsm,-p_a^\Rsm,-\beta_a,v_a,u_a)\,.
\ee
Rules in \rf{19042022-man-09-a1}-\rf{19042022-man-09-a7} are obtained by using the hermitian conjugation conditions for the generators $P^-$, $J^{-\Rsm,\Lsm}$ \rf{18042022-man-06} and the fields \rf{18042022-man-15} (for more detailed discussion, see Appendix D).%
\footnote{ Considering two terms on r.h.s. in \rf{19042022-man-09-a2}, we introduce $P_>^- \equiv \int \langle \Phi_\smpn | p_\smpn^-\rangle $ and $P_<^-\equiv \int \langle \Phi_{\smpn,\II}| \II\, p_\smpn^-\rangle $. Using the hermicity conditions \rf{18042022-man-15} and the map $\II$ \rf{19042022-man-09-a7}, we find $P_>^{-\dagger} = P_<^-$ and $P_<^{-\dagger} = P_>^-$. Therefore the hermitian $P^-$ is given by $P^- = P_>^- + P_<^-$. If the readers prefer not to use the hermicity conditions \rf{18042022-man-15}  and the map $\II$ \rf{19042022-man-09-a7}, then the hermitian $P^-$ is given by $P^- = P_>^- + P_>^{-\dagger}$. }
In \rf{19042022-man-03} and \rf{19042022-man-04}, the respective operators  $\Xbf_\smpn^\Rsm$ and $\Xbf_\smpn^\Lsm$ act only on the arguments of the fields.

The remaining kinematical restrictions are given by%
\footnote{ In \rf{19042022-man-10} and below, in place of equations for the ket-vectors $|g_\smpn\rangle$, we prefer to use equations for the densities $g_\smpn$ \rf{19042022-man-08}. This implies that, in equations for $g_\smpn$, we should replace the annihilation operators $\ub$ and $\vb$ \rf{18042022-man-25} by the respective derivatives $\partial / \partial u$ and $\partial / \partial v$.}
\vspace{-0.3cm}
\beq
&& \hspace{-2cm} \hbox{$J^{+-}$-symmetry restrictions:}%
\nonumber\\[-8pt]
\label{19042022-man-10} && \Jbf^{+-} p_\smpn^- =  0\,, \qquad \Jbf^{+-} j_\smpn^{-\Rsm,\Lsm} =  0\,, \qquad
\Jbf^{+-} \equiv \sum_{a=1}^n  \beta_a\partial_{\beta_a} \,;
\\[-6pt]
&& \hspace{-2cm} \hbox{$J^{\Rsm\Lsm}$-symmetry \ restrictions:}
\nonumber\\[-5pt]
\label{19042022-man-11} && \Jbf^{\Rsm\Lsm} p_\smpn^- = 0\,,\qquad  (\Jbf^{\Rsm\Lsm} -1) j_\smpn^{-\Rsm} = 0 \,,\qquad  (\Jbf^{\Rsm\Lsm} + 1) j_\smpn^{-\Lsm} = 0\,,
\\
\label{19042022-man-12} && \Jbf^{\Rsm\Lsm} \equiv \sum_{a=1}^n  \big( p_a^\Rsm\partial_{p_a^\Rsm} - p_a^\Lsm\partial_{p_a^\Lsm} + M_a^{\Rsm\Lsm} \big)\,;
\\[-6pt]
&& \hspace{-2cm} \hbox{$J^{+\Rsm}, J^{+\Lsm}$-symmetry  restrictions:}
\nonumber\\[-5pt]
\label{19042022-man-13} && g_\smpn = g_\smpn (\Po_{ab}^\Rsm,\Po_{ab}^\Lsm\,, \beta_a,u_a,v_a)\,,\hspace{1cm} g_\smpn= p_\smpn^-,\quad   j_\smpn^{-\Rsm},\quad  j_\smpn^{-\Lsm}\,,
\\
\label{19042022-man-14} && \Po_{ab}^\Rsm \equiv p_a^\Rsm \beta_b - p_b^\Rsm \beta_a\,, \qquad
\Po_{ab}^\Lsm \equiv p_a^\Lsm \beta_b - p_b^\Lsm \beta_a\,.
\eeq

Let us briefly comment the restrictions presented in \rf{19042022-man-02}-\rf{19042022-man-14}.

\noindent \ibf) The Poincar\'e algebra commutators between the dynamical generators $P^-$, $J^{-\Rsm}$, $J^{-\Lsm}$ and the kinematical generators $P^\Rsm$, $P^\Lsm$, $P^+$ imply the delta-functions in \rf{19042022-man-05}.

\noindent \iibf) The Poincar\'e algebra commutators between the dynamical generators $P^-$, $J^{-\Rsm}$, $J^{-\Lsm}$ and the kinematical generators $J^{+-}$, $J^{\Rsm\Lsm}$ amount to equations given in \rf{19042022-man-10}-\rf{19042022-man-12}.

\noindent \iiibf) From the Poincar\'e algebra commutators between the dynamical generators $P^-$, $J^{-\Rsm}$, $J^{-\Lsm}$ and the kinematical generators $J^{+\Rsm}$, $J^{+\Lsm}$,  we learn that the $n$-point densities $p_\smpn^-$,  $j_\smpn^{-\Rsm}$, $j_\smpn^{-\Lsm}$ depend on the momenta $\Po_{ab}^\Rsm$ and $\Po_{ab}^\Lsm$ \rf{19042022-man-14} in place of the  momenta $p_a^\Rsm$ and $p_a^\Lsm$ respectively.

\noindent \ivbf) Making use of the conservation laws for the momenta $p_a^\Rsm$, $p_a^\Lsm$, $\beta_a$, one can verify that there are only $n-2$ independent momenta $\Po_{ab}^\Rsm$ and $n-2$ independent momenta $\Po_{ab}^\Lsm$  \rf{19042022-man-14}. This implies that, for $n=3$, there is only one independent momentum $\Po^\Rsm$ and one independent momentum $\Po^\Lsm$.

\newsection{ \large Poincar\'e symmetry restrictions on cubic vertices and light-cone gauge dynamical principle } \label{sec-04}

Making use of the conservation laws for the momenta $p_a^\Rsm$, $p_a^\Lsm$, $\beta_a$,
\be  \label{20042022-man-01}
\sum_{a=1,2,3} p_a^{\Rsm,\Lsm}=0\,, \qquad \sum_{a=1,2,3}\beta_a =0 \,,
\ee
one can check that momenta $\Po_{ab}^\Rsm$, $\Po_{ab}^\Lsm$ \rf{19042022-man-14} are expressible in terms of new momenta $\Po^\Rsm$, $\Po^\Lsm$,
\beq
\label{20042022-man-02} && \hspace{4cm}  \Po_{12}^{\Rsm,\Lsm} =\Po_{23}^{\Rsm,\Lsm} = \Po_{31}^{\Rsm,\Lsm} = \Po^{\Rsm,\Lsm} \,,
\\
\label{20042022-man-03} && \hspace{-1cm} \Po^\Rsm \equiv \frac{1}{3}\sum_{a=1,2,3} \betach_a p_a^\Rsm\,, \qquad \Po^\Lsm \equiv \frac{1}{3} \sum_{a=1,2,3} \betach_a p_a^\Lsm\,, \qquad \betach_a \equiv \beta_{a+1} - \beta_{a+2}\,, \qquad\beta_a=\beta_{a+3}\,.  \qquad
\eeq
This implies that  the cubic densities $g_\smp3$ are functions of the momenta $\beta_a$, $\Po^{\Rsm,\Lsm}$, and oscillators $u_a$, $v_a$
\be \label{20042022-man-04}
g_\smp3 = g_\smp3(\Po^\Rsm,\Po^\Lsm,\beta_a,u_a,v_a)\,,\qquad
g_\smp3= p_\smp3^-,\quad   j_\smp3^{-\Rsm},\quad  j_\smp3^{-\Lsm}\,.
\ee
In other words, the momenta $p_a^\Rsm$ and $p_a^\Lsm$ enter the cubic densities $g_\smp3$ through the momenta $\Po^\Rsm$ and $\Po^\Lsm$ respectively. Thank to this feature, a study of the cubic densities is considerably simplified.

In this Section, our aims are as follows.

\noinbf{i)} To represent kinematical $J^{+-}$, $J^{\Rsm\Lsm}$ symmetry equations \rf{19042022-man-10},\rf{19042022-man-11} in terms of $\Po^{\Rsm,\Lsm}$.

\noinbf{ii)} To find restrictions imposed on the cubic vertex by the Poincar\'e algebra dynamical symmetries.

\noinbf{iii)} To formulate so called light-cone gauge dynamical principle and present equations which are required to fix all possible solutions for the cubic vertices uniquely.

\noindent {\bf Kinematical $J^{+-}$, $J^{\Rsm\Lsm}$- symmetries}.  Using \rf{20042022-man-03} and \rf{20042022-man-04}, we find that, for $n=3$,  equations \rf{19042022-man-10},\rf{19042022-man-11} take the following form:
\vspace{-0.2cm}
\beq
&& \hspace{-2cm} \hbox{$J^{+-}$-symmetry restrictions:}
\nonumber\\
\label{20042022-man-05} && \Jbf^{+-}  p_\smp3^- = 0\,,  \hspace{0.7cm} \Jbf^{+-} j_\smp3^{-\Rsm} = 0\,, \qquad \Jbf^{+-} j_\smp3^{-\Lsm} = 0\,,
\\
\label{20042022-man-06} && \Jbf^{+-} \equiv    N_{\Po^\Rsm} + N_{\Po^\Lsm}+   \sum_{a=1,2,3}  \beta_a \partial_{\beta_a} \,,
\\
\label{20042022-man-07} && N_{\Po^\Rsm} \equiv \Po^\Rsm \partial_{\Po^\Rsm}\,, \qquad N_{\Po^\Lsm} \equiv \Po^\Lsm \partial_{\Po^\Lsm}\,, \qquad   \partial_{\Po^\Rsm} \equiv \partial/\partial \Po^\Rsm\,, \qquad
\partial_{\Po^\Lsm} \equiv \partial/\partial \Po^\Lsm\,; \qquad
\\
&& \hspace{-2cm} \hbox{$J^{\Rsm\Lsm}$-symmetry restrictions:}
\nonumber\\
\label{20042022-man-08} &&  \Jbf^{\Rsm\Lsm}p_\smp3^- =0\,, \hspace{0.7cm}  (\Jbf^{\Rsm\Lsm} -1) j_\smp3^{-\Rsm} =0\,,   \hspace{1.8cm}  (\Jbf^{\Rsm\Lsm} +  1 ) j_\smp3^{-\Lsm} =0\,,\qquad
\\
\label{20042022-man-09} && \Jbf^{\Rsm\Lsm} \equiv    N_{\Po^\Rsm} -  N_{\Po^\Lsm}  +\Mbf^{\Rsm\Lsm}\,, \qquad \Mbf^{\Rsm\Lsm} \equiv \sum_{a=1,2,3} M_a^{\Rsm\Lsm}\,.
\eeq

\noindent {\bf Dynamical symmetries}. Restrictions on the densities imposed by the Poincar\'e algebra commutators between the dynamical generators \rf{18042022-man-05} we refer to as dynamical restrictions.
Dynamical restrictions on the cubic densities  are obtainable from the following commutators of the Poincar\'e algebra:
\be \label{20042022-man-10}
[P^-,J^{-\Rsm}]=0\,, \hspace{1cm} [P^-,J^{-\Lsm}]=0\,, \hspace{1cm} [J^{-\Rsm},J^{-\Lsm}]=0\,.
\ee
In the cubic approximation, the first two commutators in \rf{20042022-man-10} lead
to the relations for the cubic densities,
\beq
\label{20042022-man-12} && j_\smp3^{-\Rsm}  = -\frac{1}{ \Pbf^-} \Jbf^{-\Rsm} p_\smp3^- \,, \hspace{1cm} j_\smp3^{-\Lsm}  = - \frac{1}{ \Pbf^-} \Jbf^{-\Lsm} p_\smp3^- \,,
\eeq
where the notation $\Jbf^{-\Rsm}$, $\Jbf^{-\Lsm}$, $\Pbf^-$ is used for the following operators:
\beq
\label{20042022-man-13} && \Jbf^{-\Rsm}   \equiv   \frac{\Po^\Rsm}{\beta} \big( - \No_\beta + \Mo^{\Rsm\Lsm} \big) + \sum_{a=1,2,3}  \frac{\check\beta_a }{6\beta_a} m_a^2 \partial_{\Po^\Lsm}  - \frac{1}{\beta_a} M_a^\Rsm\,,
\\
\label{20042022-man-14} && \Jbf^{-\Lsm}   \equiv   \frac{\Po^\Lsm}{\beta} \big( - \No_\beta - \Mo^{\Rsm\Lsm} \big) + \sum_{a=1,2,3}  \frac{\check\beta_a }{6\beta_a} m_a^2 \partial_{\Po^\Rsm}  - \frac{1}{\beta_a} M_a^\Lsm\,,
\\
\label{20042022-man-15} && \Pbf^- \equiv \frac{1}{\beta}\big(\Po^\Rsm\Po^\Lsm -\half \rho^2\big) \,, \qquad \rho^2 \equiv \beta\sum_{a=1,2,3} \frac{m_a^2}{\beta_a}\,, \hspace{1cm} \beta \equiv \beta_1\beta_2\beta_3\,,
\\
\label{20042022-man-16} && \hspace{1.3cm} \No_\beta \equiv \frac{1}{3}\sum_{a=1,2,3}\betach_a \beta_a\partial_{\beta_a}\,,   \hspace{1.3cm} \Mo^{\Rsm\Lsm} \equiv \frac{1}{3}\sum_{a=1,2,3}\betach_a M_a^{\Rsm\Lsm}\,,
\eeq
and $\betach_a$ is defined in \rf{20042022-man-03}.

If kinematical equations for $p_\smp3^-$ in \rf{20042022-man-05}, \rf{20042022-man-08} are satisfied, then, using \rf{20042022-man-12}, we verify that kinematical equations for $j_\smp3^-$ in \rf{20042022-man-05}, \rf{20042022-man-08} and the third commutator in \rf{20042022-man-10} considered in the cubic approximation are also satisfied. This implies that, in the cubic approximation, the kinematical equations for $p_\smp3^-$ in \rf{20042022-man-05}, \rf{20042022-man-08}  and dynamical equations  \rf{20042022-man-12} exhaust all restrictions obtainable from the commutators of the Poincar\'e algebra.

\noindent {\bf Light-cone gauge dynamical principle}. The kinematical equations for $p_\smp3^-$ in \rf{20042022-man-05}, \rf{20042022-man-08} and the dynamical equations \rf{20042022-man-12} do not fix all solutions for the cubic vertex $p_\smp3^-$ unambiguously. To find all solutions for the cubic vertex $p_\smp3^-$ unambiguously we impose additional restrictions on the cubic vertex $p_\smp3^-$. Throughout this paper, these additional restrictions are referred to as light-cone gauge dynamical principle and we formulate this principle in the following way.

\noindent \ibf) The cubic vertex $p_\smp3^-$ and the densities $j_\smp3^{-\Rsm,\Lsm}$ should be polynomial in the momenta  $\Po^\Rsm$, $\Po^\Lsm$;

\noindent \iibf) The cubic vertex $p_\smp3^-$ should obey the following restriction:
\be \label{20042022-man-17}
p_\smp3^-  \ne  \Pbf^- W\,, \quad W \ \hbox{is polynomial in } \Po^\Rsm\,, \ \Po^\Lsm\,,
\ee
where $\Pbf^-$ is given in \rf{20042022-man-15}. We briefly comment restriction \rf{20042022-man-17}. Upon field redefinitions, the vertex $p_\smp3^-$ is changed as $p_\smp3^-\rightarrow p_\smp3^- + \Pbf^-f$ (see, e.g., Appendix B in Ref.\cite{Metsaev:2005ar}).  If $p_\smp3^- =\Pbf^- W$, then such vertex can be made trivial by using the field redefinition with $f=-W$. As we are going to find non-trivial solutions for the cubic vertex $p_\smp3^-$ we impose restriction \rf{20042022-man-17}. The collection of restrictions imposed by the Poincar\'e algebra commutators and restrictions of the light-cone dynamical principle we refer to as complete system of equations.

\noindent {\bf Complete system of equations}. The complete system of equations for the cubic vertex
\be  \label{20042022-man-18}
p_\smp3^- = p_\smp3^-(\Po^\Rsm,\Po^\Lsm,\beta_a,u_a,v_a)
\ee
takes the following  form:
\beq
\label{20042022-man-19}  && \Jbf^{+-}p_\smp3^- =0 \,, \hspace{4.7cm} \hbox{kinematical } \  J^{+-}-\hbox{ symmetry};
\\
\label{20042022-man-20} &&  \Jbf^{\Rsm\Lsm} p_\smp3^- = 0\,, \hspace{4.8cm} \hbox{kinematical } \  J^{\Rsm\Lsm}-\hbox{ symmetry};
\\
\label{20042022-man-21} && j_\smp3^{-\Rsm,\Lsm} = - \frac{1}{ \Pbf^-}\Jbf^{-\Rsm,\Lsm} p_\smp3^- \,, \hspace{2.7cm} \hbox{ dynamical } P^-, J^{-\Rsm,\Lsm} \hbox{ symmetries};\qquad
\\
&& \hspace{3cm} \hbox{ \it Light-cone gauge dynamical principle:}
\nonumber\\
\label{20042022-man-22} && p_\smp3^-\,, \ j_\smp3^{-\Rsm,\Lsm} \hspace{0.7cm} \hbox{ are polynomial in } \Po^\Rsm, \Po^\Lsm;
\\
\label{20042022-man-23} && p_\smp3^- \ne \Pbf^- W, \hspace{0.4cm} W  \hbox{ is polynomial in } \Po^\Rsm, \Po^\Lsm; \qquad
\eeq
where $\Jbf^{+-}$, $\Jbf^{\Rsm\Lsm}$, $\Jbf^{-\Rsm,\Lsm}$, $\Pbf^-$  are given in  \rf{20042022-man-06}, \rf{20042022-man-09} and  \rf{20042022-man-13}-\rf{20042022-man-16}.

Equations \rf{20042022-man-18}-\rf{20042022-man-23} constitute the complete
system of equations allowing us to fix all solutions for the cubic vertex $p_\smp3^-$ and the densities $j_\smp3^{-\Rsm,\Lsm}$ uniquely up to the freedom related to field redefinitions. Namely, if a cubic vertex $p_{\smp3\, \fix}^-$ obeys the complete system of equations, then the cubic vertex $p_\smp3^-$ which is obtained from $p_{\smp3\, \fix}^-$ by using the field redefinition, $p_{\smp3}^- = p_{\smp3\, \fix}^-  + \Pbf^- f$, also obeys the complete system of equations. To determine the cubic vertex uniquely we have to choose some representative of the cubic vertex.  After choosing a representative of the cubic vertex, the complete system of equations provides us the possibility to find all solutions for the cubic vertex $p_\smp3^-$ and the densities  $j_\smp3^{-\Rsm,\Lsm}$ uniquely.
Below we choose the representative of the cubic vertex which we refer to as harmonic cubic vertex.

\newsection{\large Equations for cubic harmonic and meromorphic vertices }\label{sec-05}

According to \rf{20042022-man-22}, the cubic vertex $p_\smp3^-$ is a degree-$K$ polynomial in the momenta $\Po^\Rsm$, $\Po^\Lsm$, $K < infty$. By using field redefinitions, we obtain various representatives of $p_\smp3^-$.  The representative of the cubic vertex which obeys the harmonic equation
\be \label{21042022-man-01}
\partial_{ \Po^\Rsm } \partial_{ \Po^\Lsm } p_\smp3^-=0
\ee
will be referred to as harmonic vertex (for more detailed discussion, see Sec.4.1 in Ref.\cite{Metsaev:2005ar}%
\footnote{ Equations for the harmonic cubic vertices of light-cone gauge fields in $R^{d-1,1}$, $d\geq4$, were obtained in Ref.\cite{Metsaev:2005ar}. Here, for the case of $d=4$, we represent equations in Ref.\cite{Metsaev:2005ar} by using the light-cone gauge helicity basis framework.}).
For the harmonic vertex $p_\smp3^-$, kinematical equations \rf{20042022-man-19},\rf{20042022-man-20} take the same form. Note also that the harmonic vertex $p_\smp3^-$ automatically obeys restriction \rf{20042022-man-23}. We then find that restrictions  \rf{20042022-man-21}, \rf{20042022-man-22} amount to the following
\vspace{-0.3cm}
\beq
&& \hspace{-2cm} \hbox{\bf Equations for harmonic vertex $p_\smp3^-$}:
\nonumber\\[-3pt]
\label{21042022-man-02a1} && \hspace{-1cm} \Big( \Jbf_\Thsm^{-\Rsm} +  \frac{1}{N_{\Po^\Lsm} +1} \sum_{a=1,2,3}\frac{m_a^2}{2\beta_a} \big(- \No_\beta  + \Mo^{\Rsm\Lsm} \big)\partial_{\Po^\Lsm} \Big)p_\smp3^- = 0 \,,
\\
\label{21042022-man-02a2} && \hspace{-1cm} \Big( \Jbf_\Thsm^{-\Lsm} +  \frac{1}{N_{\Po^\Rsm} +1} \sum_{a=1,2,3}\frac{m_a^2}{2\beta_a} \big(- \No_\beta  - \Mo^{\Rsm\Lsm} \big)\partial_{\Po^\Rsm} \Big)p_\smp3^- = 0 \,,
\\
\label{21042022-man-02a3} && \Jbf_\Thsm^{-\Rsm}   \equiv     \frac{\Po_\Thsm^\Rsm}{\beta} \big( - \No_\beta + \Mo^{\Rsm\Lsm} \big) + \sum_{a=1,2,3}  \frac{\check\beta_a }{6\beta_a} m_a^2 \partial_{\Po^\Lsm}  - \frac{1}{\beta_a} M_a^\Rsm\,,
\\
\label{21042022-man-02a4} && \Jbf_\Thsm^{-\Lsm}  \equiv     \frac{\Po_\Thsm^\Lsm}{\beta} \big( - \No_\beta - \Mo^{\Rsm\Lsm} \big) + \sum_{a=1,2,3}  \frac{\check\beta_a }{6\beta_a} m_a^2 \partial_{\Po^\Rsm}  - \frac{1}{\beta_a} M_a^\Lsm\,,
\\
\label{21042022-man-02a5} && \Po_\Thsm^\Rsm \equiv \Po^\Rsm \Pi^\Rsm\,,\hspace{2cm} \Pi^\Rsm \equiv (1 - \Po^\Lsm \frac{1}{N_{\Po^\Lsm}+1}\partial_{\Po^\Lsm}\big)\,,
\\
\label{21042022-man-02a6} && \Po_\Thsm^\Lsm \equiv \Po^\Lsm \Pi^\Lsm\,, \hspace{2cm} \Pi^\Lsm \equiv (1 - \Po^\Rsm \frac{1}{N_{\Po^\Rsm}+1}\partial_{\Po^\Rsm}\big)\,;
\\[-5pt]
&& \hspace{-2cm} \hbox{\bf Explicit local representation for densities $j_\smp3^{-\Rsm,\Lsm}$}:
\\[-5pt]
\label{21042022-man-02a7} && j_\smp3^{-\Rsm} = \frac{1}{N_{\Po^\Lsm} + 1} \big( \No_\beta - \Mo^{\Rsm\Lsm}  \big)  \partial_{\Po^\Lsm}^{\vphantom{5pt}} p_\smp3^-\,,
\\
\label{21042022-man-02a8} && j_\smp3^{-\Lsm} = \frac{1}{N_{\Po^\Rsm} + 1} \big( \No_\beta + \Mo^{\Rsm\Lsm}  \big)  \partial_{\Po^\Rsm}^{\vphantom{5pt}} p_\smp3^-\,.
\eeq
The definition of the notation we use in \rf{21042022-man-02a1}-\rf{21042022-man-02a8} may be found in \rf{20042022-man-03}, \rf{20042022-man-07}, \rf{20042022-man-15}, and \rf{20042022-man-16}. For the derivation of \rf{21042022-man-02a1}-\rf{21042022-man-02a8}, see Appendix E.

It is the explicit local representation for the densities $j_\smp3^{-\Rsm,\Lsm}$ given in  \rf{21042022-man-02a7}, \rf{21042022-man-02a8} that we consider as one of the attractive features for the use of the harmonic vertex. Other attractive feature for the use of the harmonic vertex is that all restrictions on $p_\smp3^-$ and  $j_\smp3^{-\Rsm,\Lsm}$ in \rf{20042022-man-19}-\rf{20042022-man-23} amount  to equations solely for the harmonic vertex given in \rf{20042022-man-19},\rf{20042022-man-20} and \rf{21042022-man-01}-\rf{21042022-man-02a2}. Obviously,
the harmonic equation \rf{21042022-man-01} and the kinematical equations  \rf{20042022-man-19},\rf{20042022-man-20} present no difficulties. A real difficulty is to find solution to equations \rf{21042022-man-02a1}, \rf{21042022-man-02a2}. Our method for solving equations \rf{21042022-man-02a1}, \rf{21042022-man-02a2} is realized in the following 3 steps.

\noinbf{Step 1}. Solution to harmonic equation \rf{21042022-man-01}, which is polynomial in $\Po^\Rsm$ and $\Po^\Lsm$, can be presented as
\beq
&& \label{21042022-man-03} \hspace{-1cm} p_\smp3^- = V_N(\Po^\Rsm,\beta_a,u_a,v_a) + V_0(\beta_a,u_a,v_a) + \Vb_\Nb(\Po^\Lsm,\beta_a,u_a,v_a)\,,
\\[-3pt]
\label{21042022-man-04} && V_N (\Po^\Rsm,\beta_a,u_a,v_a) \equiv \sum_{n=1}^N (\Po^\Rsm)^n V_{N,n}(\beta_a,u_a,v_a)\,,
\\[-3pt]
\label{21042022-man-05} && \Vb_\Nb(\Po^\Lsm,\beta_a,u_a,v_a) \equiv \sum_{n=1}^\Nb (\Po^\Lsm)^n \Vb_{\Nb,n}(\beta_a,u_a,v_a)\,.
\eeq
Now, for each harmonic vertex $p_\smp3^-$ \rf{21042022-man-03}, we associate a new vertex $\VVb$ which is meromorphic function of the momentum $\Po^\Lsm$ and independent of the momentum $\Po^\Rsm$,
\be
\label{21042022-man-06} \VVb \equiv V_N^\otimes(\Po^\Lsm,\beta_a,u_a,v_a) + V_0(\beta_a,u_a,v_a) + \Vb_\Nb(\Po^\Lsm,\beta_a,u_a,v_a)\,,
\ee
where a new vertex $V_N^\otimes$ is defined in terms of the vertex $V_N$ appearing in \rf{21042022-man-03}, \rf{21042022-man-04} as follows
\beq
\label{21042022-man-07} &&  \hspace{-1cm} V_N^\otimes(\Po^\Lsm,\beta_a,u_a,v_a) \equiv V_N(\Po_\otimes^\Rsm,\beta_a,u_a,v_a)\,,
\\
\label{21042022-man-08} && \Po_\otimes^\Rsm \equiv \frac{\rho^2}{2\Po^\Lsm} \,, \qquad \Po_\otimes^\Lsm \equiv \frac{\rho^2}{2\Po^\Rsm} \,, \qquad
\rho^2 \equiv \beta \sum_{a=1,2,3}\frac{m_a^2}{\beta_a}\,,\qquad \beta\equiv \beta_1\beta_2\beta_3\,.\qquad
\eeq
From \rf{21042022-man-03}-\rf{21042022-man-05}, we see that the meromorphic vertex $\VVb$ \rf{21042022-man-06} is obtained  from the harmonic vertex $p_\smp3^-$ by using the replacement $\Po^\Rsm\rightarrow \Po_\otimes^\Rsm$ in the expression for the vertex $V_N$ \rf{21042022-man-03}, where $\Po_\otimes^\Rsm$ is defined in \rf{21042022-man-08}.
Note that the vertex $V_N^\otimes$ involves terms of negative powers of $\Po^\Lsm$.
This provides us the following rule for building the harmonic vertex $p_\smp3^-$ by using the meromorphic vertex $\VVb$: given the meromorphic vertex $\VVb$, we make the replacement $\Po^\Lsm\rightarrow \Po_\otimes^\Lsm$ in the terms of negative powers of $\Po^\Lsm$,
\be \label{21042022-man-09}
p_\smp3^- = V_N^\otimes(\Po^\Lsm,\beta_a,u_a,v_a)\big|_{\Po^\Lsm \rightarrow \Po_\otimes^\Lsm}^{\vphantom{\int}} + V_0(\beta_a,u_a,v_a) + \Vb_\Nb(\Po^\Lsm,\beta_a,u_a,v_a)\,,
\ee
where $\Po_\otimes^\Lsm$ is defined in \rf{21042022-man-08}. In other words, in view of relations \rf{21042022-man-03}-\rf{21042022-man-09}, there is one-to-one correspondence between the harmonic vertex $p_\smp3^-$, which is polynomial in $\Po^\Rsm$ and $\Po^\Lsm$, and the vertex $\VVb$, which is meromorphic function of  $\Po^\Lsm$ and independent of $\Po^\Rsm$.

\noinbf{Step 2}. In Appendix E, we show that, in terms of the meromorphic vertex $\VVb$, equations \rf{21042022-man-02a1},\rf{21042022-man-02a2} take the form
\beq
\label{21042022-man-14} && \sum_{a=1,2,3}\Big( \frac{m_a^2}{2\beta_a} \big(  \No_\beta - \Mo^{\Rsm\Lsm}) - \frac{\check\beta_a }{6\beta_a} m_a^2 N_{\Po^\Lsm} + \frac{\Po^\Lsm}{\beta_a} M_a^\Rsm   \Big) \VVb = 0 \,,
\\
\label{21042022-man-15} && \Big( \frac{\Po^\Lsm}{\beta} \big(  \No_\beta + \Mo^{\Rsm\Lsm} \big) + \sum_{a=1,2,3}    \frac{1}{\beta_a} M_a^\Lsm\Big) \VVb  =  0 \,,
\eeq
while kinematical equations \rf{20042022-man-19}, \rf{20042022-man-20} can be represented as
\beq
\label{21042022-man-16}  && \big( \Mbf^{\Rsm\Lsm} +   \sum_{a=1,2,3}  \beta_a \partial_{\beta_a} \big) \VVb = 0  \,,
\\
\label{21042022-man-17} && \big( N_{\Po^\Lsm}  - \Mbf^{\Rsm\Lsm} ) \VVb= 0\,.
\eeq
Note that, for the derivation of equation \rf{21042022-man-16}, we used equation \rf{21042022-man-17}.

\noinbf{Step 3}. In Appendix E, we find that the general solution of equations \rf{21042022-man-14}-\rf{21042022-man-17} can be presented as
\beq
\label{21042022-man-18}  && \VVb = E_m E_\beta \VVb^{(2)}\,,\hspace{0.8cm} E_m\equiv \prod_{a=1,2,3} E_{ma}\,,\hspace{0.5cm} E_\beta\equiv \prod_{a=1,2,3} E_{\beta a}\,, %
\\
\label{21042022-man-19} && E_{ma} \equiv \exp\Big(\frac{1}{\Po^\Lsm} f_a
M_a^\Lsm\Big)\,, \hspace{1cm} E_{\beta a} \equiv \Big(\frac{\Po^\Lsm}{\beta_a}\Big)^{M_a^{\Rsm\Lsm}} \,,
\nonumber\\
&&  f_a \equiv -\frac{1}{2}\betach_a - \beta_a c_a\,, \qquad \hspace{1cm} c_a = \frac{m_{a+1}^2 -m _{a+2}^2}{2m_a^2}\,,\hspace{1cm} \for \ \ m_a\ne 0\,,
\\
\label{21042022-man-20} && E_{ma} \equiv 1\,, \hspace{1cm} E_{\beta a} \equiv \Big(\frac{\Po^\Lsm}{\beta_a}\Big)^{\lambda_a} \,, \hspace{4.5cm} \for \ \ m_a = 0\,,
\eeq
where a new vertex $\VVb^{(2)}$ is independent of the momenta $\beta_1$, $\beta_2$, $\beta_3$, and $\Po^\Lsm$,
\be \label{21042022-man-21}
\VVb^{(2)} = \VVb^{(2)}(u_a, v_a)\,,
\ee
and satisfies the following equation:
\beq
\label{21042022-man-22} && \sum_{a=1,2,3}\left\{2  M_a^\Rsm + \Bigl(2c_a m_a^2
-m_{a+1}^2 + m_{a+2}^2\Bigr) M_a^{\Rsm\Lsm}\right.
\nonumber\\
&& + \left.  \Bigl(m_a^2(c_a^2
-\frac{1}{4}) - (c_a-\half)m_{a+1}^2
+(c_a+\half)m_{a+2}^2\Bigr) M_a^\Lsm\right\}\VVb^{(2)}=0\,.
\eeq
To summarize, to each harmonic vertex $p_\smp3^-$ we associated the meromorphic vertex $\VVb$. The harmonic vertex and the meromorphic vertex are in one-to-one correspondence and they are related to each other by simple transformations given in \rf{21042022-man-03}-\rf{21042022-man-09}. For the meromorphic vertex $\VVb$, we find the representation  given in \rf{21042022-man-18}-\rf{21042022-man-20}, where  $\VVb^{(2)}$ satisfies equation \rf{21042022-man-22}. Relations \rf{21042022-man-18}-\rf{21042022-man-20} show the dependence of the vertex $\VVb$ on the momenta $\beta_1$, $\beta_2$, $\beta_3$, and $\Po^\Lsm$, while the equation for $\VVb^{(2)}$ in \rf{21042022-man-22} involves only the oscillators $u_a$, $v_a$. In other words, the problem of finding all cubic vertices is reduced to the problem of finding all solutions to equation \rf{21042022-man-22}. Finding all solutions to equation \rf{21042022-man-22} turns out to be a simple problem.

\newsection{\large Classification of cubic vertices  } \label{sec-06}

A structure of the solutions to equation \rf{21042022-man-22} depends crucially on the masses. Therefore, before discussing solutions for the meromorphic vertex  $\VVb$, we explain our classification of cubic vertices. To this end we find it convenient to introduce a notion of critical and non-critical masses.

\noinbf{Critical and non-critical masses}. Consider a cubic vertex for massive and massless fields which have masses $m_1$, $m_2$, $m_3$. We introduce quantities $D$, $\Pbf_{\epsilon m}$ defined in the following way:
\beq
\label{23042022-man-01} && D \equiv m_1^4 + m_2^4 + m_3^4 - 2m_1^2m_2^2 -2 m_2^2 m_3^2 - 2 m_3^2 m_1^2\,,
\\
\label{23042022-man-02} && \Pbf_{\epsilon m}\equiv \sum_{a=1,2,3} \epsilon_a m_a  \,,
\\
\label{23042022-man-03} && \epsilon_1^2 =1\,, \quad \epsilon_2^2 =1\,, \quad \epsilon_3^2 =1\,,
\\
\label{23042022-man-04} && \hspace{1cm}D = (m_1 + m_2 + m_3)(m_1 + m_2 - m_3)(m_1 - m_2 + m_3)(m_1 - m_2 - m_3)\,. \qquad
\eeq
Note that relation \rf{23042022-man-04} gives the alternative representation for $D$ \rf{23042022-man-01}. If, for given masses $m_1$, $m_2$, $m_3$, we find that $D=0$,  then we refer to such masses as critical masses, while if, for given masses $m_1$, $m_2$, $m_3$, we find that $D\ne 0$, then such masses are referred to  as non-critical masses. From \rf{23042022-man-04}, we see, that the restriction $\Pbf_{\epsilon m}=0$ implies the restriction $D=0$. Also, from \rf{23042022-man-04}, we see that the restriction $D=0$ implies that there exist $\epsilon_1$, $\epsilon_2$, $\epsilon_3$ \rf{23042022-man-03} such that the restriction  $\Pbf_{\epsilon m}=0$ holds true.%
\footnote{For example, if the masses satisfy the relation $m_1+m_2-m_3=0$, then the corresponding $\epsilon_a$ are given by $\epsilon_1=1$, $\epsilon_2=1$, $\epsilon_3=-1$. In $R^{2,1}$ space, some cubic vertices for massive fields with the critical masses can be obtained via dimensional reduction from vertices for massless fields in $R^{3,1}$ space (see Refs.\cite{Metsaev:2020gmb,Skvortsov:2020pnk}). It remains to be understood about whether or not, and in what ways, the cubic vertices for massive fields with the critical masses in $R^{3,1}$ space can be obtained via dimensional reduction from cubic vertices of massless fields in $R^{4,1}$ space.}
This is to say that we use the definition
\beq
\label{23042022-man-05} && D \ne 0\,,\hspace{4.3cm} \hbox{ for non-critical masses;}
\\
\label{23042022-man-06} && D=0\,, \qquad \Pbf_{\epsilon m} =0\,,\hspace{1.7cm}  \hbox{ for critical masses.}
\eeq

All cubic vertices for three massless fields were obtained in Ref.\cite{Bengtsson:1986kh} (see \rf{11032022-39a1} in Appendix F of our paper). In our paper, we study cubic vertices that involve at least one massive field. This is to say that, depending on masses of fields involved in the cubic vertex, we introduce the following
\vspace{-0.1cm}
\beq
&& \hspace{-1.5cm} \hbox{\bf Classification of cubic vertices:}%
\nonumber\\[-5pt]
&& \hspace{-0.5cm} \hbox{\bf Two massless and one massive fields},
\nonumber\\[-4pt]
\label{23042022-man-07} && m_1 = 0\,, \qquad m_2 = 0\,, \qquad m_3 \ne 0\,, \hspace{1cm} D > 0\,;
\\
&& \hspace{-0.5cm} \hbox{\bf Two massive and one massless fields},
\nonumber\\[-4pt]
\label{23042022-man-09} && m_1 \ne 0\,, \hspace{0.7cm} m_2\ne 0\,, \hspace{0.6cm} m_1 \ne m_2\,,  \hspace{0.5cm} m_3 = 0\,, \hspace{1cm} D > 0\,;
\\
\label{23042022-man-08} && m_1 = m\,, \hspace{0.5cm} m_2=m \,, \hspace{0.5cm} m\ne 0 \hspace{1.2cm} m_3 = 0\,, \hspace{1cm} D = 0\,, \hspace{0.5cm} \Pbf_{\epsilon m}=0\,; \hspace{1cm}
\\
&& \hspace{-0.5cm} \hbox{\bf Three massive fields},
\nonumber\\[-5pt]
\label{23042022-man-10} && m_1 \ne 0\,, \qquad m_2\ne 0\,, \qquad m_3 \ne 0\,, \hspace{1cm} D > 0;
\\
\label{23042022-man-11} && m_1 \ne 0\,, \qquad m_2\ne 0\,, \qquad m_3 \ne 0\,, \hspace{1cm} D < 0;
\\
\label{23042022-man-12} && m_1 \ne 0\,, \qquad m_2\ne 0\,, \qquad m_3\ne 0  \hspace{1.3cm} D = 0\,, \hspace{0.5cm}  \Pbf_{\epsilon m}=0\,.
\eeq

The following remarks are in order.

\noinbf{i}) In \rf{23042022-man-07}-\rf{23042022-man-08}, the restrictions on $D$ follow from the restrictions on the masses shown explicitly in \rf{23042022-man-07}-\rf{23042022-man-08}.
Namely, for $m_1=0$, $m_2=0$ \rf{23042022-man-07}, the restriction $m_3\ne 0$ in  \rf{23042022-man-07} leads  not only to the restriction $D\ne 0$ but also to the restriction $D > 0$. For $m_3=0$ \rf{23042022-man-09}, the restriction $m_1\ne m_2$ in \rf{23042022-man-09} implies not only the restriction $D\ne 0$, but also the restriction $D > 0$. For $m_3=0$ \rf{23042022-man-08}, the restriction $D=0$ follows from the restriction $m_1=m_2$ in \rf{23042022-man-08}.

\noinbf{ii}) For three massive fields, the restrictions on $D$ in  \rf{23042022-man-10}-\rf{23042022-man-12} provide additional restrictions on masses. We note that our use of $D$ for the classification of cubic vertices of massive fields is related to the following two reasons: 1) a structure of the cubic vertices and hermitian conjugation rules for coupling constants depend on the restrictions on $D$ and 2) depending on value of $D$, all processes of a decay of massive particle into two particles can be classified as follows,
\beq
\label{23042022-man-12a1}  && D >  0 \qquad \hbox{ for real processes with non-zero transfer of momentum};
\\
\label{23042022-man-12a2} && D =  0 \qquad  \hbox{ for real processes with zero transfer of momentum};
\\
\label{23042022-man-12a3} && D <  0 \qquad  \hbox{ for virtual processes}.
\eeq
We see that the classification of the processes  \rf{23042022-man-12a1}-\rf{23042022-man-12a3} match with the classification we are going to use for cubic vertices in \rf{23042022-man-10}-\rf{23042022-man-12}. For other comments, see Appendix G.

\noinbf{iii}) Classification in \rf{23042022-man-07}-\rf{23042022-man-08} is well known and was already used in Refs.\cite{Metsaev:2005ar,Metsaev:2007rn} upon building the cross-interactions between massive and massless fields in the framework of light-cone gauge approach. To our knowledge, classification \rf{23042022-man-10}-\rf{23042022-man-12} has not been used in the earlier literature.

In what follows we use the shortcut $(0,\lambda)$ for the massless field $\phi_\lambda^\dagger$ and the shortcut $(m,\lambda)$ for the massive field $\langle \phi_{m,s}|$
\beq
\label{23042022-man-14} && (0,\lambda) \hspace{1.5cm} \lambda \in \Zo\,, \hspace{3.1cm} \hbox{ for massless field $\phi_\lambda^\dagger$}\,;
\\
\label{23042022-man-15} && (m,s) \hspace{1cm} \hspace{0.5cm} m\ne 0\,, \hspace{0.5cm} s \in \No_0\,, \hspace{1cm} \hbox{ for massive field $\langle\phi_{m,s}|$}\,.
\eeq

\noinbf{On-shell cubic vertex and 3-point amplitudes}. Detailed analysis of 3-point amplitudes is beyond the scope of our study. However because our results provide quick and easy access to 3-point amplitudes we briefly comment the light-cone frame representation for the 3-point amplitudes. In the light-cone frame, the energy conservation law amounts to the equation $\Pbf^-=0$, where $\Pbf^-$ is given in \rf{20042022-man-15}. We will refer to this equation as on-shell condition. From \rf{20042022-man-15}, we see that the on-shell condition amounts to the equation for the momenta $\Po^\Rsm$, $\Po^\Lsm$ and $\beta_1$, $\beta_2$, $\beta_3$,
\be
\label{23042022-man-16}  \Po^\Rsm \Po^\Lsm - \frac{\rho^2}{2}  = 0 \,, \qquad \hbox{ on-shell condition},
\ee
where $\rho^2$ is defined in \rf{20042022-man-15}. Note that relation \rf{23042022-man-16} implies the on-shell restriction $\rho^2\geq 0$.
We are interested in a decay of a massive particle into two particles with non-collinear momenta for incoming and outgoing particles. For such decay, we have the restrictions $\Po^\Rsm\ne0$, $\Po^\Lsm\ne0$ and hence we can represent the on-shell condition \rf{23042022-man-16} as the following
\vspace{-10pt}
\beq
&& \hspace{-3cm} \hbox{ \bf{On-shell conditions for non-collinear momenta}:}
\nonumber\\
\label{23042022-man-18} && \Po^\Rsm = \frac{\rho}{\sqrt{2}}e^{\irm \varphi}\,, \qquad \Po^\Lsm = \frac{\rho}{\sqrt{2}} e^{-\irm \varphi}\,, \qquad  \rho > 0\,.
\eeq
Plugging the on-shell values of $\Po^\Rsm$, $\Po^\Lsm$ \rf{23042022-man-18} into the vertices $p_\smp3^-$ \rf{21042022-man-03} and $\VVb$ \rf{21042022-man-06}, we get the respective on-shell values of $p_\smp3^-$ and  $\VVb$.
From relations \rf{21042022-man-03}-\rf{21042022-man-09}, we then see that the on-shell values of the harmonic vertex $p_\smp3^-$ and the meromorphic vertex $\VVb$ coincide
\be \label{23042022-man-20}
p_\smp3^{-\,\on-sh} = \VVb^\on-sh\,.
\ee
Up to the sign, the 3-point invariant amplitude is equal to $p_\smp3^{-\,\on-sh}$ (for some details, see Appendix F)\,. Relation \rf{23042022-man-20} tells us then that the meromorphic vertex $\VVb$ provides us quick and easy access to the 3-point invariant amplitudes. All that is required is to plug the on-shell value of  $\Po^\Lsm$ \rf{23042022-man-18} into the meromorphic vertex $\VVb$.

\newsection{\large Vertex $\VVb$ for two massless and one massive
fields} \label{sec-07}

Using notation given in \rf{23042022-man-14},\rf{23042022-man-15}, we start with the meromorphic vertex for two massless fields and one massive field \rf{23042022-man-07},
\be \label{26042022-man-01}
(0,\lambda_1)-(0,\lambda_2)-(m_3,s_3)\,, \qquad  m_3 \ne  0\,,
\ee
i.e. two massless fields carry external line indices $a=1,2$,
while one massive field carries external line index $a=3$.
For this particular case, the general expression for the meromorphic vertex $\VVb$ \rf{21042022-man-18} takes the following form (for the derivation, see Appendix H):
\beq
\label{26042022-man-02} && \hspace{-1.4cm} \VVb = C_{\lambda_1,\lambda_2} N_{\lambda_1,\lambda_2}\VVb_{\lambda_1,\lambda_2}^\bas\,,
\\
\label{26042022-man-03} && \VVb_{\lambda_1,\lambda_2}^\bas \equiv  \beta_1^{-\lambda_1}\beta_2^{-\lambda_2}\beta_3^{s_3}
(\Po^\Lsm)^{\lambda_1+\lambda_2 -s_3}  L_{3+}^{s_3 -\lambda_1 + \lambda_2}
L_{3-}^{s_3 + \lambda_1 - \lambda_2}\,,
\\
\label{26042022-man-04} && L_{3 \pm } \equiv \frac{\Po^\Lsm}{\beta_3}u_3 +  \frac{ g_{3\pm}  }{\sqrt{2}}
v_3  \,, \hspace{1cm} g_{3+} \equiv - \frac{\beta_2 m_3}{\beta_3}\,, \hspace{1cm} g_{3-} \equiv  \frac{\beta_1 m_3}{\beta_3}\,,\qquad
\\
\label{26042022-man-06} && N_{\lambda_1,\lambda_2} \equiv 2^{(\lambda_1+\lambda_2+s_3)/2} m_3^{-\lambda_1-\lambda_2}\,,
\eeq
where coupling constant $C_{\lambda_1,\lambda_2}$, the helicities $\lambda_1$, $\lambda_2$, and the spin $s_3$ satisfy the restrictions
\beq
\label{26042022-man-07} && C_{\lambda_1,\lambda_2}^* = C_{-\lambda_1,-\lambda_2}\,,
\\
\label{26042022-man-07a1} && s_3 \geq |\lambda_1-\lambda_2|\,.
\eeq
Vertex $\VVb_{\lambda_1,\lambda_2}^\bas$ \rf{26042022-man-03} is a unique solution to the equations for the vertex $\VVb$. In the expression for the vertex $\VVb$ \rf{26042022-man-02}, we inserted the normalization factor  $N_{\lambda_1,\lambda_2}$ \rf{26042022-man-03} and the coupling constant $C_{\lambda_1,\lambda_2}$.%
\footnote{ The normalization factor $N_{\lambda_1,\lambda_2}$ is used for the convenience (see relations \rf{14022022-01a1}, \rf{14022022-01a2} in Appendix H).}
In general, the coupling constant $C_{\lambda_1,\lambda_2}$ depends not only on $\lambda_1$, $\lambda_2$ but also on the spin $s_3$ and the mass $m_3$. We show explicitly the labels $\lambda_1$, $\lambda_2$ because only these labels are not inert under the complex conjugation of the coupling constant \rf{26042022-man-07}. The following remarks are in order.

\noinbf{i}) Restriction \rf{26042022-man-07a1} is obtained by requiring that the powers of $L_1$ and $L_2$ in \rf{26042022-man-03} be non--negative integers.
Given values of $\lambda_1$, $\lambda_2$, and $s_3$ subjected to restriction \rf{26042022-man-07a1} there is only one vertex $\VVb$.%
\footnote{ By using the helicity-spinor language, 3-point amplitudes for arbitrary spin massless and massive particles were studied in Refs.\cite{Conde:2016vxs,Arkani-Hamed:2017jhn}. In these references, it was noted that, for the case under consideration in this Section, there is only one 3-point amplitude.}

We note then that the numbers of cubic vertices given in \rf{27042022-man-09c2}-\rf{27042022-man-09c4} coincide with the numbers of 3-point amplitudes given in (5.32) in Ref.\cite{Conde:2016vxs}.%

\noinbf{ii}) As seen from \rf{19042022-man-09-a2}, to get the hermitian $P_\smp3^-$, we need not only the vertex $\VVb$ \rf{26042022-man-02} but also the vertex $\II\, \VVb$ which is associated with $\II\, p_\smp3^-$. Realization of the operator $\II$ on the vertex $\VVb$ is given in \rf{24042022-man-10} in Appendix D. Using \rf{26042022-man-02} and \rf{24042022-man-10}, we get the relation
\be  \label{26042022-man-07a1-b1}
\II\, \VVb =
C_{\lambda_1,\lambda_2}^* N_{-\lambda_1,-\lambda_2}\VVb_{-\lambda_1,-\lambda_2}^\bas\,,
\ee
which tells us that $\II\,\VVb$ is associated with $\VVb_{-\lambda_1,-\lambda_2}^\bas$. This motivates us to represent $\II\,\VVb$ as
\be  \label{26042022-man-07a1-b2}
\II\, \VVb =
C_{-\lambda_1,-\lambda_2} N_{-\lambda_1,-\lambda_2}\VVb_{-\lambda_1,-\lambda_2}^\bas\,.
\ee
Comparison of \rf{26042022-man-07a1-b1} and \rf{26042022-man-07a1-b2} gives then  restriction \rf{26042022-man-07}.

\noinbf{iii}) Explicit expression for the harmonic vertex $p_\smp3^-$ associated with the meromorphic vertex $\VVb$ \rf{26042022-man-02}, \rf{26042022-man-03} can be obtained by expanding the meromorphic vertex $\VVb$ \rf{26042022-man-02} in Laurent series in $\Po^\Lsm$ and using \rf{21042022-man-06}-\rf{21042022-man-09}. As the explicit expression for $p_\smp3^-$ is not illuminating let us make comment on the general structure of $p_\smp3^-$. To this end we note that the Laurent series  expansion of the meromorphic vertex $\VVb$ \rf{26042022-man-02}, \rf{26042022-man-03} in $\Po^\Lsm$ can be presented as
\be \label{26042022-man-07a2}
\VVb = \sum_{n= \lambda_1+\lambda_2 -s_3}^{\lambda_1+\lambda_2 + s_3}(\Po^\Lsm)^n \VVb_n\,.
\ee
Using \rf{26042022-man-07a2}, it is easy to see that, depending on the values $\lambda_1$, $\lambda_2$, $s_3$, a general form of the harmonic vertex $p_\smp3^-$ \rf{21042022-man-09} can be presented as
\beq
\label{26042022-man-07a3} && p_\smp3^- = V_{s_3 - \lambda_1 - \lambda_2}(\Po^\Rsm)  + V_0
+  \Vb_{\lambda_1+\lambda_2 +s_3}(\Po^\Lsm)\,,
\nonumber\\
&& \hspace{5.2cm} \for \ \ \lambda_1+\lambda_2-s_3 < 0\,, \hspace{0.8cm} \lambda_1+\lambda_2+s_3 > 0\,;
\\
\label{26042022-man-07a4} && p_\smp3^- = V_{s_3 - \lambda_1 - \lambda_2}(\Po^\Rsm)  + V_0\,, \hspace{2cm}  \for \ \ \lambda_1+\lambda_2+s_3 \leq  0\,;\qquad
\\
\label{26042022-man-07a5} && p_\smp3^- = V_0 +  \Vb_{\lambda_1+\lambda_2 +s_3}(\Po^\Lsm)\,, \hspace{2cm} \for \ \ \lambda_1+\lambda_2-s_3 \geq 0\,;
\eeq
where, in \rf{26042022-man-07a3}-\rf{26042022-man-07a5}, the dependence of the vertices $V_N$, $V_0$, $\Vb_\Nb$ on the $\beta$-momenta and the oscillators is implicit. In \rf{26042022-man-07a4}, $V_0=0$ for $\lambda_1 + \lambda_2 + s_3<0$, while, in \rf{26042022-man-07a5}, $V_0=0$ for $\lambda_1 + \lambda_2 - s_3 > 0$. Comparing \rf{21042022-man-03} and \rf{26042022-man-07a5}, we see that, for $p_\smp3^-$ \rf{26042022-man-07a5}, we have $V_N=0$ and hence $V_N^\otimes=0$. Using \rf{21042022-man-06}, \rf{21042022-man-09}, we get then the relation
\be \label{26042022-man-07a6}
p_\smp3^- = \VVb \,, \qquad \for \quad \lambda_1+\lambda_2\geq s_3\,.
\ee
Thus, for $\lambda_1$, $\lambda_2$, $s_3$ that satisfy the restriction in  \rf{26042022-man-07a6}, the vertices $\VVb$ and $p_\smp3^-$ coincide. We recall that the restrictions in \rf{26042022-man-07a1} and \rf{26042022-man-07a6} are the well known triangle inequalities.

\noinbf{On-shell cubic vertex}. 3-point invariant amplitudes are expressed in terms of on-shell cubic vertices. For the decay processes with non-collinear momenta, the on-shell cubic vertex is obtained from off-shell vertex \rf{26042022-man-02} by using the on-shell conditions (see \rf{23042022-man-18}, \rf{23042022-man-20}). Consider the decay of the massive particle ($a=3$) into the two massless particles ($a=1,2$),
\be \label{26042022-man-08}
3 \rightarrow 1 + 2\,.
\ee
On-shell value of the off-shell vertex \rf{26042022-man-02} is found to be
\beq
\label{26042022-man-09} && \hspace{-2cm} \VVb^\on-sh =  (-)^{s_3} C_{\lambda_1\lambda_2} e^{-\irm(\lambda_1+\lambda_2)\varphi}  \Lbf_{3+}^{s_3 -\lambda_1 + \lambda_2}
\Lbf_{3-}^{s_3 + \lambda_1 - \lambda_2}\,,
\\
\label{26042022-man-10}  &&  \Lbf_{3+} \equiv  - \sqrt{-g_{3-}} e^{-\frac{\irm }{2}\varphi} u_3  + \sqrt{g_{3+}} e^{\frac{\irm }{2}\varphi} v_3 \,,
\\
\label{26042022-man-11} &&   \Lbf_{3-} \equiv  - \sqrt{g_{3+}} e^{-\frac{\irm }{2}\varphi} u_3  -  \sqrt{-g_{3-}} e^{\frac{\irm }{2}\varphi} v_3 \,,
\\
\label{26042022-man-12} && g_{3+} =  \frac{m_3}{1+r}\,, \hspace{0.5cm} g_{3-} =  -\frac{r m_3}{1+r}\,,
\\
\label{26042022-man-13} && r \equiv \beta_1/\beta_2\,, \qquad 0< r <\infty\,,
\eeq
where an angle variable $\varphi$ appearing in \rf{26042022-man-09}-\rf{26042022-man-11} is related to $\Po^\Lsm$ as
\be \label{26042022-man-14}
\Po^\Lsm = \frac{\rho}{\sqrt{2}} e^{-\irm \varphi}\,, \qquad \rho =  m_3 \beta_2\sqrt{r}\,.
\ee
Relations \rf{26042022-man-14} are obtained from \rf{23042022-man-18} and the definitions of $\rho^2$ and $r$ in \rf{20042022-man-15}, \rf{26042022-man-13}.  The allowed values of $r$ in \rf{26042022-man-13} are fixed by the condition $\rho^2>0$. We note the helpful inequalities,
\beq
\label{26042022-man-14a1}  && \beta_1 > 0\,, \qquad \beta_2 > 0\,, \qquad \beta_3 < 0\,,
\\
\label{26042022-man-14a2} && g_{3+}  > 0\,, \hspace{1cm} g_{3-} < 0\,,
\eeq
where in \rf{26042022-man-14a1}, we show the allowed values of  the $\beta$-momenta for the process \rf{26042022-man-08}, while restrictions \rf{26042022-man-14a2} are  obtained from definitions \rf{26042022-man-12}.
For the reader convenience, we note the following relations which are helpful for the derivation of the on-shell vertex \rf{26042022-man-09} from the off-shell vertex \rf{26042022-man-02}:
\be
L_{3\pm}\big|_\on-sh =   \sqrt{  \pm  \frac{g_{3\pm}}{2} } e^{-\frac{\irm }{2}\varphi}  \Lbf_{3\pm}\,, \qquad \frac{\rho}{\beta_3} = -  \sqrt{- g_{3+}g_{3-}}\,.
\ee

\newsection{\large Vertex $\VVb$ for two massive fields with non-equal masses and one massless field} \label{sec-08}

Using notation given in \rf{23042022-man-14},\rf{23042022-man-15}, we now consider the meromorphic vertex for two massive fields with non-equal masses and one massless field \rf{23042022-man-09},
\be \label{27042022-man-01}
(m_1,s_1)-(m_2,s_2)-(0,\lambda_3)\,, \qquad  m_1 \ne m_2\,,
\ee
i.e. two massive fields carry external line indices $a=1,2$,
while one massless field carries external line index $a=3$.
For this particular case, the general expression for the meromorphic vertex $\VVb$ \rf{21042022-man-18} takes the following form (for the derivation, see Appendix H):
\beq
\label{27042022-man-02} && \hspace{-1.2cm}\VVb = C_{n_1,n_2,\lambda_3}  N_{n_1,n_2,\lambda_3} \VVb_{n_1,n_2,\lambda_3}^\bas\,,
\\
\label{27042022-man-03} && \VVb_{n_1,n_2,\lambda_3}^\bas = \beta_1^{s_1}\beta_2^{s_2} \beta_3^{-\lambda_3} (\Po^\Lsm)^{-s_1 - s_2 +\lambda_3}
\prod_{a=1,2} L_{a+}^{s_a+n_a } L_{a-}^{s_a-n_a }\,,
\\
\label{27042022-man-04} && L_{a\pm } \equiv \frac{\Po^\Lsm}{\beta_a} u_a +  \frac{g_{a\pm}^{\vphantom{5pt}}}{\sqrt{2} }v_a\,, \hspace{1cm} g_{a\pm} \equiv g_a \pm \frac{\gamma}{m_a}\, \qquad g_a \equiv \frac{\betach_a}{2\beta_a}m_a + c_a m_a\,,\qquad
\nonumber\\
\label{27042022-man-06} && c_1 = \frac{m_2^2}{2m_1^2}\,, \qquad c_2 = -\frac{m_1^2}{2m_2^2}\,,\hspace{0.5cm} \gamma \equiv \half( m_1^2 -m_2^2)\,,
\\
\label{27042022-man-07} &&  N_{n_1,n_2,\lambda_3} \equiv 2^{(s_1+s_2+\lambda_3)/2} m_1^{-n_1}m_2^{-n_2}\,,
\eeq
where coupling constants $C_{n_1,n_2,\lambda_3}$ and integers $n_1,n_2\in \Zo$ satisfy the restrictions
\beq
\label{27042022-man-09} && C_{n_1,n_2,\lambda_3}^* = C_{-n_1,-n_2,-\lambda_3}\,,
\\
\label{27042022-man-09a1} && n_1 + n_2 =\lambda_3\,, \hspace{1.2cm}  n_1,n_2 \in \Zo\,,
\\
\label{27042022-man-09a2} && - s_1 \leq n_1 \leq s_1\,, \qquad - s_2 \leq n_2 \leq s_2\,.
\eeq

The vertices $\VVb_{n_1,n_2,\lambda_3}^\bas$ \rf{27042022-man-03} constitute  a basis of all solutions for the vertex $\VVb$. In the expression for vertex $\VVb$ \rf{27042022-man-02}, we inserted the normalization factors $N_{\lambda_1,\lambda_2}$ \rf{27042022-man-07} and the coupling constants $C_{n_1,n_2,\lambda_3}$. In general, the coupling constants depend not only on  $\lambda_3$ and the integers $n_1$, $n_2$ but also on the spins $s_1$, $s_2$ and the masses $m_1$, $m_2$. Only $\lambda_3$ and the integers $n_1$, $n_2$ are not inert under the complex conjugation of the coupling constants \rf{27042022-man-09}. The following remarks are in order.

\noinbf{i}) Restriction \rf{27042022-man-09a1} is obtained by solving equation \rf{21042022-man-22}, while restrictions \rf{27042022-man-09a2} are obtained by requiring that the powers of $L_1$ and $L_2$ in \rf{27042022-man-03}
be non--negative integers.

\noinbf{ii}) Two integers $n_1$, $n_2$ subjected to restrictions  \rf{27042022-man-09a1}, \rf{27042022-man-09a2} express the freedom of the solution for $\VVb$. These two integers label all possible vertices $\VVb$ that can be built for the fields in \rf{27042022-man-01}. Using restrictions \rf{27042022-man-09a1}, \rf{27042022-man-09a2}, we can find a number of the cubic vertices. Let us first note the following general restriction obtained from \rf{27042022-man-09a1}, \rf{27042022-man-09a2}:
\be \label{27042022-man-09c1}
-s_1-s_2\leq \lambda_3 \leq s_1 + s_2\,.
\ee
Now using restrictions \rf{27042022-man-09a1}, \rf{27042022-man-09a2}, we find the following numbers of the cubic vertices:
\beq
\label{27042022-man-09c2} && \nbf =  s_1 + s_2 - \lambda_3 + 1 \hspace{1.5cm} \qquad \for \qquad |s_1-s_2| \leq \lambda_3 \leq s_1 + s_2\,;
\\
\label{27042022-man-09c3} && \nbf =  s_1 + s_2 - |s_1-s_2| + 1 \hspace{0.4cm} \qquad \for\qquad  -|s_1-s_2| \leq \lambda_3  \leq |s_1-s_2|\,;
\\
\label{27042022-man-09c4} && \nbf =  s_1+s_2 + \lambda_3 + 1 \hspace{1.5cm} \qquad \for\qquad -s_1-s_2 \leq \lambda_3  \leq - |s_1-s_2|\,. \qquad
\eeq
We note then that the numbers of cubic vertices given in \rf{27042022-man-09c2}-\rf{27042022-man-09c4} coincide with the numbers of 3-point amplitudes given in (5.32) in Ref.\cite{Conde:2016vxs}.%
\footnote{ In Ref.\cite{Conde:2016vxs}, the authors found it convenient to assume $s_1\leq s_2$. We do not use such assumption.}

\noinbf{iii}) From \rf{19042022-man-09-a2} we learn that, to get the hermitian $P_\smp3^-$, we need not only the vertex $\VVb$ \rf{27042022-man-02} but also the vertex $\II\, \VVb$ which is associated with $\II\, p_\smp3^-$. Realization of the operator $\II$ on the vertex $\VVb$ is given  in \rf{24042022-man-10} in Appendix D. Using \rf{27042022-man-02}, \rf{24042022-man-10}, and \rf{24042022-man-11}, we find the expression for $\II\,\VVb$,
\be  \label{27042022-man-07a1-b1}
\II\, \VVb = C_{n_1,n_2,\lambda_3}^* N_{-n_1,-n_2,-\lambda_3} \VVb_{-n_1,-n_2,-\lambda_3}^\bas\,,\qquad
\ee
which tells us that $\II\,\VVb$ is associated with $\VVb_{-n_1,-n_2,-\lambda_3}^\bas$. This motivates us to represent $\II\,\VVb$ as
\be  \label{27042022-man-07a1-b2}
\II\, \VVb = C_{-n_1,-n_2,-\lambda_3} N_{-n_1,-n_2,-\lambda_3} \VVb_{-n_1,-n_2,-\lambda_3}^\bas\,.\qquad
\ee
The comparison of \rf{27042022-man-07a1-b1} and \rf{27042022-man-07a1-b2} gives restriction \rf{27042022-man-09}.

\noinbf{iv}) Explicit expression for the harmonic vertex $p_\smp3^-$ associated with the meromorphic vertex $\VVb$ \rf{27042022-man-02} can be obtained by expanding the meromorphic vertex $\VVb$ \rf{27042022-man-02} in Laurent series in $\Po^\Lsm$ and using \rf{21042022-man-06}-\rf{21042022-man-09}. As the explicit expression for  $p_\smp3^-$ is not illuminating let us make comment on the general structure of $p_\smp3^-$. To this end we note that the Laurent series expansion of the meromorphic vertex $\VVb$ \rf{27042022-man-02}, \rf{27042022-man-03} in $\Po^\Lsm$ can be presented as
\be \label{27042022-man-08a2}
\VVb = \sum_{n= -s_1-s_2 +\lambda_3}^{s_1+s_2 + \lambda_3}(\Po^\Lsm)^n \VVb_n\,.
\ee
Using \rf{27042022-man-09c1}, \rf{27042022-man-08a2}, it is easy to see that, a general form of the harmonic vertex $p_\smp3^-$ \rf{21042022-man-09} can be presented as
\be \label{27042022-man-08a3}
p_\smp3^- = V_{s_1+s_2-\lambda_3}(\Po^\Rsm)  + V_0
+  \Vb_{s_1+s_2 +\lambda_3}(\Po^\Lsm)\,,
\ee
where, in \rf{27042022-man-08a3}, the dependence of the vertices $V_N$, $V_0$, $\Vb_\Nb$ on the $\beta$-momenta and the oscillators is implicit. Note also that, for $\lambda_3 = \pm (s_1+s_2$), the vertex \rf{27042022-man-08a3} takes the form
\beq
\label{27042022-man-08a5} && p_\smp3^- =  V_{2s_1+2s_2}(\Po^\Rsm) + V_0\,, \hspace{1cm} \for \ \ \lambda_3 = - s_1- s_2\,;
\\
\label{27042022-man-08a4} && p_\smp3^- = V_0 +  \Vb_{2s_1+2s_2}(\Po^\Lsm)\,, \hspace{1cm} \for \ \ \lambda_3 = s_1+s_2\,.
\eeq
Comparing \rf{21042022-man-03} and \rf{27042022-man-08a4}, we see, that, for the vertex $p_\smp3^-$ \rf{27042022-man-08a3}, we have $V_N=0$ and hence $V_N^\otimes=0$. Using \rf{21042022-man-06}, \rf{21042022-man-09}, we get then the relation
\be \label{27042022-man-08a6}
p_\smp3^- = \VVb \,, \qquad \for \ \  \lambda_3 = s_1+s_2\,.
\ee
Thus, for $\lambda_3 = s_1+s_2$, the vertices $\VVb$ and $p_\smp3^-$ coincide.

\noinbf{v}) The quantities $g_{a\pm}$ defined in \rf{27042022-man-04} can alternatively be represented in a more explicit form as
\be
\label{27042022-man-10}   g_{1+} = -\frac{\beta_3}{\beta_1} m_1\,, \quad  g_{1-} = \frac{\beta_2}{m_1} \sum_{a=1,2}\frac{m_a^2}{\beta_a}\,, \qquad g_{2+} =  \frac{\beta_3}{\beta_2} m_2\,, \quad g_{2-} = -\frac{\beta_1}{m_2} \sum_{a=1,2}\frac{m_a^2}{\beta_a}\,.
\ee

\noinbf{On-shell cubic vertex}. 3-point amplitudes are expressed in terms of the on-shell cubic vertices. The on-shell cubic vertex is obtained from off-shell vertex \rf{27042022-man-02} by using on-shell conditions \rf{23042022-man-18}. Consider the decay of the massive particle ($a=1$) into the massive particle ($a=2$) and the massless particle ($a=3$)
\be \label{27042022-man-12}
1 \rightarrow 2 + 3\,, \hspace{1cm} m_1 > m_2\,.
\ee
The restriction on the masses in \rf{27042022-man-12} implies that momenta of the particles are non-collinear. For the non-collinear momenta, the on-shell value of the off-shell vertex \rf{27042022-man-02} is found to be
\beq
\label{27042022-man-13} && \hspace{-2cm} \VVb^\on-sh =   C_{n_1n_2\lambda_3} (-)^{s_1}  e^{-\irm \lambda_3 \varphi}
\prod_{a=1,2} \Lbf_{a+}^{s_a+n_a } \Lbf_{a-}^{s_a-n_a }\,,
\\
\label{27042022-man-14} && \Lbf_{a+}=  \varepsilon_a \sqrt{-g_{a-}} e^{-\frac{\irm }{2}\varphi} u_a  + \sqrt{g_{a+}} e^{\frac{\irm }{2}\varphi} v_a \,,
\nonumber\\
&& \Lbf_{a-}=  \varepsilon_a \sqrt{g_{a+}} e^{-\frac{\irm }{2}\varphi} u_a  -  \sqrt{-g_{a-}} e^{\frac{\irm }{2}\varphi} v_a \,,  \qquad \varepsilon_a =  {\rm sign} \beta_a\,, \qquad a=1,2;
\\
\label{27042022-man-15} && g_{1+} =   \frac{m_1 r}{1+r}\,, \hspace{1cm} g_{1-} =  - \frac{m_2^2(r_+-r)}{m_1(1+r)}\,,
\\
\label{27042022-man-16} && g_{2+} =  m_2 r\,, \hspace{1cm} g_{2-} =  - m_2 (r_+-r)\,,
\\
\label{27042022-man-17} && r \equiv \beta_3/\beta_2\,, \qquad r_+ \equiv \frac{m_1^2-m_2^2}{m_2^2}\,,\hspace{1cm} 0 < r < r_+\,,
\eeq
where an angle variable $\varphi$ appearing in \rf{27042022-man-14} is related to $\Po^\Lsm$ as
\be  \label{27042022-man-18}
\Po^\Lsm = \frac{\rho}{\sqrt{2}} e^{-\irm \varphi}\,, \hspace{1cm} \rho =  m_2\beta_2 \sqrt{r(r_+-r)}\,.
\ee
Relations \rf{27042022-man-18} are obtained from \rf{23042022-man-18}
and the definitions of $\rho^2$ and $r$ in \rf{20042022-man-15}, \rf{27042022-man-17}. The allowed values of $r$ in \rf{27042022-man-17} are fixed by requiring $\rho^2>0$. We note the following helpful inequalities
\beq
\label{27042022-man-18a1}  && \beta_1 < 0\,, \qquad \beta_2 > 0\,, \qquad \beta_3 > 0\,,
\\
\label{27042022-man-18a2} && g_{a+} >  0\,, \hspace{1cm} g_{a-} < 0\,, \qquad a=1,2\,,
\eeq
where in \rf{27042022-man-18a1}, we show the allowed values of  the $\beta$-momenta for the process in \rf{27042022-man-12}, while restrictions \rf{27042022-man-18a2} follow from restriction \rf{27042022-man-12} and the definitions in \rf{27042022-man-15},\rf{27042022-man-16}.

For the reader convenience, we note the following relations which are helpful for the derivation of the on-shell vertex \rf{27042022-man-13} from the off-shell vertex \rf{27042022-man-02}:
\be
L_{a\pm}\big|_\on-sh= \sqrt{ \pm \frac{g_{a\pm}}{2} } e^{-\frac{\irm }{2}\varphi} \Lbf_{a\pm}\,, \hspace{1cm} \frac{\rho}{\beta_a} = \varepsilon_a \sqrt{-g_{a+}g_{a-}}\,, \qquad a=1,2\,.
\ee

\newsection{\large Vertex $\VVb$ for two massive fields with equal masses and one massless field}  \label{sec-09}

Using notation given in \rf{23042022-man-14},\rf{23042022-man-15}, we now consider the meromorphic vertex for two massive fields with equal masses and one massless field \rf{23042022-man-08},
\be \label{28042022-man-01}
(m_1,s_1)-(m_2,s_2)-(0,\lambda_3)\,,\qquad  m_1=m\,, \quad m_2=m\,, \quad m\ne 0\,,
\ee
i.e. two massive fields carry external line indices $a=1,2$,
while one massless field carries external line index $a=3$.
For this particular case, the general expression for the meromorphic vertex $\VVb$ \rf{21042022-man-18} takes the following form (for the derivation, see Appendix H):
\beq
\label{28042022-man-02} && \hspace{-1.3cm} \VVb = C_{n,\lambda_3}  N_{n,\lambda_3} \VVb_{n,\lambda_3}^\bas \,,
\\
\label{28042022-man-03} && \VVb_{n,\lambda_3}^\bas   =  \beta_1^{s_1}\beta_2^{s_2}\beta_3^{-\lambda_3} (\Po^\Lsm)^{-s_1-s_2+\lambda_3} L_1^{2s_1-n} L_2^{2s_2-n} Q^n \,,
\\
\label{28042022-man-04} && L_1\equiv \frac{\Po^\Lsm}{\beta_1} u_1 -\frac{m \beta_3}{\sqrt{2}\beta_1} v_1\,,  \hspace{1cm} L_2 \equiv \frac{\Po^\Lsm}{\beta_2} u_2 + \frac{m \beta_3}{\sqrt{2}\beta_2} v_2\,,
\\
\label{28042022-man-06}  && Q = v_1 L_2 - v_2 L_1 - \frac{1}{\sqrt{2} m}L_1 L_2\,,
\\
\label{28042022-man-07}  && N_{n,\lambda_3} \equiv \irm^n \big(\frac{\sqrt{2}}{m}\big)^{\lambda_3}\,,
\eeq
where coupling constants $C_{n,\lambda_3}$ and integer $n\in \No_0$ satisfy the restrictions
\beq
\label{28042022-man-08} && C_{n,\lambda_3}^* = C_{n,-\lambda_3}\,,
\\
\label{28042022-man-08a1} && 0 \, \leq \,  n \, \leq 2 s_\minrm\,, \qquad s_\minrm \equiv \min_{a=1,2} s_a\,.
\eeq
The vertices $\VVb_{n,\lambda_3}^\bas$ \rf{28042022-man-03} constitute  a basis of all solutions for the vertex $\VVb$. In the full expression for the vertex $\VVb$ \rf{28042022-man-02}, we inserted the normalization factors $N_{n,\lambda_3}$ \rf{28042022-man-07} and the coupling constants $C_{n,\lambda_3}$. In general, the coupling constants depend not only on $n$, $\lambda_3$ but also on the spins $s_1$, $s_2$ and the mass $m$. We note that only $\lambda_3$ is not inert under the complex conjugation of the coupling constants \rf{28042022-man-08}. The following comments are in order.

\noinbf{i)} The restrictions \rf{28042022-man-08a1} are obtained by requiring that the powers of $L_1$, $L_2$, and $Q$ in \rf{28042022-man-03}
be non--negative integers.

\noinbf{ii}) The integer $n$ subjected to restrictions  \rf{28042022-man-08a1} expresses the freedom of the solution for $\VVb$. This   integer labels all possible vertices $\VVb$ that can be built for the fields in \rf{28042022-man-01}. Now using restrictions \rf{28042022-man-08a1}, we find the following number of the cubic vertices:
\be \label{28042022-man-08c1}
\nbf = 2s_\minrm +1\,,
\ee
where $s_\minrm$ is given in \rf{28042022-man-08a1}. We note then that the number of cubic vertices given in \rf{28042022-man-08c1} coincides with the number of 3-point amplitudes given in  Ref.\cite{Arkani-Hamed:2017jhn}.%
\footnote{ In Ref.\cite{Arkani-Hamed:2017jhn}, the authors use the relation $2s_\minrm = s_1 + s_2 - |s_1-s_2|$.}

\noinbf{iii}) The Laurent series expansion of the meromorphic vertex $\VVb$ \rf{28042022-man-02} in $\Po^\Lsm$ takes the form
\be \label{28042022-man-08c1b0}
\VVb = \sum_{n= -s_1-s_2 +\lambda_3}^{s_1+s_2 + \lambda_3}(\Po^\Lsm)^n \VVb_n\,.
\ee
Using \rf{28042022-man-08c1b0} , it is easy to see that, depending on the values $s_1$, $s_2$, $\lambda_3$, a general form of the harmonic vertex $p_\smp3^-$ \rf{21042022-man-09} can be presented as
\beq
\label{28042022-man-08c1b1} && p_\smp3^- = V_{s_1 + s_2 -\lambda_3}(\Po^\Rsm)  + V_0 +  \Vb_{s_1+s_2 +\lambda_3}(\Po^\Lsm)\,,
\nonumber\\
&& \hspace{5.2cm} \for \ \ \lambda_3 - s_1-s_2   < 0\,, \hspace{0.8cm} s_1+s_2+\lambda_3 > 0\,;
\\
\label{28042022-man-08c1b2} && p_\smp3^- = V_{s_1 + s_2 - \lambda_3}(\Po^\Rsm)  + V_0\,, \hspace{1.5cm}  \for  \ \ s_1 + s_2 + \lambda_3 \leq  0\,;\qquad
\\
\label{28042022-man-08c1b3} && p_\smp3^- = V_0 +  \Vb_{s_1+s_2 +\lambda_3}(\Po^\Lsm)\,, \hspace{1.5cm} \for \ \ \lambda_3 - s_1 - s_2 \geq 0\,.
\eeq
In \rf{28042022-man-08c1b2}, $V_0=0$ for $s_1 + s_2 + \lambda_3<0$, while, in \rf{28042022-man-08c1b3}, $V_0=0$ for $\lambda_3 - s_1 - s_2 > 0$. Comparing \rf{21042022-man-03} and \rf{28042022-man-08c1b3}, we see that, for $p_\smp3^-$ \rf{28042022-man-08c1b3}, we have $V_N=0$ and hence $V_N^\otimes=0$. Using \rf{21042022-man-06}, \rf{21042022-man-09}, we get then the relation
\be \label{28042022-man-08c1b4}
p_\smp3^- = \VVb \,, \qquad \for \quad \lambda_3 - s_1 - s_2 \geq 0\,.
\ee

\noinbf{iv}) From \rf{19042022-man-09-a2}, we see that, to get the hermitian $P_\smp3^-$, we need not only the vertex $\VVb$ \rf{28042022-man-02} but also the vertex $\II\, \VVb$ which is associated with $\II\, p_\smp3^-$. Realization of the operator $\II$ on the vertex $\VVb$ is given in \rf{24042022-man-10} in Appendix D. Using \rf{28042022-man-02}, \rf{24042022-man-10}, \rf{15022022-26}, we find the expression for $\II\,\VVb$,
\be  \label{28042022-man-07a1-b1}
\II\, \VVb  =  C_{n,\lambda_3}^* N_{n,-\lambda_3}\VVb_{n,-\lambda_3}^\bas\,,
\ee
which tells us that $\II\,\VVb$ is associated with the vertex $\VVb_{n,-\lambda_3}^\bas$. This motivates us to represent $\II\,\VVb$ as
\be  \label{28042022-man-07a1-b2}
\II\, \VVb  =  C_{n,-\lambda_3} N_{n,-\lambda_3}\VVb_{n,-\lambda_3}^\bas\,.
\ee
Comparison of \rf{28042022-man-07a1-b1} and \rf{28042022-man-07a1-b2} gives restriction \rf{28042022-man-08}.

\noinbf{v}) Decay of the massive particle with the mass $m$ into one massive particle with the same mass $m$ and one massless particle is prohibited by the energy-momentum conservation laws. We skip therefore a discussion of the on-shell reduction of vertices \rf{28042022-man-02}.

\newsection{\large Vertex $\VVb$ for three massive fields with non-critical masses, $D\ne 0$}
\label{sec-10}

Using notation \rf{23042022-man-14}, we consider the meromorphic vertex for three massive fields \rf{23042022-man-10}, \rf{23042022-man-11},
\be \label{29042022-man-01}
(m_1,s_1)-(m_2,s_2)-(m_3,s_3)\,,\qquad  D \ne 0\,,\quad m_a\ne 0\,, \qquad a=1,2,3\,, \qquad
\ee
i.e. three massive fields carry external line indices $a=1,2,3$.
The general expression for the meromorphic vertex $\VVb$ \rf{21042022-man-18} takes the following form (for the derivation, see Appendix H):
\beq
\label{29042022-man-02} && \hspace{-1cm} \VVb  =  C_{n_1,n_2,n_3} N_{n_1,n_2,n_3} \VVb_{n_1,n_2,n_3}^\bas + C_{-n_1,-n_2,-n_3} N_{-n_1,-n_2,-n_3} \VVb_{-n_1,-n_2,-n_3}^\bas\,, \hspace{0.3cm} \for \ \ D>0\,,\qquad
\\
\label{29042022-man-03} && \hspace{-1cm} \VVb  =  C_{n_1,n_2,n_3} N_{n_1,n_2,n_3} \VVb_{n_1,n_2,n_3}^\bas\,,\hspace{7.3cm} \for \ \ D <  0 \,,
\\
\label{29042022-man-04} && \VVb_{n_1,n_2,n_3}^\bas  \equiv   \prod_{a=1,2,3}
L_{a+}^{s_a+n_a} L_{a-}^{s_a - n_a} \Big(\frac{\Po^\Lsm}{\beta_a}\Big)^{-s_a}\,,
\\
\label{29042022-man-06} && L_{a\pm } \equiv \frac{\Po^\Lsm }{\beta_a}u_a+
\frac{g_{a\pm}}{\sqrt{2}} v_a\,,
\\
\label{29042022-man-07} && g_{a\pm} \equiv g_a \pm \frac{\gamma}{m_a}\,,\qquad   g_a \equiv \frac{\betach_a}{2\beta_a}m_a + c_a m_a\,, \qquad c_a = \frac{m_{a+1}^2 - m_{a+2}^2}{2m_a^2}\,,\qquad
\\
\label{29042022-man-08} && \gamma \equiv \half \sqrt{D}   \hspace{0.5cm} \hbox{ for } \ D>0\,; \hspace{1cm}  \gamma \equiv \frac{\irm}{2} \sqrt{-D}  \hspace{0.5cm} \hbox{ for } \ D<0\,,
\\
\label{29042022-man-09} && N_{n_1,n_2,n_3} \equiv  \prod_{a=1,2,3} 2^{s_a/2} \kappa_a^{(n_{a+1}-n_{a+2})/6}\,,
\nonumber\\
&& \kappa_a \equiv \frac{Y_{a+}}{Y_{a-}}\,, \qquad
Y_{a\pm} \equiv m_a^2 - m_{a+1}^2 - m_{a+2}^2  \pm 2\gamma  \,,
\eeq
where coupling constants $C_{n_1,n_2,n_3}$ and integers $n_1,n_2,n_3 \in \Zo$ satisfy the restrictions
\beq
\label{29042022-man-11} && C_{n_1,n_2,n_3}^* = C_{-n_1,-n_2,-n_3}\,, \hspace{1.5cm}  \for \ \ D >0\,;
\\
\label{29042022-man-12} && C_{n_1,n_2,n_3}^* = C_{n_1,n_2,n_3}\,, \hspace{2.2cm} \for \ \ D < 0\,;
\\
\label{29042022-man-12a1} &&  n_1 + n_2 + n_3 = 0\,,\qquad n_1,n_2,n_3 \in \Zo\,,
\\
\label{29042022-man-12a2} && - s_a \leq n_a \leq s_a\,, \hspace{1.3cm} a=1,2,3\,.
\eeq
The vertices $\VVb_{n_1,n_2,n_3}^\bas$ \rf{29042022-man-04} constitute a basis of all solutions for the vertex $\VVb$. In the full expressions for the vertices $\VVb$  \rf{29042022-man-02} and \rf{29042022-man-03}, we inserted the normalization factors $N_{n_1,n_2,n_3}$ \rf{29042022-man-09} and the coupling constants $C_{n_1,n_2,n_3}$. In general, the coupling constants depend not only on the integers $n_1$, $n_2$, $n_3$ but also on the spins $s_a$ and the masses $m_a$, $a=1,2,3$. For $D>0$, only the integers $n_1$, $n_2$, $n_3$ are not inert under the complex conjugation of the coupling constants \rf{29042022-man-11}.
The following remarks are in order.

\noinbf{i}) Restriction \rf{29042022-man-12a1} is obtained by solving equation \rf{21042022-man-22}, while restrictions \rf{29042022-man-12a2} are obtained by requiring that the powers of $L_1$, $L_2$, and $L_3$ in \rf{29042022-man-04} be non--negative integers.

\noinbf{ii}) The integers $n_1$, $n_2$, $n_3$ subjected to restrictions  \rf{29042022-man-12a1}, \rf{29042022-man-12a2} express the freedom of the solution for $\VVb$. These  integers label all possible vertices $\VVb$ that can be built for the fields in \rf{29042022-man-01}. Now using the restrictions \rf{29042022-man-12a1}, \rf{29042022-man-12a2}, we find the following number of cubic vertices:
\beq
\label{29042022-man-12c1} && \hspace{-2cm} \nbf = \big( s_1+s_2 - |s_1-s_2|+1 \big)  (2s_3 + 1)\,, \hspace{2cm}  \for \ \ s_3 \leq |s_1-s_2|\,;
\\
\label{29042022-man-12c2} && \hspace{-2cm}  \nbf = (2s_1 +1) (2s_2+1) - (s_1+s_2-s_3)(s_1+s_2-s_3+1)\,,
\nonumber\\
&& \hspace{7cm}  \for \ \ |s_1-s_2| \leq s_3 \leq s_1+s_2\,;
\\
\label{29042022-man-12c3} && \hspace{-2cm}  \nbf = (2s_1 +1) (2s_2+1)\,, \hspace{4.7cm}  \for \ \ s_1+s_2 \leq s_3\,.
\eeq
To our knowledge our result in \rf{29042022-man-12c1}-\rf{29042022-man-12c3} has not been discussed in the earlier literature.

\noinbf{iii}) Explicit expressions for the harmonic vertices $p_\smp3^-$ can be obtained by expanding the meromorphic vertices $\VVb$ \rf{29042022-man-02}, \rf{29042022-man-03} in Laurent series in $\Po^\Lsm$ and using \rf{21042022-man-06}-\rf{21042022-man-09}. As the explicit Taylor series expansions for the harmonic vertices $p_\smp3^-$ in $\Po^\Rsm$ and $\Po^\Lsm$ are not illuminating we briefly comment on the general structure of $p_\smp3^-$. To this end we note that the Laurent series expansion of the meromorphic  vertices $\VVb$ \rf{29042022-man-02}, \rf{29042022-man-03} in  $\Po^\Lsm$ can be presented as
\be \label{29042022-man-12a3}
\VVb = \sum_{n= -s_1-s_2 -s_3}^{s_1+s_2 + s_3}(\Po^\Lsm)^n \VVb_n\,.
\ee
Using \rf{29042022-man-12a3}, it is easy to see that, for all values $s_1$, $s_2$, $s_3$, a general form of the harmonic vertex $p_\smp3^-$ \rf{21042022-man-09} can be presented as
\be \label{29042022-man-12a4}
p_\smp3^- = V_{s_1 + s_2 + s_3}(\Po^\Rsm)  + V_0
+  \Vb_{s_1+s_2 +s_3}(\Po^\Lsm)\,,
\ee
where, in \rf{29042022-man-12a4}, the dependence of the vertices $V_{s_1+s_2+s_3}$, $V_0$, $\Vb_{s_1+s_2+s_3}$ on the $\beta$-momenta and the oscillators is implicit.

\noinbf{On-shell values of cubic vertex for $D>0$}. 3-point  amplitudes are expressed in terms of on-shell cubic vertices obtained from off-shell vertices \rf{29042022-man-02} by using on-shell conditions \rf{23042022-man-18}. Consider the decay of the massive particle ($a=1$) into the two massive particles ($a=2,3$),
\be \label{29042022-man-13}
1 \rightarrow 2 + 3\,, \qquad  m_1 > m_2 + m_3\,,
\ee
where in \rf{29042022-man-13}, for the decay with non-collinear momenta, we recall the restriction on the masses.

On-shell value of the off-shell vertex \rf{29042022-man-02} is found to be
\beq
\label{29042022-man-14} && \hspace{-1cm} \VVb^\on-sh =  C_{n_1,n_2,n_3} \bar\Vbf_{n_1,n_2,n_3}^\bas + C_{-n_1,-n_2,-n_3} \bar\Vbf_{-n_1,-n_2,-n_3}^\bas\,,
\\
\label{29042022-man-15} && \bar\Vbf_{n_1,n_2,n_3}^\bas \equiv  (-)^{s_1} \prod_{a=1,2,3} \Lbf_{a+}^{s_a+n_a} \Lbf_{a-}^{s_a - n_a}\,,
\\
\label{29042022-man-16} && \Lbf_{a+} \equiv  \varepsilon_a\sqrt{-g_{a-}} e^{-\frac{\irm }{2}\varphi} u_a  + \sqrt{g_{a+}} e^{\frac{\irm }{2}\varphi} v_a\,,
\nonumber\\
&& \Lbf_{a-} \equiv  \varepsilon_a  \sqrt{g_{a+}} e^{-\frac{\irm }{2}\varphi} u_a  -  \sqrt{-g_{a-}} e^{\frac{\irm }{2}\varphi} v_a \,, \qquad \varepsilon_a = \sign \beta_a\,, \qquad a=1,2,3\,; \qquad
\\
\label{29042022-man-17} && g_{1\pm} = \frac{Y_{3\mp}}{2 m_1 (1+r)} (r_\mp-r)\,, \hspace{0.5cm} g_{2\pm} = m_2(r-r_\mp)\,, \hspace{0.5cm} g_{3\pm} = \frac{m_3}{r_\mp r} (r-r_\mp)\,,\qquad
\\
\label{29042022-man-18} && r \equiv \beta_3/\beta_2\,, \hspace{1cm} r_\pm \equiv \big( m_1^2 -m_2^2 - m_3^2 \pm \sqrt{D}\big)/ 2m_2^2\,,\qquad r_- < r < r_+\,,
\\
\label{29042022-man-18a1} && Y_{3\pm} \equiv  m_3^2 -m_1^2 - m_2^2 \pm \sqrt{D}\,,
\eeq
where an angle variable $\varphi$ appearing in \rf{29042022-man-16} is related to $\Po^\Lsm$ as
\be  \label{29042022-man-19}
\Po^\Lsm = \frac{\rho}{\sqrt{2}} e^{-\irm \varphi}\,,\hspace{1cm}  \rho =   m_2 \beta_2 \sqrt{(r_+-r)(r-r_-)}\,.
\ee
Relations \rf{29042022-man-19} are obtained from \rf{23042022-man-18} and the definitions of $\rho^2$ and $r$ in \rf{20042022-man-15}, \rf{29042022-man-18}. The allowed values of $r$ in \rf{29042022-man-18} are fixed by requiring $\rho^2>0$.
We note the helpful inequalities,
\beq
\label{29042022-man-20}  && \beta_1 < 0\,, \qquad \beta_2 > 0 \,, \qquad \beta_3 > 0\,,
\\
\label{29042022-man-20a1} && g_{a+} >  0, \qquad g_{a-} < 0, \qquad a=1,2,3; \qquad Y_{3 \pm } < 0\,,\qquad r_->0\,, \qquad
\eeq
where in \rf{29042022-man-20}, we show the allowed values of  the $\beta$-momenta for the process in \rf{29042022-man-13}, while restrictions \rf{29042022-man-20a1} follow from restriction \rf{29042022-man-13} and the definitions in \rf{29042022-man-17}-\rf{29042022-man-18a1}. For the reader convenience, we note also the following relations which are helpful for the derivation of the on-shell vertex \rf{29042022-man-14} from the off-shell vertex \rf{29042022-man-02},
\be
L_{a\pm}\big|_\on-sh= \sqrt{\pm \frac{g_{a\pm}}{2} } e^{-\frac{\irm }{2}\varphi} \Lbf_{a\pm}\,, \hspace{1cm} \frac{\rho}{\beta_a} = \varepsilon_a \sqrt{-g_{a+}g_{a-}}\,,  \qquad a=1,2,3\,.\qquad
\ee

\newsection{\large Vertex $\VVb$ for three massive fields  with critical masses, $D = 0$}
\label{sec-11}

Using notation \rf{23042022-man-14}, we finish with the meromorphic vertex for three massive fields \rf{23042022-man-12},
\be \label{30042022-man-01}
(m_1,s_1)-(m_2,s_2)-(m_3,s_3)\,, \qquad  D =0\,, \quad \Pbf_{\epsilon m}=0\,, \quad m_a\ne 0\,, \quad a=1,2,3\,,\quad
\ee
i.e. three massive fields carry external line indices $a=1,2,3$.
For this particular case, the general expression for the vertex $\VVb$ \rf{21042022-man-18} takes the following form (for the derivation, see Appendix H):
\beq
\label{30042022-man-02} && \hspace{-1cm} \VVb  =  C_{n,l} N_{n,l} \VVb_{n,l}^\bas\,,
\\
\label{30042022-man-03} && \VVb_{n,l}^\bas  =   Q_X^n Q_Y^l \prod_{a=1,2,3} L_a^{2s_a-n-l} \big(\frac{\Po^\Lsm}{\beta_a}\big)^{-s_a}
\,,
\\
\label{30042022-man-04} && L_a \equiv  \frac{\Po^\Lsm}{\beta_a} u_a + \frac{\epsilon_a \Po_{\epsilon m}}{\sqrt{2} \beta_a} v_a\,,
\hspace{1cm}  \Po_{\epsilon m}\equiv \frac{1}{3}\sum_{a=1,2,3} \betach_a\epsilon_a m_a\,, \quad \betach_a \equiv \beta_{a+1} - \beta_{a+2}\,,\qquad
\\
\label{30042022-man-05}  && Q_X \equiv \sum_{a=1,2,3} c_a^X v_a L_{a+1}L_{a+2}\,,\hspace{3cm} c_a^X = \epsilon_a \,,
\\
\label{30042022-man-07} && Q_Y \equiv \frac{1}{\sqrt{2}} L_1 L_2 L_3  + \sum_{a=1,2,3} c_a^Y  v_a L_{a+1}L_{a+2}\,, \hspace{0.5cm}  c_a^Y = \frac{1}{3} \epsilon_a (\epsilon_{a+1} m_{a+1} - \epsilon_{a+2} m_{a+2})\,,\qquad
\\
\label{30042022-man-08}  && N_{n,l} \equiv \irm^{n+l}\,,
\eeq
where coupling constants $C_{n,l}$ and integers $n,l \in \No_0$ satisfy the restrictions
\beq
\label{30042022-man-09} && C_{n,l}^* = C_{n,l}\,,
\\
\label{30042022-man-10} && n \geq 0\,, \qquad l\geq 0\,, \qquad n + l\, \leq 2 s_\minrm\,, \qquad s_\minrm \equiv \min_{a=1,2,3} s_a\,.
\eeq
The vertices $\VVb_{n,l}^\bas$ \rf{30042022-man-03} constitute a basis of all solutions for the vertex $\VVb$. In the full expression for the vertex $\VVb$ \rf{30042022-man-02}, we inserted the normalization factors $N_{n,l}$ \rf{30042022-man-08} and the coupling constants $C_{n,l}$. In general, the coupling constants depend not only on the integers $n$ and $l$ but also on the spins $s_a$ and the masses $m_a$, $a=1,2,3$.
We note that restrictions \rf{30042022-man-10} are obtained by requiring that the powers of $Q_X$, $Q_Y$, $L_1$, $L_2$, and $L_3$ in \rf{30042022-man-03}
be non--negative integers. Two integers $n$, $l$ subjected to restrictions  \rf{30042022-man-10} express the freedom of the solution for $\VVb$. These  integers label all possible vertices $\VVb$ that can be built for the fields in \rf{30042022-man-01}. Using the restrictions \rf{30042022-man-10}, we find the following number of cubic vertices:
\be \label{30042022-man-10a1}
\nbf =  (2s_\minrm + 1)(s_\minrm + 1)\,,
\ee
where $s_\minrm$ is defined in \rf{30042022-man-10}. To our knowledge our result in \rf{30042022-man-10a1} has not been discussed in the earlier literature.

The Laurent series expansion of the meromorphic vertex $\VVb$ \rf{30042022-man-02} in  $\Po^\Lsm$ and the corresponding Taylor series expansion of the harmonic vertex $p_\smp3^-$ in $\Po^\Rsm$ and $\Po^\Lsm$ can schematically be presented as
\be
\label{30042022-man-11}  \VVb = \sum_{n= -s_1-s_2 -s_3}^{s_1+s_2 + s_3}(\Po^\Lsm)^n \VVb_n\,, \qquad  p_\smp3^- = V_{s_1 + s_2 + s_3}(\Po^\Rsm)  + V_0
+  \Vb_{s_1+s_2 +s_3}(\Po^\Lsm)\,,
\ee
where, in \rf{30042022-man-11}, the dependence of the vertices $V_{s_1+s_2+s_3}$, $V_0$, $\Vb_{s_1+s_2+s_3}$ on the $\beta$-momenta and the oscillators is implicit.

For $D=0$, the decay of one massive particle into two massive particles is allowed only for collinear momenta of the particles. The collinear momenta lead to the restrictions $\Po^\Rsm=0$, $\Po^\Lsm=0$. For such $\Po^\Rsm$, $\Po^\Lsm$, the on-shell cubic vertices are governed by $V_0$ \rf{30042022-man-11}. We skip a discussion of such on-shell cubic vertices.

\newsection{ \large Conclusions}\label{concl}

In this paper, we used the light-cone gauge helicity basis formalism for the investigation of cubic interactions of arbitrary integer spin massive and massless fields in $R^{3,1}$.
We studied cross-interactions between massive and massless fields and interactions between massive fields. Depending on masses of the fields, we introduced the classification of cubic vertices. As convenient representatives of cubic vertices we used the harmonic vertices. To each harmonic vertex we associated the meromorphic vertex. The harmonic vertex and the meromorphic vertex are in one-to-one correspondence and they are related to each other by simple transformation rule. We found simple explicit expressions for all meromorphic vertices. On-shell expressions for the cubic vertices and the meromorphic vertices coincide.
We expect that the methods and the results obtained in this paper might have the  following applications and generalizations.

\smallskip
\noindent \ibf) In this paper, we restricted our study to the bosonic fields. The light-cone gauge helicity basis formalism is well adapted also for a study of fermionic fields. Application of our method for a study of Fermi-Bose couplings would be interesting.
The Fermi-Bose couplings of the massive and massless light-cone gauge tensor (tensor-spinor) fields in $R^{d-1,1}$, $d>4$, were studied in Ref.\cite{Metsaev:2007rn}. The study of Fermi-Bose couplings of massless fields in $R^{3,1}$ by using the light-cone gauge helicity basis formalism may be found in Ref.\cite{Akshay:2015kxa}.
The investigation of electromagnetic and gravitational couplings of fermionic fields by using BRST approach may be found in Ref.\cite{Henneaux:2012wg}.%
\footnote{ Recent discussion of the interesting formulation of free fermionic fields may be found in Ref.\cite{Najafizadeh:2018cpu}.}

\smallskip
\noinbf{ii}) In Ref.\cite{Metsaev:2012uy}, we used the bosonic vector-like oscillators to build all BRST-BV parity-even cubic vertices for interacting massive and massless fields in $R^{d-1,1}$, $d\geq 4$.%
\footnote{ BRST-BV cubic vertex for the massive gravity in $AdS$ space was discussed in Ref.\cite{Boulanger:2018dau}. BRST-BV parity-even cubic vertices for arbitrary spin massless fields in flat space were discussed in Refs.\cite{Metsaev:2012uy,Fotopoulos:2010ay,Buchbinder:2021xbk}. In the earlier literature, BRST-BV cubic vertex for spin-3 massless fields was considered in Ref.\cite{Bekaert:2005jf}. For the interesting recent development, see Ref.\cite{Sakaguchi:2020sxi}. For various metric-like formulations of cubic vertices for arbitrary spin massless fields, see Refs.\cite{Manvelyan:2010jr}, while, for the metric-like formulation of cubic vertices for spin-2 massive  fields, see Ref.\cite{Zinoviev:2013hac}.}
However our research in this paper convinced us that the bosonic spinor-like oscillators are more convenient for the study of interacting massive and massless fields in $R^{3,1}$.  Use of the bosonic spinor-like oscillators for the investigation of BRST-BV free fields in $R^{3,1}$ may be found in Ref.\cite{Buchbinder:2015kca}. We think that building of BRST-BV counterparts of our light-cone gauge cubic vertices of massive fields by using spinor-like oscillators could be of interest. Also note that the bosonic spinor-like oscillators will provide us the possibility to treat BRST-BV parity-even and parity-odd vertices for fields in $R^{3,1}$ on an equal footing. In Ref.\cite{Metsaev:2012uy}, we studied BRST-BV parity-even cubic vertices for arbitrary spin fields. To our knowledge, BRST-BV parity-odd vertices for arbitrary spin fields have not been studied in the literature. For fields of some particular values of spins (so called Curtright fields) in $R^{4,1}$ and $R^{6,1}$, the discussion of BRST-BV parity-odd and parity-even cubic vertices may be found in Ref.\cite{Brandt:2020pry}. Light-cone gauge parity-odd and parity-even cubic vertices for arbitrary spin massless fields in $R^{4,1}$ and $R^{5,1}$ were considered in Refs.\cite{Metsaev:1993gx,Metsaev:1993mj}.
Lorentz covariant parity-odd (and parity even) cubic vertices for arbitrary spin on-shell massless TT fields in $R^{3,1}$ were built in Ref.\cite{Conde:2016izb}. Parity-even and parity-odd cubic vertices of higher-spin massless fields in $R^{2,1}$ were considered in Refs.\cite{Mkrtchyan:2017ixk} (see also \cite{Zinoviev:2021cmi}).

\smallskip
\noinbf{iii}) Light-cone gauge approach turns out to be convenient for the study of supersymmetric higher-spin theories. For arbitrary spin massless supermultiplets in  $R^{3,1}$ and massive supermultiplets in $R^{2,1}$, the light-cone gauge approach provides us the possibility  for the use of the unconstrained light-cone gauge superfields (see, e.g., Ref.\cite{Bengtsson:1983pg}-\cite{Metsaev:2021bjh}). It would be interesting to extend results and methods in this paper to the case of  massive and massless supermultiplets in $R^{3,1}$. The discussion of various methods for building interaction vertices in supersymmetric theories may be found in Refs.\cite{Alkalaev:2002rq}-\cite{Buchbinder:2022kzl}.

\smallskip
\noinbf{iv})  In this paper, we studied cubic vertices of massive and massless fields. Extension of our study  to quartic vertices of  massless and massive fields along the lines of the light-cone gauge methods in Refs.\cite{Metsaev:1991mt}-\cite{Ponomarev:2016lrm} is of great interest.
Use of other various methods for the study of quartic vertices of higher-spin fields in flat and AdS spaces may be found in Refs.\cite{Taronna:2011kt}-\cite{Karapetyan:2021wdc}.

\smallskip
\noinbf{v}) In this paper, we considered interacting light-cone gauge massive and massless fields in the flat space. Extension of our results and methods to the case of light-cone gauge fields in AdS space is very interesting problem. In this respect, we note that the light-cone gauge formulation of free arbitrary spin massive and massless fields in AdS space was developed in Refs.\cite{Metsaev:1999ui,Metsaev:2003cu}, while, in Ref.\cite{Metsaev:2018xip}, we considered interacting arbitrary spin massless fields in $AdS_4$.%
\footnote{ The use of the light-cone approach in AdS for the study of light-front bootstrap of Chern-Simons matter theories may be found in Ref.\cite{Skvortsov:2018uru}. For the study of bilocal holography by using the light-cone approach in AdS, see Ref.\cite{deMelloKoch:2021cni}.}
We expect that a generalization of the results in Ref.\cite{Metsaev:2018xip} to the case of  interacting massive fields should be relatively straightforward. Use of frame-like approach for building interaction vertices for fields in AdS space may be found, e.g., in Refs.\cite{Fradkin:1987ks}-\cite{Khabarov:2021xts}.%
\footnote{ The frame-like approach for free massive fields in AdS was developed in Refs.\cite{Zinoviev:2008ze,Ponomarev:2010st} (see also Ref.\cite{Khabarov:2019dvi}). The use of the frame-like cubic vertices in AdS space for deriving the cubic vertices in flat space may be found in Ref.\cite{Khabarov:2020bgr}.}

\smallskip
\noinbf{vi}) As noted in Ref.\cite{Metsaev:2007rw}, the ordinary-derivative formulation of conformal fields and the gauge invariant formulation of massive
fields share certain common features. The light-cone gauge formulations of massive fields and conformal fields in Ref.\cite{Metsaev:2013kaa} also share some common features. For this reason we expect that the method for the study of interacting massive fields we developed in this paper can be adopted for a study of interacting conformal fields. We note also the interesting proposal for building action of interacting conformal fields in Ref.\cite{Vasiliev:2009ck}. Various interesting recent developments in the topic of conformal fields may be found, e.g., in Refs.\cite{Basile:2017mqc}-\cite{Chekmenev:2020lkb}.

\smallskip
\noinbf{vii}) In the recent time, the interesting investigations for the use of the twistor method in Lagrangian formulation of interacting higher-spin massless fields were carried out in Refs.\cite{Krasnov:2021nsq}-\cite{Steinacker:2022jjv}. Application of the twistor method in Refs.\cite{Krasnov:2021nsq}-\cite{Steinacker:2022jjv} for the study of Lagrangian formulation of interacting massless and massive fields in four dimensions seems to be interesting avenue to go.

\medskip

{\bf Acknowledgments}. This work was supported by the RFBR Grant No.20-02-00193.

\setcounter{section}{0}\setcounter{subsection}{0}
\appendix{ \large Notation and useful identities }

Throughout this paper we use the following notation:
\beq
\label{01052022-man-01}  && D \equiv m_1^4 + m_2^4 + m_3^4 - 2m_1^2m_2^2 -2 m_2^2 m_3^2 - 2 m_3^2 m_1^2\,,
\\
\label{01052022-man-02} && \gamma \equiv \half \sqrt{D}\,,  \hspace{0.5cm} \for \ D>0\,; \hspace{1cm} \gamma \equiv \frac{\irm}{2} \sqrt{-D}\,,  \hspace{0.5cm} \for \ D<0\,;
\\
\label{01052022-man-03} && g_a \equiv \frac{\betach_a}{2\beta_a}m_a + c_a m_a\,,\qquad c_a \equiv \frac{m_{a+1}^2 -m _{a+2}^2}{2m_a^2}\,,\qquad \betach_a \equiv \beta_{a+1} - \beta_{a+2}\,,\qquad
\\
\label{01052022-man-04}  && g_{a\pm} \equiv g_a \pm \frac{\gamma}{m_a}\,,
\\
\label{01052022-man-05} && \rho^2 \equiv  \beta\sum_{a=1,2,3} \frac{m_a^2}{\beta_a} \,, \hspace{1cm}  \beta\equiv \beta_1\beta_2\beta_3\,,
\\
\label{01052022-man-05a1} && \Po_{\epsilon m}\equiv \frac{1}{3}\sum_{a=1,2,3} \betach_a\epsilon_a m_a\,,  \qquad \Pbf_{\epsilon m}\equiv \sum_{a=1,2,3} \epsilon_a m_a\,, \quad \epsilon_1^2=1\,,\quad \epsilon_2^2=1\,, \quad \epsilon_3^2=1\,,\qquad
\\
\label{01052022-man-06}  && Y_{a\pm} \equiv m_a^2 - m_{a+1}^2 - m_{a+2}^2  \pm 2\gamma\,, \hspace{1cm} \kappa_a \equiv \frac{Y_{a+}}{Y_{a-}}\,,
\eeq
where $a=1,2,3$. Using the definitions in \rf{01052022-man-03}-\rf{01052022-man-05a1}, we find the relations
\be
c_a = \frac{\epsilon_a}{2m_a}(\epsilon_{a+2} m_{a+2} - \epsilon_{a+1} m_{a+1} )\,, \quad g_a =  \frac{\epsilon_a}{\beta_a}\Po_{\epsilon m}\,,
\quad \rho^2 = - \Po_{\epsilon m}^2\,, \quad \for \quad \Pbf_{\epsilon m}=0\,.\qquad
\ee
Using the definitions in \rf{01052022-man-06}, we find the relations,
\beq
\label{01052022-man-08}  &&  Y_{a+} Y_{a-} = 4 m_{a+1}^2m_{a+2}^2 \,,\qquad  Y_{a\pm} = \frac{1}{2m_a^2} Y_{a+1\,\mp} Y_{a+2\,\mp}\,,\qquad a=1,2,3\,;
\\
\label{01052022-man-09} &&  Y_{1+} Y_{2+} Y_{3+}= 8m_1^2m_2^2m_3^2 \,,\qquad  Y_{1-} Y_{2-} Y_{3-} = 8m_1^2m_2^2m_3^2\,,
\\
\label{01052022-man-10} &&  \kappa_1\kappa_2\kappa_3 =1 \,, \qquad \kappa_a = \frac{Y_{a+}Y_{a+}}{4m_{a+1}^2m_{a+2}^2}\,,\qquad a=1,2,3\,.
\eeq
For $D<0$, the quantities $\kappa_a$ and $Y_{a\pm}$ are complex-valued. The relations for $\kappa_a$ in \rf{01052022-man-10} imply
\be \label{01052022-man-11}
\kappa_a > 0 \qquad \hbox{ for } \ D > 0\,, \qquad a=1,2,3;
\ee
while the definitions in \rf{01052022-man-02}, \rf{01052022-man-06} imply the following relations:
\beq
\label{01052022-man-12} && \kappa_a^*=\kappa_a\,, \ \qquad Y_{a\pm}^* = Y_{a\pm}\,,\qquad \for \ \ D> 0\,; \qquad
\\
\label{01052022-man-13} && \kappa_a^*= \kappa_a^{-1}\,, \qquad Y_{a\pm}^* = Y_{a\mp}\,, \qquad \for \ \ D < 0\,.
\eeq
Using definitions in \rf{01052022-man-03}-\rf{01052022-man-05}, we get the relations
\beq
\label{01052022-man-14} && g_{a+} g_{a-} = - \frac{\rho^2}{\beta_a^2} \,, \hspace{2cm} \frac{ g_{a+} g_{a+1\,-} }{ g_{a\,-}
g_{a+1\,+}} =\kappa_{a+2}^{-1}\,,\hspace{1cm}  a=1,2,3\,,
\\
\label{01052022-man-15} && \prod_{a=1,2,3} \Big( \frac{ g_{a+} }{ g_{a-} } \Big)^{n_a} =
\prod_{a=1,2,3} \kappa_a^{-(n_{a+1}-n_{a+2})/3 }\,,\hspace{3.3cm}  n_1 + n_2+ n_3=0\,,
\\
\label{01052022-man-16} && \prod_{a=1,2,3}  g_{a+}^{s_a+n_a} g_{a-}^{s_a-n_a} =
\prod_{a=1,2,3} \big(-\frac{\rho^2}{\beta_a^2}\big)^{s_a} \kappa_a^{-(n_{a+1}-n_{a+2})/3 }\,,\hspace{1cm}  n_1 + n_2+ n_3=0\,.\qquad
\eeq

\appendix{ \large Various realizations of operators $M^\Rsm$, $M^\Lsm$, $M^{\Rsm\Lsm}$ and helpful formulas for operators $E_m$, $E_\beta$ \rf{21042022-man-19}}

\noinbf{$u,v$-realization}. Sometimes, in place of $u,v$-oscillators \rf{18042022-man-10}, we find it convenient to use c-number complex-valued variables $u,v$. In terms of such variables, the generating function of massive field is defined as
\be \label{25042022-man-01}
\phi_{m,s}(x^+,p,u,v) = \sum_{n=-s}^s \frac{ u^{s+n} v^{s-n} }{ \sqrt{(s+n)!(s-n)!} } \phi_{m,s;n}(x^+,p)\,.
\ee
On space of generating function \rf{25042022-man-01}, the spin operators are realized as follows
\be \label{25042022-man-02}
M^\Rsm = \frac{m}{\sqrt{2}} u \partial_v \,,\hspace{0.5cm}
M^\Lsm = - \frac{m}{\sqrt{2}} v \partial_u \,,\hspace{0.5cm}
M^{\Rsm\Lsm} = \half (u\partial_u - v\partial_v)\,,
\ee
where $\partial_u\equiv \partial/\partial u$, $\partial_v\equiv \partial/\partial v$.
Using the shortcut $\phi(u,v)$ for the generating function \rf{25042022-man-01}, we note that the scalar product for the generating functions $\phi(u,v)$ and $\varphi(u,v)$ takes the form
\be  \label{25042022-man-03}
(\phi,\varphi) \equiv \int d^2u d^2v e^{-u\ub - v\vb} (\phi(u,v))^\dagger \varphi(u,v)\,,\qquad  \int d^2u e^{-u\ub} \equiv 1\,,
\ee
where $\phi(u,v)$ and $\varphi(u,v)$ are degree-$2s$ homogeneous polynomials in $u$, $v$,
\be  \label{25042022-man-04}
(u\partial_u + v\partial_v)\phi(u,v) = 2s \phi(u,v)\,,\qquad  (u\partial_u + v\partial_v)\varphi(u,v) = 2s \varphi(u,v)\,.
\ee

\noinbf{ Projective realization}. Introducing a projective variable $\alpha$ and a generating function $\phi_\prj(\alpha)$,
\be \label{25042022-man-05}
\alpha = \frac{v}{u}\,,  \hspace{1cm} \phi(u,v) = u^{2s} \phi_\prj(\alpha)\,,
\ee
we find that on space of $\phi_\prj(z)$ the spin operators \rf{25042022-man-02} are realized as
\be  \label{25042022-man-07}
M^\Rsm = \frac{m}{\sqrt{2}} \partial_\alpha\,,
\hspace{0.7cm} M^\Lsm = \frac{m}{\sqrt{2}}\big( \alpha^2 \partial_\alpha - 2 s \alpha\big)\,,\hspace{0.7cm} M^{\Rsm\Lsm} =  s - \alpha\partial_\alpha\,, \qquad \partial_\alpha\equiv \partial/\partial \alpha\,.\qquad
\ee
The scalar product for the generating functions $\phi_\prj(\alpha)$ and $\varphi_\prj(\alpha)$ takes the form
\be \label{25042022-man-08}
(\phi_\prj,\varphi_\prj) = \int d\sigma_s(\alpha,\alphab)  (\phi_\prj(\alpha))^\dagger \varphi_\prj(\alpha)\,,\hspace{1cm} d\sigma_s(\alpha,\alphab)\equiv \frac{(2s+1)!}{(1 + \alpha\alphab)^{ 2s+2 } }d^2\alpha\,. \qquad
\ee
Scalar products \rf{25042022-man-03} and \rf{25042022-man-08} are related as
$(\phi,\varphi) =  (\phi_\prj,\varphi_\prj)$. Relations \rf{25042022-man-05} imply that the $u,v$-realization and the projective realization of vertices are related as
\be \label{25042022-man-08a1}
V(u_1,v_1;u_2,v_2;u_3,v_3) = \prod_{a=1,2,3} u_a^{2s_a} V_\prj(z_1;z_2;z_3)\,.
\ee

\noinbf{ Action of operators $E_m$, $E_\beta$ \rf{21042022-man-19}}. Using $M^\Lsm$, $M^{\Rsm\Lsm}$ given in \rf{25042022-man-07}, we find the relations
\beq
\label{25042022-man-10} && e^{tM^\Lsm}\alpha e^{-tM^\Lsm} = \frac{\alpha}{1-\frac{t m}{\sqrt{2}}\alpha}\,, \hspace{1.4cm} t^{M^{\Rsm\Lsm}}\alpha t^{-M^{\Rsm\Lsm}} = \frac{1}{t} \alpha\,,
\\
\label{25042022-man-11} && e^{tM^\Lsm}|0\rangle = \big(1-\frac{tm}{\sqrt{2}}\alpha\big)^{2s}|0\rangle\,,\hspace{1cm} t^{M^{\Rsm\Lsm}}|0\rangle = t^s|0\rangle\,,
\eeq
where $|0\rangle \equiv 1$. Using definitions in \rf{21042022-man-19}, \rf{01052022-man-03} and relations \rf{25042022-man-10}, \rf{25042022-man-11}, we get the following relations for massive fields:
\beq
\label{25042022-man-12} && E_{\beta a} \alpha_a E_{\beta a}^{-1} = \frac{\beta_a}{\Po^\Lsm} \alpha_a\,, \hspace{1cm}  E_{ma} \alpha_a E_{ma}^{-1} = \frac{\Po^\Lsm\alpha_a}{\beta_a L_{\alpha,a}}\,,
\\
\label{25042022-man-12a1} && E_{ma} E_{\beta a} \alpha_a E_{\beta a}^{-1} E_{m a}^{-1}
=\frac{\alpha_a}{L_{\alpha,a}}\,,\hspace{1cm} L_{\alpha,a} \equiv \frac{\Po^\Lsm}{\beta_a} + \frac{g_a}{\sqrt{2}} \alpha_a\,,
\\
\label{25042022-man-13} && E_{m a} |0\rangle  =  L_{\alpha,a}^{2s_a} \big(\frac{\Po^\Lsm}{\beta_a}\big)^{-2s_a} \,, \hspace{1cm} E_{\beta a} |0\rangle = \big(\frac{\Po^\Lsm}{\beta_a}\big)^{s_a} \,.
\eeq
In turn, for arbitrary function $F(\alpha_a)$, relations \rf{25042022-man-12a1}, \rf{25042022-man-13}, \rf{21042022-man-20} lead to the relations
\beq
\label{25042022-man-14} && E_{ma} E_{\beta a} F(\alpha_a)|0\rangle = F\big(\frac{\alpha_a}{L_{\alpha,a}}\big)   L_{\alpha,a}^{2s_a} \big(\frac{\Po^\Lsm}{\beta_a}\big)^{-s_a}| 0\rangle\,, \hspace{1cm} \for \ \ m_a\ne 0\,;
\\
\label{25042022-man-15} && E_{ma} E_{\beta a}| 0\rangle =  \big(\frac{\Po^\Lsm}{\beta_a}\big)^{\lambda_a}| 0\rangle\,, \hspace{4.6cm} \for \ \ m_a  = 0\,.
\eeq

\appendix{ \large Incorporation of internal $o(\Nsf)$ symmetry }

For example, consider an internal $o(\Nsf)$ symmetry. Incorporation of the internal $o(\Nsf)$ symmetry into our treatment of massive and massless fields can be realized in the following four steps.%
\footnote{ Discussions of the Chan-Paton gauging for the $U(N)$ and $Usp(N)$ symmetries  may  be found in Refs.\cite{Skvortsov:2020wtf}.}

\noinbf{Step 1}. Let $\asf,\bsf$,$\csf$ be the matrix indices of the $o(\Nsf)$ algebra, $\asf,\bsf,\csf=1,\ldots,\Nsf$. In place of the singlet massless field $\phi_\lambda$, we introduce the colored massless fields $\phi_\lambda^{\asf\bsf}$, while, in place of the singlet massive fields
$\phi_{m,s;n}$, we introduce the colored massive fields $\phi_{m,s;n}^{\asf\bsf}$.
By definition, these colored fields obey the relations
\beq
&& \phi_\lambda^{\asf\bsf} = (-)^\lambda\phi_\lambda^{\bsf\asf}\,, \qquad
\phi_{m,s;n}^{\asf\bsf} = (-)^s \phi_{m,s;n}^{\bsf\asf}\,,
\\
&& (\phi_\lambda^{\asf\bsf}(x^+,p))^\dagger = \phi_{-\lambda}^{\asf\bsf}(x^+,-p)\,, \hspace{1cm}
(\phi_{m,s;n}^{\asf\bsf}(x^+,p))^\dagger = \phi_{m,s;-n}^{\asf\bsf}(x^+,-p)\,.
\eeq

\noinbf{ Step 2}. In the scalar products entering actions of free fields \rf{18042022-man-35}, we make the replacements
\be
\phi_s^\dagger \phi_s \rightarrow \phi_s^{\asf\bsf\dagger}  \phi_s^{\asf\bsf},\qquad \langle\phi_{m,s}|\phi_{m,s}\rangle \rightarrow \langle \phi_{m,s}^{\asf\bsf} |\phi_{m,s}^{\asf\bsf}\rangle\,,
\ee
while, in the cubic vertices, the usual products should be replaced by the traced products,
\beq
&& \prod_{a=1,2,3}\phi_{\lambda_a}^\dagger \rightarrow \phi_{\lambda_1}^{\asf\bsf\dagger}
\phi_{\lambda_2}^{\bsf\csf\dagger} \phi_{\lambda_3}^{\csf\asf\dagger}\,, \hspace{1cm}  \langle\phi_{m_1,s_1}| \langle\phi_{m_2,s_2}| \phi_{\lambda_3}^\dagger \rightarrow \langle\phi_{m_1,s_1}^{\asf\bsf}| \langle\phi_{m_2,s_2}^{\bsf\csf}| \phi_{\lambda_3}^{\csf\asf\dagger}\,,
\\
&& \phi_{\lambda_1}^\dagger\phi_{\lambda_2}^\dagger \langle\phi_{m_3,s_3}| \rightarrow \phi_{\lambda_1}^{\asf\bsf\dagger}
\phi_{\lambda_2}^{\bsf\csf\dagger} \langle\phi_{m_3,s_3}^{\csf\asf}|\,,
\hspace{0.5cm}
\prod_{a=1,2,3}\langle\phi_{s_a,m_a}| \rightarrow
\langle\phi_{m_1,s_1}^{\asf\bsf}| \langle\phi_{m_2,s_2}^{\bsf\csf}| \langle\phi_{m_3,s_3}^{\csf\asf}|\,.\qquad
\eeq

\noinbf{ Step 3}. In place of the equal-time commutator \rf{18042022-man-31}, \rf{18042022-man-32}, we use the respective commutators
\beq
&& [\phi_\lambda^{\asf\bsf}(x^+,p),\phi_{\lambda'}^{\asf'\bsf'}(x^+,p')] = \frac{1}{2\beta}\delta^{(3)}(p+p') \delta_{\lambda+\lambda',0} \Pi_\lambda^{\asf\bsf,\asf'\bsf'}\,,
\\
&& [\phi_{m,s;n}^{\asf\bsf}(x^+,p),\phi_{m,s';n'}^{\asf'\bsf'}(x^+,p')] = \frac{1}{2\beta}\delta^{(3)}(p+p') \delta_{ss'}  \delta_{n+n',0} \Pi_s^{\asf\bsf,\asf'\bsf'}\,,
\\
&& \Pi_X^{\asf\bsf,\asf'\bsf'} \equiv \half\big( \delta^{\asf\asf'} \delta^{\bsf\bsf'} + (-)^X \delta^{\asf\bsf'} \delta^{\bsf\asf'} \big)\,, \qquad \Pi_X^{\asf\bsf,\asf'\bsf'} \Pi_X^{\asf'\bsf',\csf\esf} = \Pi_X^{\asf\bsf,\csf\esf}\,.
\eeq

\noinbf{Step 4}. The realization of the $o(\Nsf)$ algebra generators $J^{\asf\bsf}$ on space of fields is given by
\beq
&& J^{\asf\bsf} = 4 \int d^3p \beta \big( \phi_s^{\asf\csf\dagger} \phi_s^{\bsf\csf} - \phi_s^{\bsf\csf\dagger}\phi_s^{\asf\csf} \big)\,, \hspace{2.1cm} \hbox{for massless field};
\\
&& J^{\asf\bsf} =  2 \int d^3p \beta \big( \langle\phi_{m,s}^{\asf\csf} |\phi_{m,s}^{\bsf\csf}\rangle - \langle\phi_{m,s}^{\bsf\csf} |\phi_{m,s}^{\asf\csf}\rangle \big)\,, \hspace{1cm} \hbox{for massive field}.
\eeq
Generators of the $o(\Nsf)$ algebra, $J^{\asf\bsf}= - J^{\bsf\asf}$, and fields $\chi^{\asf\bsf}=\phi_\lambda^{\asf\bsf}, |\phi_{m,s}^{\asf\bsf}\rangle$ satisfy the commutators
\be
[J^{\asf\bsf},J^{\csf\esf} ] = \delta^{\bsf\csf} J^{\asf\esf} + 3 \hbox{ terms}, \qquad [\chi^{\asf\bsf},J^{\csf\esf} ] = \delta^{\bsf\csf} \chi^{\asf\esf} + 3 \hbox{ terms}.
\ee

\appendix{ \large Hermitian conjugation rules for vertices }

To derive hermitian conjugation rule for vertices, we find it convenient to use the c-number complex-valued variables $u$ and $v$ and the generating function given in \rf{25042022-man-01}. In terms of generating function \rf{25042022-man-01}, the hermicity condition for the massive fields $\phi_{m,s;n}(x^+,p)$ in \rf{18042022-man-15} takes the following form:
\be \label{24042022-man-02}
(\phi_{m,s}(x^+,p,u,v))^\dagger = \phi_{m,s}(x^+,-p,\vb,\ub)\,, \qquad u^*=\ub\,,\qquad v^*=\vb\,.
\ee
For the case of three massive fields entering the cubic vertex, the generator $P_\smp3^-$ \rf{19042022-man-02}  can be represented in terms of $\phi_{m,s}$ \rf{25042022-man-01} in the following way:
\beq
\label{24042022-man-03}  && P_\smp3^- = \int d\Gamma_\smp3 d\Gamma_\smp3^{u,v} \Phi_\smp3^\dagger\,\,  p_\smp3^-(\Po^\Rsm,\Po^\Lsm,\beta_a,u_a,v_a)\,,
\\
\label{24042022-man-04}  && \hspace{1cm} \Phi_\smp3^\dagger \equiv \prod_{a=1,2,3} (\phi_{m_a,s_a}(x^+,p_a,u_a,v_a))^\dagger\,, \hspace{0.5cm}  d\Gamma_\smp3^{u,v} \equiv \prod_{a=1,2,3} e^{-u_a \ub_a - v_a\vb_a}d^2u_a d^2v_a\,.\qquad \qquad
\eeq
Using \rf{24042022-man-02}, \rf{24042022-man-03}, it is easy to verify that the hermitian conjugation of $P_\smp3^-$ can be presented as
\be
\label{24042022-man-06} \hspace{-1.3cm} P^{-\dagger} = \int   d\Gamma_\smp3  d\Gamma_\smp3^{u,v}  \Phi_\smp3^\dagger\,\, \II p_\smp3^{-}(\Po^\Rsm,\Po^\Lsm,\beta_a,u_a,v_a)\,,
\ee
where we introduce
\vspace{-0.3cm}
\beq
&& \hspace{-1cm} \hbox{ \bf Realization of operator $\II$ on cubic vertex $p_\smp3^-$}:
\nonumber\\
\label{24042022-man-07} && \II p_\smp3^{-}(\Po^\Rsm,\Po^\Lsm,\beta_a,u_a,v_a) =  p_\smp3^{-*}(\II\,\Po^\Rsm,\II\,\Po^\Lsm,\II\,\beta_a,\II\,u_a,\II\,v_a)\,,
\nonumber\\
&& \II\,\Po^\Rsm \equiv \Po^\Lsm\,, \quad \II\,\Po^\Lsm \equiv \Po^\Rsm\,, \quad\II\,\beta_a \equiv -\beta_a\,, \quad \II\,u_a \equiv v_a\,,  \quad \II\,v_a \equiv u_a\,.
\eeq
In terms of $p_\smp3^-$, the hermicity condition $P_\smp3^- = P_\smp3^{-\dagger}$ is represented as
\be \label{24042022-man-08}
p_\smp3^-= \II\,p_\smp3^-\,, \quad \hbox{or explicitly as} \quad p_\smp3^{-}(\Po^\Rsm,\Po^\Lsm,\beta_a,u_a,v_a) =  p_\smp3^{-*}(\Po^\Lsm,\Po^\Rsm,-\beta_a,v_a,u_a)\,.
\ee
Using formulas in \rf{21042022-man-03}-\rf{21042022-man-09}, we find
\vspace{-0.3cm}
\beq
&& \hspace{-1cm} \hbox{ \bf Realization of operator $\II$ on meromorphic vertex  $\VVb$}:
\nonumber\\
&& \II\, \VVb(\Po^\Lsm,\beta_a,u_a,v_a) =  \VVb^*(\II\,\Po^\Lsm\,,\II\,\beta_a,\II\,u_a,\II\,v_a)\,,
\nonumber\\
\label{24042022-man-10}  && \II\,\Po^\Lsm \equiv \frac{\rho^2}{2\Po^\Lsm}\,, \quad \II\,\beta_a\equiv -\beta_a\,, \quad \II\,u_a\equiv v_a\,,  \quad \II\,v_a \equiv u_a\,,\qquad
\eeq
where $\rho^2$ is given in \rf{01052022-man-05}. In terms of $\VVb$, the hermicity condition \rf{24042022-man-08} is realized as
\be \label{24042022-man-09}
\VVb = \II\, \VVb \,, \quad \hbox{or explicitly as} \quad
\VVb(\Po^\Lsm,\beta_a,u_a,v_a) =  \VVb^*(\frac{\rho^2}{2\Po^\Lsm},-\beta_a,v_a,u_a)\,.
\ee
For the reader convenience, we present the transformations of $L_{a\pm}$ \rf{29042022-man-06} under the action of the operator $\II$ given in \rf{24042022-man-10},
\be
\label{24042022-man-11} \II\,L_{a\pm} = \frac{\beta_a}{\sqrt{2}\Po^\Lsm} g_{a\pm} L_{a\mp}\,, \hspace{0.3cm} \hbox{ for } \ D>0\,; \hspace{0.7cm} \II\,L_{a\pm} = \frac{\beta_a}{\sqrt{2}\Po^\Lsm} g_{a\mp}  L_{a\pm}\,,\hspace{0.3cm} \hbox{ for } \ D<0\,.
\ee

\appendix{ \large Derivation of equations for $p_\smp3^-$, $\VVb$, $\VVb^{(2)}$ in  \rf{21042022-man-02a1}, \rf{21042022-man-14}, \rf{21042022-man-15}, \rf{21042022-man-22} and representation for $j_\smp3^{-\Rsm}$ \rf{21042022-man-02a7}}

\noinbf{ Derivation of \rf{21042022-man-02a1} and \rf{21042022-man-02a7}}.
We represent equation \rf{20042022-man-21} as
\be \label{26012022-30}
\Jbf_\Thsm^{-\Rsm} p_\smp3^- + \frac{\Po^\Rsm\Po^\Lsm}{\beta} \big( - \No_\beta + \Mo^{\Rsm\Lsm})\frac{1}{N_\Po +1} \partial_{\Po^\Lsm} p_\smp3^- = \Big( -  \frac{\Po^\Rsm\Po^\Lsm}{\beta} + \sum_{a=1,2,3} \frac{m_a^2}{2\beta_a} \Big) j_\smp3^{-\Rsm}\,,
\ee
where $\Jbf_\Thsm^{-\Rsm}$ is given in \rf{21042022-man-02a3}. For the harmonic function $p_\smp3^-$, the $\Jbf_\Thsm^{-\Rsm}p_\smp3^-$ is also the harmonic function.  Taking this into account and comparting the non-harmonic $(\Po^\Rsm\Po^\Lsm)^n$-terms in \rf{26012022-30}, we get the representation for $j_\smp3^{-\Rsm}$ \rf{21042022-man-02a7}. Plugging $j_\smp3^{-\Rsm}$ \rf{21042022-man-02a7} into \rf{26012022-30}, we get the equation for $p_\smp3^-$  \rf{21042022-man-02a1}. Derivation of the equation for $p_\smp3^-$ \rf{21042022-man-02a2} and the representation for $j_\smp3^{-\Lsm}$ \rf{21042022-man-02a8} is in complete analogy with the derivation of  \rf{21042022-man-02a1} and \rf{21042022-man-02a7}
above presented.

\noinbf{Derivation of equations for vertex $\VVb$ \rf{21042022-man-14},\rf{21042022-man-15}}. We split our derivation in the three steps.

\noinbf{Step 1}. Plugging \rf{21042022-man-03} into \rf{21042022-man-02a1},\rf{21042022-man-02a2}, we find that, in terms of the vertices $V_N$, $V_0$, $\Vb_\Nb$, equations \rf{21042022-man-02a1},\rf{21042022-man-02a2} can be represented as
\beq
\label{26012022-13} && \Big( \frac{\Po^\Rsm}{\beta} \big( - \No_\beta + \Mo^{\Rsm\Lsm} \big) - \sum_{a=1,2,3}    \frac{1}{\beta_a} M_a^\Rsm\Big) (V_N + V_0) %
\nonumber\\
&& + \sum_{a=1,2,3}\Big(  \frac{\check\beta_a }{6\beta_a} m_a^2 \partial_{\Po^\Lsm}   + \frac{m_a^2}{2\beta_a \Po^\Lsm} \big( - \No_\beta + \Mo^{\Rsm\Lsm}) -\frac{1}{\beta_a} M_a^\Rsm \Big)\Vb_\Nb
= 0 \,,\qquad
\\[10pt]
\label{26012022-14} && \Big( \frac{\Po^\Lsm}{\beta} \big( - \No_\beta - \Mo^{\Rsm\Lsm} \big) - \sum_{a=1,2,3}    \frac{1}{\beta_a} M_a^\Lsm\Big) (\Vb_\Nb + V_0)
\nonumber\\
&&+ \sum_{a=1,2,3}\Big(  \frac{\check\beta_a }{6\beta_a} m_a^2 \partial_{\Po^\Rsm}   + \frac{m_a^2}{2\beta_a \Po^\Rsm} \big( - \No_\beta - \Mo^{\Rsm\Lsm}) - \frac{1}{\beta_a} M_a^\Lsm \Big)V_N
= 0 \,.\qquad
\eeq

\noinbf{Step 2}. Equations \rf{26012022-13},\rf{26012022-14} amount to the following equations:
\beq
\label{26012022-17} && \Big( \frac{\Po^\Rsm}{\beta} \big(  \No_\beta - \Mo^{\Rsm\Lsm} \big) + \sum_{a=1,2,3}    \frac{1}{\beta_a} M_a^\Rsm\Big) V_N  + \frac{\Po^\Rsm}{\beta} \big(  \No_\beta - \Mo^{\Rsm\Lsm} \big) V_0 =  0\,,
\\
\label{26012022-18} && \sum_{a=1,2,3}\Big( \frac{m_a^2}{2\beta_a} \big(  \No_\beta - \Mo^{\Rsm\Lsm}) - \frac{\check\beta_a }{6\beta_a} m_a^2 N_{\Po^\Lsm} + \frac{\Po^\Lsm}{\beta_a} M_a^\Rsm \Big)\Vb_\Nb
+ \sum_{a=1,2,3} \frac{\Po^\Lsm}{\beta_a} M_a^\Rsm V_0  = 0 \,,\qquad
\eeq

\beq
\label{26012022-19} && \Big( \frac{\Po^\Lsm}{\beta} \big(  \No_\beta + \Mo^{\Rsm\Lsm} \big) + \sum_{a=1,2,3}    \frac{1}{\beta_a} M_a^\Lsm\Big) \Vb_\Nb + \frac{\Po^\Lsm}{\beta} \big(  \No_\beta + \Mo^{\Rsm\Lsm} \big)V_0= 0 \,,
\\
\label{26012022-20} && \sum_{a=1,2,3}\Big(  \frac{m_a^2}{2\beta_a} \big(  \No_\beta + \Mo^{\Rsm\Lsm}) -  \frac{\check\beta_a }{6\beta_a} m_a^2 N_{\Po^\Rsm} + \frac{\Po^\Rsm}{\beta_a} M_a^\Lsm  \Big)V_N
+ \sum_{a=1,2,3} \frac{\Po^\Rsm}{\beta_a} M_a^\Lsm   V_0  = 0 \,.\qquad
\eeq
Equations \rf{26012022-17}-\rf{26012022-20} are obtained from equations \rf{26012022-13},\rf{26012022-14} in the following way. Considering terms of the powers $(\Po^\Rsm)^n$, $n > 0$, we see that equation \rf{26012022-13} amounts to equation \rf{26012022-17}, while considering terms of the powers $(\Po^\Lsm)^n$, $n \geq  0$, we see that equation \rf{26012022-13} amounts to equation \rf{26012022-18}.
Considering terms of the powers $(\Po^\Lsm)^n$, $n > 0$, we see that equation \rf{26012022-14} amounts to equation \rf{26012022-19}, while considering terms of the powers $(\Po^\Rsm)^n$, $n \geq  0$, we see that equation \rf{26012022-14} amounts to equation \rf{26012022-20}.

\noinbf{Step 3}. We note that, in terms of $V_N^\otimes$ defined in \rf{21042022-man-07}, equation \rf{26012022-17} can be represented as the equation given below in \rf{26012022-24}, while equation \rf{26012022-20} can be represented as the equation given below in \rf{26012022-25},
\beq
\label{26012022-24} &&  \sum_{a=1,2,3} \Big( \frac{m_a^2}{2\beta_a} \big(  \No_\beta - \Mo^{\Rsm\Lsm} \big) - \frac{\betach_a m_a^2}{6\beta_a}N_{\Po^\Lsm}  +    \frac{\Po^\Lsm}{\beta_a} M_a^\Rsm\Big) V_N^\otimes  +   \frac{m_a^2}{2\beta_a} \big(  \No_\beta - \Mo^{\Rsm\Lsm} \big) V_0 =  0\,,\qquad\quad
\\
\label{26012022-25} && \frac{\Po^\Lsm}{\beta} \sum_{a=1,2,3} \big(  \No_\beta + \Mo^{\Rsm\Lsm}) V_N^\otimes +  \sum_{a=1,2,3} \frac{1}{\beta_a} M_a^\Lsm  (V_N^\otimes + V_0)  = 0 \,.
\eeq
Finally, we note that, by combining equations \rf{26012022-18} and \rf{26012022-24}, we get equation \rf{21042022-man-14}, while, by combining equations \rf{26012022-19} and \rf{26012022-25}, we get equation \rf{21042022-man-15}.

\noinbf{Derivation of \rf{21042022-man-22}}. What is required is to prove that the vertex $\VVb^{(2)}$ is independent of $\beta_1$, $\beta_2$, $\beta_3$, and $\Po^\Lsm$, and satisfies equation \rf{21042022-man-22}. We do this in the following two steps.

\noinbf{Step 1}. Plugging \rf{21042022-man-18} into equation \rf{21042022-man-15} and using the relation
\be \label{08022022-09}
(\No_\beta-\frac{1}{3}\betach_a)f_a = -\frac{\beta}{\beta_a}\,,
\ee
we find that equation \rf{21042022-man-15} leads to the equation
\be \label{08022022-10}
\No_\beta \VVb^{(2)} = 0\,.
\ee
Plugging $\VVb$ \rf{21042022-man-18} into equations \rf{21042022-man-16},\rf{21042022-man-17}, we find that equations \rf{21042022-man-16} and \rf{21042022-man-17}  lead to the following respective equations for $\VVb^{(2)}$:
\beq
\label{08022022-11}  &&  \sum_{a=1,2,3}    \beta_a\partial_{\beta_a} \VVb^{(2)}  =  0 \,,  \hspace{1cm} N_{\Po^\Lsm}  \VVb^{(2)} = 0 \,.
\eeq
Equation \rf{08022022-10} and the 1st equation in \rf{08022022-11} imply that the vertex $\VVb^{(2)}$ does not depend on $\beta_1$, $\beta_2$, $\beta_3$, while from the 2nd equation \rf{08022022-11}, we learn that the vertex $\VVb^{(2)}$ is independent of $\Po^\Lsm$. In other words, as it is stated in \rf{21042022-man-21}, the vertex $\VVb^{(2)}$ depends only on the oscillators $u_a$, $v_a$.

\noinbf{Step 2}. Plugging \rf{21042022-man-18} into equation \rf{21042022-man-14} and using the fact that  the vertex $\VVb^{(2)}$ is independent of $\beta_1$, $\beta_2$, $\beta_3$, and $\Po^\Lsm$, we find that equation \rf{21042022-man-14} leads to  equation \rf{21042022-man-22}.

\appendix{ \large Invariant amplitudes in light-cone frame }

First, we explain our notation and conventions for the $S$-matrix and amplitudes in the light-cone frame. Second, we explain how our cubic vertex is related to 3-point invariant amplitude.

For simplicity of the presentation, we consider massless fields. To discuss amplitudes we use fields in the Dirac (interaction) picture,
\be \label{11032022-04}
\phi_\lambda(x^+,p) = e^{{\rm i}x^+p^-}\phi_\lambda(p)\,,\qquad p^- = - \frac{p^\Rsm p^\Lsm}{\beta}\,,
\ee
where the $\phi_\lambda(p)$ is expressed in terms of annihilation $\ab_\lambda(p)$ and creation $a_\lambda(p)$ operators as
\beq
\label{11032022-07} && \phi_\lambda(p) = \frac{\theta(\beta)}{\sqrt{2\beta}}\, \ab_\lambda (p)
+ \frac{\theta(-\beta)}{\sqrt{-2\beta}}\, a_{-\lambda}(-p)\,,
\\
\label{11032022-11} && \theta(\beta) = 1 \ \ \for \ \ \beta>0\,, \qquad \theta(\beta) = 0 \ \ \for \ \ \beta < 0\,,
\\
\label{11032022-13} && [ \ab_\lambda(p), a_{\lambda'}(p')] = \delta_{\lambda,\lambda'} \delta^{(3)}(p-p')\,,
\\
\label{11032022-10} && \ab_\lambda(p)|0\rangle = 0 \,, \qquad (\ab_\lambda(p))^\dagger = a_\lambda(p)\,.
\eeq

The matrix elements of the $S$-matrix are defined as
\be \label{11032022-22} S_{fi} \equiv \langle f |S|i\rangle\,,\qquad
|i\rangle = \prod_{a_i} a_{-\lambda_{a_i}}(-p_{a_i}) |0\rangle \,,\hspace{1cm}  \langle f|  = \langle 0 | \prod_{a_f} \ab_{\lambda_{a_f} }(p_{a_f})\,,\qquad
\ee
where $|i\rangle$ and $\langle f|$ stand for the respective in- and out-states, while the indices $a_i$ and $a_f$ label external lines of in-coming and out-going particles respectively. In terms of an invariant amplitude denoted as $\AA_{fi}$, the matrix elements $S_{fi}$ \rf{11032022-22} can be presented as
\beq
\label{11032022-38} &&  \hspace{-1cm} S_{fi} = - \irm (2\pi)^4 \Nbf_{fi}^{-1}\, \delta^{(4)}(\sum_{a_i} p_{a_i} + \sum_{a_f} p_{a_f}) \AA_{fi}\,, %
\\
&& \Nbf_{fi} \equiv \prod_{a_f} N_{a_f}\prod_{a_i} N_{a_i}\,, \hspace{0.5cm}
 N_{a_i} \equiv (2\pi)^{3/2}\sqrt{-2\beta_{a_i}}\,,\qquad N_{a_f} \equiv (2\pi)^{3/2}\sqrt{2\beta_{a_f}}\,,\qquad
\qquad
\eeq
where, for in-coming particles, $\beta_{a_i}<0$ , while, for out-going particles, $\beta_{a_f}>0$. To cubic approximation, the $S$-matrix and the corresponding 3-point invariant amplitude for different particles are given by
\beq
\label{11032022-21} && \hspace{-1cm} S = 1 + \irm  \int dx^+ P_\smp3^- \,,
\\
\label{11032022-39} && \hspace{-1cm} \AA_{fi} = - \big(p_\smp3^-(\lambda_{a_i}, \lambda_{a_f},p_{a_i},p_{a_f}) + \II\,p_\smp3^-(-\lambda_{a_i}, -\lambda_{a_f},p_{a_i},p_{a_f})\big)|^{\on-sh}\,,
\eeq
where the operator $\II$ is defined in \rf{19042022-man-09-a7}. Formula \rf{11032022-39} explains how our cubic vertex $p_\smp3^-$ is related to the 3-point invariant amplitude. For the reader convenience, we present $p_\smp3^-$ for the case when all three fields in the vertex are massless,
\be \label{11032022-39a1}
p_\smp3^- = C_{\lambda_1,\lambda_2,\lambda_3} \beta^{-\lambda_1}\beta_2^{-\lambda_2}\beta_3^{-\lambda_3}\, \Po^{\lambda_1+\lambda_2+\lambda_3}\,, \qquad  \lambda_1+ \lambda_2+\lambda_3>0\,,
\ee
where $C_{\lambda_1,\lambda_2,\lambda_3}$ are coupling constants. Cubic vertex $p_\smp3^-$ \rf{11032022-39a1} was obtained in Ref.\cite{Bengtsson:1986kh}.

\noinbf{ Internal $o(\Nsf)$  symmetry and massless fields}. Generalization of above given formulas is straightforward. In place of relations \rf{11032022-04}-\rf{11032022-10}, we use their obvious generalization given by
\beq
\label{11032022-04a1} && \phi_\lambda^{\asf\bsf}(x^+,p) = e^{{\rm i}x^+p^-}\phi_\lambda^{\asf\bsf}(p)\,,\qquad p^- = - \frac{p^\Rsm p^\Lsm}{\beta}\,,
\\
\label{12032022-08} && \phi_\lambda^{\asf\bsf} (p) = \sum_\Asf \frac{\theta(\beta)}{\sqrt{2\beta}}\, t_\Asf^{\asf\bsf}\ab_{\lambda, \Asf} (p)
+ \frac{\theta(-\beta)}{\sqrt{-2\beta}}\, t_\Asf^{\asf\bsf} a_{-\lambda, \Asf}(-p)\,,
\\
\label{12032022-14} && [ \ab_{\lambda,\Asf}(p), a_{\lambda',\Asf'}(p')] = \delta_{\lambda,\lambda'} \delta_{\Asf\Asf'}\delta^{(3)}(p-p')\,,
\\
&& \ab_{\lambda,\Asf}(p)|0\rangle = 0 \,, \qquad (\ab_{\lambda,\Asf}(p))^\dagger = a_{\lambda,\Asf}(p)\,,
\\
\label{12032022-17} && \Asf, \Bsf  = 1,2,\ldots, \Nsf_\lambda\,, \qquad N_\lambda \equiv \half \Nsf (\Nsf+ (-)^\lambda)\,,
\eeq
where real-valued quantities $t_\Asf^{\asf\bsf}$ \rf{12032022-08} satisfy the following relations:
\be \label{12032022-18}
t_\Asf^{\asf\bsf} = (-)^\lambda t_\Asf^{\bsf\asf}\,,
\hspace{0.7cm} \sum_\Asf t_\Asf^{\asf\bsf} t_\Asf^{\asf'\bsf'} = \half \big(\delta^{\asf\asf'} \delta^{\bsf\bsf'} + (-)^\lambda \delta^{\asf\bsf'} \delta^{\bsf\asf'}\big)\,, \hspace{0.7cm}  \sum_{\asf,\bsf} t_\Asf^{\asf\bsf} t_\Bsf^{\asf\bsf} = \delta_{\Asf\Bsf}\,.
\ee
For the case of different colored massless particles, the 3-point invariant amplitude and the respective states of in-coming and out-going particles are given by
\beq
&& \AA_{fi} = - \tr(t_{\Asf_1} t_{\Asf_2}t_{\Asf_3}) \Big(p_\smp3^-(\lambda_{a_i}, \lambda_{a_f},p_{a_i},p_{a_f}) +  \II\, p_\smp3^-(-\lambda_{a_i}, -\lambda_{a_f},p_{a_i},p_{a_f})\Big)|^{\on-sh} \,,\qquad \quad
\\
&& |i\rangle = \prod_{a_i} a_{-\lambda_{a_i},\Asf_{a_i}}(-p_{a_i}) |0\rangle \,, \hspace{1cm} \langle f|  = \langle 0 | \prod_{a_f} \ab_{ \lambda_{a_f},\Asf_{a_f} }(p_{a_f})\,,
\eeq
where the operator $\II$ is defined in \rf{19042022-man-09-a7}.

\noinbf{ Internal $o(\Nsf)$ symmetry and massive fields}. Generalization of the above given formulas to the massive fields that respect the internal $o(\Nsf)$ symmetry is straightforward. In place of relations \rf{11032022-04}-\rf{11032022-10}, we use their generalization given by
\beq
\label{11032022-04msv} && \phi_{m,s;n}^{\asf\bsf}(x^+,p) = e^{{\rm i}x^+p^-}\phi_{m,s;n}^{\asf\bsf}(p)\,,\qquad p^- = - \frac{2p^\Rsm p^\Lsm + m^2}{2\beta}\,,
\\
\label{12032022-08masv} && \phi_{m,s;n}^{\asf\bsf} (p) = \sum_\Asf \frac{\theta(\beta)}{\sqrt{2\beta}}\, t_\Asf^{\asf\bsf}\ab_{m,s;n, \Asf} (p)
+ \frac{\theta(-\beta)}{\sqrt{-2\beta}}\, t_\Asf^{\asf\bsf} a_{m,s;-n, \Asf}(-p)\,,
\\
\label{12032022-14msv} && [ \ab_{m,s;n,\Asf}(p), a_{m,s';n',\Asf'}(p')] = \delta_{s,s'}\delta_{n,n'} \delta_{\Asf\Asf'}\delta^{(3)}(p-p')\,,
\\
\label{12032022-12msv} &&\ab_{m,s;n, \Asf}(p) |0\rangle = 0\,, \hspace{1cm}  (\ab_{m,s;n, \Asf}(p))^\dagger = a_{m,s;n, \Asf}(p)\,,
\eeq
where $\Asf,\Bsf$ take values as in \rf{12032022-17} with the replacement $\lambda\rightarrow s$. The $t_\Asf^{\asf\bsf}$ appearing in \rf{12032022-08masv} satisfy the relations as in \rf{12032022-18} with the replacement $\lambda\rightarrow s$.

\appendix{\large Comments on classification \rf{23042022-man-12a1}-\rf{23042022-man-12a3} }

\noinbf{ Classification \rf{23042022-man-12a1}-\rf{23042022-man-12a3} via instant form of relativistic dynamics}. Consider the decay of the massive particle ($a=1$) into the two massive particles ($a=2,3)$. In the rest frame of the decaying particle, we can use the following expressions for the particles momenta:
\be \label{02052022-man-05}
p_1^\mu = (m_1,0)\,,\hspace{0.4cm}  p_2^\mu = (E_2,\pbf)\,, \hspace{0.4cm} p_3^\mu = (E_3,-\pbf)\,,\hspace{0.4cm}  E_a = \sqrt{m_a^2 + \pbf^2}\,, \quad a=1,2\,.\qquad
\ee
Using relations \rf{02052022-man-05} and the energy conservation law $m_1 = E_2 + E_3$, we get the well known restrictions on the masses and a square of the momentum $\pbf$:%
\footnote{ For $|\pbf|\ne 0$, relation \rf{02052022-man-09} is also valid for a decay of massive particle into two massless particles and a decay of massive particle into one massive particle and one massless particle.}
\beq
\label{02052022-man-08} && m_1 > m_2 + m_3\,, \hspace{1cm} \for \ \ \pbf \ne 0\,;
\\
\label{02052022-man-08a} && m_1 = m_2 + m_3\,, \hspace{1cm} \for \ \ \pbf = 0\,;
\\
\label{02052022-man-09}  && \pbf^2 = \frac{D}{4m_1^2}\,.
\eeq
It is the relation \rf{02052022-man-09} that, among other things, motivates us to use the quantity $D$ for the classification in \rf{23042022-man-12a1}-\rf{23042022-man-12a3}. Namely, the use of relation \rf{02052022-man-09} allows us to split all processes into three groups shown in \rf{23042022-man-12a1}-\rf{23042022-man-12a3}. We recall also that, for $m_1=m_2+m_3$, i.e., $D=0$, the particles momenta turn out to be collinear\,,%
\footnote{ Note that the equation $D=0$ has two solutions: $m_1=m_2 + m_3$ and $m_1=|m_2-m_3|$.}
\be
p_1^\mu = \frac{m_1}{m_2} p_2^\mu\,, \qquad p_3^\mu = \frac{m_3}{m_2} p_2^\mu\,, \qquad \for \ \ m_1 = m_2 + m_3\,.
\ee

\noinbf{Classification \rf{23042022-man-12a1}-\rf{23042022-man-12a3} via light-cone form of relativistic dynamics}. The $\rho^2$ \rf{20042022-man-15} can be represented as
\beq
\label{02052022-man-14}  && \rho^2 =   m_2^2 \beta_2^2 (r_+-r)(r-r_-)\,,
\\
\label{02052022-man-15} && r \equiv \beta_3/\beta_2\,,\hspace{1cm} r_\pm \equiv \big( m_1^2 -m_2^2 - m_3^2 \pm \sqrt{D}\big)/ 2m_2^2\,,
\\
\label{02052022-man-18}
&& D = (m_1^2 - m_2^2 - m_3^2)^2 - 4m_2^2m_3^2\,,
\eeq
where in \rf{02052022-man-18}, we represent $D$ defined in \rf{23042022-man-01}. It is the expression for $\rho^2$ in \rf{02052022-man-14} that motivates us to use $D$ for the classification \rf{23042022-man-12a1}-\rf{23042022-man-12a3}.
Namely, from \rf{23042022-man-16} we see that, for the real processes, the $\rho^2$ should be non-negative.  From \rf{02052022-man-14}, \rf{02052022-man-15}, we find then the restriction $D\geq 0$ for the real processes and the restriction $D<0$ for virtual processes.
For the real processes with non-collinear momenta, the on-shell condition $\rho>0$ \rf{23042022-man-18} and formula \rf{02052022-man-14} lead to the restrictions
\be \label{02052022-man-16}
r_- < r < r_+\,, \qquad   D > 0\,.
\ee
For out-going particles, in view of $\beta_2>0$, $\beta_3>0$ and the definition of $r$ \rf{02052022-man-15},  we get the restriction $r >0$. In view of $r>0$ and  $r_+>r$ \rf{02052022-man-16} we get the restriction $r_+>0$.
In turn, the restrictions $r_+>0$ and $D\geq0$ lead to the restriction
\be \label{02052022-man-17}
m_1^2 > m_2^2 + m_3^2\,.
\ee
Now using \rf{02052022-man-18}, we see that the restrictions \rf{02052022-man-17} and $D>0$ lead to the restriction for the masses in \rf{02052022-man-08}.
For $D=0$, we get $r_+=r_-\equiv r_0$ \rf{02052022-man-15}. Requirement $\rho^2\geq 0$ is then realized only for $r=r_0$. However, for $r=r_0$, we get $\rho^2=0$ and this leads to the collinear momenta. Note also that $D=0$ and \rf{02052022-man-17} imply the restriction on the masses in \rf{02052022-man-08a} as it should be for the collinear momenta.

\appendix{ \large Derivation of meromorphic vertices $\VVb$ }

We start with the following comment on equation \rf{21042022-man-22}. Let $\VVb_\sol^{(2)}$ be some solution of equation \rf{21042022-man-22}. Making the transformation
\be \label{31052022-man-01}
\VVb_\sol^{(2)} = \exp(\sum_{a=1,2,3} \omega_a M_a^\Lsm) \VVb_\sol^{(2)}(\omega)\,,
\ee
we find that a vertex $\VVb_\sol^{(2)}(\omega)$ should satisfy the following equation:
\beq
\label{31052022-man-02} && \sum_{a=1,2,3}\left\{2 M_a^\Rsm + \Bigl(2c_{\omega,a} m_a^2
-m_{a+1}^2 + m_{a+2}^2\Bigr) M_a^{\Rsm\Lsm}\right.\nonumber
\nonumber\\
&& + \left. \Bigl(m_a^2(c_{\omega,a}^2
-\frac{1}{4}) - (c_{\omega,a}-\half)m_{a+1}^2
+ (c_{\omega,a}+\half) m_{a+2}^2\Bigr) M_a^\Lsm\right\} \VVb_\sol^{(2)}(\omega)=0\,,\qquad
\\
\label{31052022-man-03} && c_{\omega,a} \equiv c_a - \omega_a \,.
\eeq
We see that equation \rf{31052022-man-02} is obtained from equation \rf{21042022-man-22} by the replacement $c_a\rightarrow c_{\omega,a}$ \rf{31052022-man-03}. We use this freedom in the choice of $c_a$. Namely, we find it convenient to use $c_a$ given in \rf{21042022-man-19}.

\noinbf{Derivation of \rf{26042022-man-03}}. For $m_1=0$, $m_2=0$, using $c_3=0$ \rf{21042022-man-19} and $M_a^\Rsm$, $M_a^\Lsm$, $M_a^{\Rsm\Lsm}$, $a=1,2$, given in \rf{18042022-man-22}, we see that equation \rf{21042022-man-22} takes the form
\be\label{12022022-81}
\Bigl( 2  M_3^\Rsm  +  m_3^2 (M_1^{\Rsm\Lsm} - M_2^{\Rsm\Lsm})
- \frac{m_3^2}{4}  M_3^\Lsm \Bigr) \VVb^{(2)}=0\,, \qquad  M_1^{\Rsm\Lsm}\equiv\lambda_1\,,\quad M_2^{\Rsm\Lsm}\equiv\lambda_2\,.
\ee
For the treatment of equation \rf{12022022-81}, we find it convenient to use the $\alpha$-representation for the operators $M_3^\Rsm$, $M_3^\Lsm$ given in \rf{25042022-man-07}. Doing so, we represent equation \rf{12022022-81} as
\be \label{12022022-84}
\Big(\big(\sqrt{2}  - \frac{m_3^2}{4\sqrt{2} } \alpha_3^2\big)\partial_{\alpha_3} + \frac{m_3^2s_3}{2\sqrt{2} }\alpha_3 + m_3 (\lambda_1 - \lambda_2)\Bigr) \VVb^{(2)} = 0\,.
\ee
General solution to equation \rf{12022022-84} is found to be
\be \label{12022022-82}
\VVb_{\lambda_1,\lambda_2}^{(2)} = y_{3+}^{s_3 -\lambda_1 +
\lambda_2}  y_{3-}^{s_3 + \lambda_1 - \lambda_2}\,,\hspace{1cm} y_{3\pm}\equiv 1 \pm \frac{m_3}{2\sqrt{2}}\alpha_3\,.
\ee
Using \rf{21042022-man-18}, \rf{25042022-man-14}, \rf{25042022-man-15} and making the transformation from the $\alpha$-representation to the $u,v$-representation \rf{25042022-man-05}, \rf{25042022-man-08a1}, we find that \rf{12022022-82} leads to \rf{26042022-man-03}.

Let us comment on the normalization factor $N_{\lambda_1\lambda_2}$ in \rf{26042022-man-02}. Under action of the operator $\II$, we get
\be  \label{14022022-01a1}
\II\, \VVb_{\lambda_1,\lambda_2}^\bas =  m_3^{2\lambda_1+2\lambda_2} 2^{-\lambda_1-\lambda_2}  \VVb_{-\lambda_1,-\lambda_2}^\bas\,.
\ee
Using \rf{14022022-01a1}, we note the more attractive relation
\be  \label{14022022-01a2}
\II\, (n_{\lambda_1,\lambda_2} \VVb_{\lambda_1,\lambda_2}^\bas) =   n_{-\lambda_1,-\lambda_2} \VVb_{-\lambda_1,-\lambda_2}^\bas\,,\qquad  n_{\lambda_1,\lambda_2} \equiv 2^{(\lambda_1+\lambda_2)/2} m_3^{-\lambda_1-\lambda_2}\,.
\ee
This is to say that using the normalization factor $n_{\lambda_1\lambda_2}$, we get the complex conjugation rule for the coupling constant in \rf{26042022-man-07}. Note that $N_{\lambda_1\lambda_2} = 2^{s_3/2} n_{\lambda_1\lambda_2} $. We inserted the extra factor $2^{s_3/2}$ to get   simple overall factor for the on-shell vertex  in \rf{26042022-man-09}.

\noinbf{Derivation of \rf{27042022-man-03}}. We use $c_a$, $a=1,2$ \rf{21042022-man-19} when $m_3=0$, and $M_3^{\Rsm\Lsm}$, $M_3^\Rsm$, $M_3^\Lsm$ given in \rf{18042022-man-22},
\be \label{14022022-01}
c_1 = \frac{m_2^2}{2m_1^2}\,, \qquad c_2 = -\frac{m_1^2}{2m_2^2}\,, \qquad M_3^\Rsm=0\,, \qquad  M_3^\Lsm=0\,, \qquad M_3^{\Rsm\Lsm}\equiv\lambda_3\,.
\ee
Using \rf{14022022-01} in equation \rf{21042022-man-22}, we obtain the equation
\be \label{14022022-02}
\Bigl( 2 M_1^\Rsm + 2 M_2^\Rsm - (m_1^2-m_2^2)
M_3^{\Rsm\Lsm} -\frac{\gamma^2}{m_1^2} M_1^\Lsm - \frac{\gamma^2}{m_2^2} M_2^\Lsm\Bigr)  \VVb^{(2)} =0\,,\qquad \gamma\equiv \half(m_1^2-m_2^2)\,.\qquad
\ee
For the treatment of equation \rf{14022022-02}, we use the operators $M_a^\Rsm$, $M_a^\Lsm$, $a=1,2$, given in \rf{25042022-man-07}. Doing so, we represent equation \rf{14022022-02} as
\be \label{14022022-02a}
\Big( (m_2^2-m_1^2) \lambda_3 + \sum_{a=1,2}  (\sqrt{2} m_ a
- \frac{\gamma^2}{\sqrt{2}m_a} \alpha_a^2 \big) \partial_{\alpha_a}    + \frac{\sqrt{2}\,s_a \gamma^2}{m_a} \alpha_a \Big)  \VVb^{(2)} =0 \,.
\ee
General solution to equation \rf{14022022-02a} is found to be
\be \label{14022022-04}
\VVb_{n_1,n_2,\lambda_3}^{(2)} =  \prod_{a=1,2}
y_{a+}^{s_a+n_a } y_{a-}^{s_a-n_a }\,, \hspace{1cm}  y_{a\pm} \equiv 1 \pm \frac{\gamma}{\sqrt{2} m_a}\alpha_a\,, \hspace{1cm} n_1 + n_2 = \lambda_3\,,\qquad
\ee
where $n_1,n_2\in \Zo$, while $\gamma$ is given in \rf{14022022-02}. Using \rf{21042022-man-18}, \rf{25042022-man-14}, \rf{25042022-man-15} and making the transformation from the $\alpha$-representation to the $u,v$-representation \rf{25042022-man-05}, \rf{25042022-man-08a1}, we find that \rf{14022022-04} leads to \rf{27042022-man-03}.

\noinbf{Derivation of \rf{28042022-man-03}}. This case is the particular case of the one above considered. Namely, plugging $m_1=m_2$
into equation \rf{14022022-02}, we see that equation \rf{14022022-02} is considerably simplified as
\be \label{15022022-39}
\sum_{a=1,2} M_a^\Rsm \VVb^{(2)} = 0 \,, \qquad \hbox{or equivalently as}  \qquad
\sum_{a=1,2} \partial_{\alpha_a} \VVb^{(2)} =0 \,,
\ee
where we use operator $M^\Rsm$ \rf{25042022-man-07}.  All solutions to equation \rf{15022022-39}, which are polynomial in $\alpha$, can be chosen as
\be  \label{15022022-40}
\VVb_n^{(2)} = X_\alpha^n \,, \hspace{1cm} X_\alpha \equiv \alpha_1 - \alpha_2 + c_0\,, \qquad n \in \No_0\,,
\ee
where $c_0$ is a real-valued constant. The freedom in the choice of $c_0$ corresponds to the freedom in the choice of a basis of the polynomials $X_\alpha^n$. We fix $c_0$ by looking for the simplest transformation of the vertex $\VVb_{n,\lambda_3}^\bas$ \rf{28042022-man-03} under the action of the operator $\II$. To this end, using \rf{21042022-man-18}, \rf{25042022-man-14} and making the transformation from the $\alpha$-representation to the $u,v$-representation \rf{25042022-man-05}, \rf{25042022-man-08a1}, we find that \rf{15022022-40} leads to
\beq \label{25042022-man-14a1}
&& \VVb_{n,\lambda_3}^\bas = X^n  \big(\frac{\Po^\Lsm}{\beta_3}\big)^{\lambda_3} \prod_{a=1,2} L_a^{2s_a}  \big(\frac{\Po^\Lsm}{\beta_a}\big)^{-s_a}\,, \qquad X \equiv \frac{v_1}{L_1} -  \frac{v_2}{L_3}  + c_0 \,,
\eeq
where $L_1$, $L_2$ are given in \rf{28042022-man-04}. By acting with the operator $\II$ on $X$, we get
\be \label{15022022-28}
\II\, X =  - X + 2c_0 + \frac{\sqrt{2}}{m}\,.
\ee
From \rf{15022022-28}, we see that the choice $c_0 = -1/\sqrt{2} m$ leads to the simplest transformation, $\II\, X = - X$. Using such $c_0$, we get $\VVb_{n,\lambda_3}^\bas$ in \rf{28042022-man-03}. Note that, in \rf{28042022-man-03}, in place of $X$ \rf{25042022-man-14a1}, we prefer to use $Q\equiv L_1 L_2 X$ \rf{28042022-man-06}. We note also the following helpful relations:
\be
\label{15022022-26}  \II\,L_1 = - \frac{m\beta_3}{\sqrt{2}\Po^\Lsm} L_1\,, \hspace{0.5cm} \II\,L_2 =  \frac{m\beta_3}{\sqrt{2}\Po^\Lsm} L_2\,,\hspace{0.5cm}
\II\, Q = \frac{m^2 \beta_3^2}{2 \Po^\Lsm\Po^\Lsm} Q\,.
\ee

\noinbf{Derivation of  \rf{29042022-man-04}}. Plugging $c_a$ \rf{21042022-man-19}
into equation \rf{21042022-man-22}, we get the following equation:

\be \label{16022022-02}
\sum_{a=1,2,3} (2  M_a^\Rsm
-\frac{D}{4m_a^2 }M_a^\Lsm )\VVb^{(2)} = 0 \,.
\ee
For the treatment of equation \rf{16022022-02}, we find it convenient to use the $\alpha$-representation for the operators $M^\Rsm$, $M^\Lsm$ given in  \rf{25042022-man-07}. Doing so, we cast equation \rf{16022022-02} into the form
\be \label{16022022-05}
\sum_{a=1,2,3} \Big( (\sqrt{2}  m_a
- \frac{D}{4\sqrt{2}m_a } \alpha_a^2 \big) \partial_{\alpha_a}  + \frac{D s_a }{2\sqrt{2}m_a } \alpha_a \Big)  \VVb^{(2)} =0 \,.
\ee
Making the transformation to a new vertex $\VVb^{(3)}$,
\be \label{16022022-19}
\VVb^{(2)} = \prod_{a=1,2,3} \Big(1 - \frac{D\alpha_a^2}{8m_a^2 }\Big)^{s_a}\, \VVb^{(3)}\,,
\ee
we find that equation \rf{16022022-05} amounts to the following equation for $\VVb^{(3)}$:
\be \label{16022022-20}
\sum_{a=1,2,3} \big(\sqrt{2} m_ a
- \frac{D}{4\sqrt{2}m_a} \alpha_a^2 \big) \partial_{\alpha_a}     \VVb^{(3)} =0 \,.
\ee
General solution to equation \rf{16022022-20} is found to be
\be \label{16022022-21}
\VVb_{n_1,n_2,n_3}^{(3)} = \prod_{a=1,2,3} y_{a+}^{n_a}y_{a-}^{-n_a}\,, \qquad n_1+n_2+n_3=0\,, \qquad y_{a\pm} \equiv 1 \pm \frac{\gamma}{\sqrt{2} m_a}\alpha_a \,,
\ee
where $n_1,n_2,n_3\in \Zo$, while $\gamma$ is defined in \rf{01052022-man-02}.
Using \rf{16022022-19}, we find then
\be \label{16022022-07}
\VVb_{n_1,n_2,n_3}^{(2)} = \prod_{a=1,2,3} y_{a+}^{s_a+n_a}y_{a-}^{s_a-n_a}\,.
\ee
Using \rf{21042022-man-18}, \rf{25042022-man-14} and making the transformation from the $\alpha$-representation to the $u,v$-representation \rf{25042022-man-05}, \rf{25042022-man-08a1}, we find that \rf{16022022-07} leads to \rf{29042022-man-04}.

\noinbf{Derivation of \rf{29042022-man-11}, \rf{29042022-man-12}}. Using relations for the action of the operator $\II$ given in \rf{24042022-man-11} and expressions for the normalization factors $N_{n_1,n_2,n_3}$ \rf{29042022-man-09}, we find
\beq
\label{30052022-man-01} && \hspace{-1cm} \II\,N_{n_1,n_2,n_3} \VVb_{n_1,n_2,n_3}^\bas =
N_{-n_1,-n_2,-n_3} \VVb_{-n_1,-n_2,-n_3}^\bas \,,   \hspace{0.9cm} \hbox{ for } \ D>0\,;\qquad
\\
\label{30052022-man-02}  && \hspace{-1cm} \II\,N_{n_1,n_2,n_3} \VVb_{n_1,n_2,n_3}^\bas =
N_{n_1,n_2,n_3} \VVb_{n_1,n_2,n_3}^\bas\,,  \hspace{2.3cm} \hbox{ for } \ D<0\,.
\eeq
From \rf{30052022-man-01}, \rf{30052022-man-02}, we see that in order to get the vertices $\VVb$ that satisfy the hermicity condition \rf{24042022-man-09}, we should use the vertices $\VVb$ given in \rf{29042022-man-02} and \rf{29042022-man-03}, where the coupling constants should satisfy the respective hermicity conditions \rf{29042022-man-11} and \rf{29042022-man-12}.

\noinbf{Derivation of \rf{30042022-man-03}}.  Plugging $c_a$ \rf{21042022-man-19}
into equation \rf{21042022-man-22}, we get equation \rf{16022022-02}. Using $D=0$ in equation \rf{16022022-02}, we get the equation
\be \label{17022022-02}
\sum_{a=1,2,3} M_a^\Rsm \VVb^{(2)} = 0 \,, \qquad \hbox{or equivalently}  \qquad
\sum_{a=1,2,3} m_ a \partial_{\alpha_a} \VVb^{(2)} =0 \,,
\ee
where we use the representation for $M^\Rsm$ given in \rf{25042022-man-07}. All solutions to equation \rf{17022022-02}, which are polynomial in $\alpha$, can be chosen as
\beq
\label{17022022-04} && \hspace{-1.2cm} \VVb_{n,l}^{(2)} = X_\alpha^n Y_\alpha^l\,, \qquad  X_\alpha  \equiv c_0^X  + \sum_{a=1,2,3} c_a^X \alpha_a\,, \qquad  Y_\alpha \equiv  c_0^Y +  \sum_{a=1,2,3} c_a^Y \alpha_a\,,
\\
\label{17022022-06} && \sum_{a=1,2,3} c_a^X m_a = 0 \,,\qquad  \sum_{a=1,2,3} c_a^Y m_a = 0 \,,
\eeq
where $n,l \in \No_0$, while $c_0^{X,Y}$, $c_a^{X,Y}$  are real-valued constants. We fix these constants in the following three steps.

\noinbf{Step 1}. Using \rf{21042022-man-18}, \rf{25042022-man-14} and making the transformation from the $\alpha$-representation to the $u,v$-representation \rf{25042022-man-05}, \rf{25042022-man-08a1}, we find that \rf{17022022-04} leads to
\beq
&& \label{17022022-06a1}  \VVb_{n,l}^\bas  = X_\Esm^n Y_\Esm^l \prod_{a=1,2,3} L_a^{2s_a} \big(\frac{\Po^\Lsm}{\beta_a}\big)^{-s_a}
\,,
\\
&& \label{17022022-06a2} X_\Esm \equiv c_0^X + \sum_{a=1,2,3} c_a^X \frac{v_a}{L_a}\,, \qquad  Y_\Esm \equiv c_0^Y + \sum_{a=1,2,3} c_a^Y \frac{v_a}{L_a}
\,,
\eeq
where $L_a$ are given in \rf{30042022-man-04}.

\noinbf{Step 2}. Introducing the notation $\XX = X_\Esm, Y_\Esm$,
\be \label{17022022-41}
\XX \equiv c_0^\XX +  \sum_{a=1,2,3} c_a^\XX \frac{v_a}{L_a}\,,
\ee
and by acting with the operator $\II$ on $\XX$, we get
\be \label{17022022-42}
\II\, \XX = - \XX + 2 c_0^\XX +  \frac{\sqrt{2}}{\Po_{\epsilon m}}  \sum_{a=1,2,3} c_a^\XX \epsilon_a \beta_a\,.
\ee
Requiring $\II\, \XX = - \XX$, we get the following equation
\be \label{17022022-44}
c_0^\XX +   \frac{1}{\sqrt{2}\,\Po_{\epsilon m}}  \sum_{a=1,2,3} c_a^\XX \epsilon_a \beta_a = 0 \,.
\ee
By using $\Pbf_{\epsilon m}=0$, we find that equation \rf{17022022-44} amounts to the two equations given by
\beq
\label{17022022-45} && c_1^\XX = \epsilon_1\epsilon_3 c_3^\XX + \sqrt{2} \epsilon_1\epsilon_2 m_2 c_0^\XX\,, \hspace{1cm} c_2^\XX = \epsilon_2\epsilon_3 c_3^\XX - \sqrt{2} \epsilon_1\epsilon_2 m_1 c_0^\XX\,.
\eeq

\noinbf{Step 3}. Equations \rf{17022022-45} have two solutions. We denote the 1st solution and the 2nd solution as $c_0$, $c_a^X$ and $c_0^Y$, $c_a^Y$ respectively. The 1st solution is fixed by the choice $c_0^X=0$. From \rf{17022022-45}, we get
\be \label{17022022-47}
c_1^X = \epsilon_1 \epsilon_3 c_3^X\,, \qquad c_2^X = \epsilon_2 \epsilon_3 c_3^X\,.
\ee
Choosing $c_3^X =  \epsilon_3$, we get $c_a^X$ given in \rf{30042022-man-05}. For the 1st solution, we find $\sum_{a=1,2,3}\epsilon_a c_a^X=3$.
We note then that the 2nd solution can be fixed by imposing the additional restriction
\be \label{17022022-49}
\sum_{a=1,2,3} \epsilon_a c_a^Y = 0 \,.
\ee
The additional restriction \rf{17022022-49} can be satisfied by the following replacement in equation \rf{17022022-44}:
\be \label{17022022-50}
c_a^Y \rightarrow c_a^Y - \frac{1}{3} \epsilon_a \sum_{b=1,2,3} c_b^Y \epsilon_b\,.
\ee
Note that equations \rf{17022022-06}, \rf{17022022-44}, \rf{17022022-45} are invariant under  replacement \rf{17022022-50}. Using \rf{17022022-45} and \rf{17022022-49},  we get
\be \label{17022022-51}
c_a^Y = \frac{\sqrt{2}}{3}c_0^Y \epsilon_a (\epsilon_{a+1}m_{a+1} - \epsilon_{a+2}m_{a+2})\,.
\ee
Choosing $c_0^Y = 1/\sqrt{2}$, we get $c_0^Y$, $c_a^Y$ given in \rf{30042022-man-07}. Note that $c_a^X$, $c_a^Y$ \rf{30042022-man-05},\rf{30042022-man-07} satisfy equations \rf{17022022-06}. Also we note that, in \rf{30042022-man-03}, in place of $X_\Esm$ and $Y_\Esm$ \rf{17022022-06a2}, we prefer to use $Q_X \equiv L_1 L_2 L_3 X_\Esm$ \rf{30042022-man-05} and $Q_Y \equiv L_1 L_2 L_3 Y_\Esm$ \rf{30042022-man-07} respectively.

\noinbf{Derivation of \rf{30042022-man-09}}. Using relations for the action of the operator $\II$ given in \rf{24042022-man-10} and the relations
\be \label{30052022-man-03}
\II\, L_a = \frac{\epsilon_a \Po_{\epsilon m}}{\sqrt{2} \Po^\Lsm} L_a\,, \qquad  \II\,\Po^\Lsm = -\frac{\Po_{\epsilon m}^2}{2 \Po^\Lsm}\,,\qquad \II\, X_\Esm = - X_\Esm\,,\qquad \II\, Y_\Esm = - Y_\Esm\,,
\ee
we find
\be \label{30052022-man-04}
\II\, N_{n,l} \VVb_{n,l}^\bas = N_{n,l} \VVb_{n,l}^\bas\,.
\ee

From \rf{30052022-man-04}, we see that in order to get the vertex $\VVb$ that satisfies the hermicity condition \rf{24042022-man-09}, we should use  the vertex  $\VVb$ given in \rf{30042022-man-02}, where the coupling constants should satisfy the hermicity condition given in \rf{30042022-man-09}. Note that the relation for $\II\,\Po^\Lsm$ in \rf{30052022-man-03} is the particular case of the relation for $\II\, \Po^\Lsm $ in \rf{24042022-man-10}, when $D=0$, $\Pbf_{\epsilon m}=0$. This can be seen by noticing the relation $\rho^2 = -\Po_{\epsilon m}^2$ when $D=0$, $\Pbf_{\epsilon m}=0$.


\end{document}